\documentclass[a4paper,11pt]{article}
\pdfoutput=0 
\usepackage{jheppub} 
\usepackage[section]{placeins}
\usepackage[T1]{fontenc} 
\usepackage{graphicx}
\usepackage{epsfig}
\usepackage{amsmath}
\usepackage{bm}
\usepackage{subfigure}
\usepackage{amssymb}
\usepackage{mathrsfs}
\usepackage{braket}
\usepackage{color}
\usepackage{comment}
\newcommand{\SCL}{0.3}
\newcommand{\comments}[1]{}

\title{\boldmath   Circuit-based digital adiabatic  quantum simulation and pseudoquantum simulation as new approaches to lattice gauge theory }

\author[a]{Xiaopeng Cui}
\author[a,1]{Yu Shi \note{Corresponding author.}}
\author[a,b]{Ji-Chong Yang}

\affiliation[a]{Department of Physics \&  State Key Laboratory of Surface Physics, Fudan University, Shanghai  200433, China}
\affiliation[b]{Department of Physics, Liaoning Normal University, Dalian 116029, China}

\emailAdd{16110190045@fudan.edu.cn}
\emailAdd{yushi@fudan.edu.cn}
\emailAdd{yangjichong@fudan.edu.cn}
\abstract{
Gauge theory is the framework of the Standard Model of particle physics and is also important in condensed matter physics.  As its major  non-perturbative approach, lattice gauge theory is traditionally implemented using Monte Carlo simulation, consequently it usually suffers such problems as the Fermion sign problem and the lack of real-time dynamics.  Hopefully they can be avoided by using quantum simulation, which simulates quantum systems by using controllable true quantum processes. The field of quantum simulation is under rapid development. Here we present a circuit-based digital scheme of quantum simulation of quantum $\mathbb{Z}_2$ lattice  gauge theory in $2+1$ and $3+1$ dimensions, using  quantum adiabatic algorithms implemented in terms of universal quantum gates.  Our algorithm generalizes the Trotter and symmetric decompositions to the case that the Hamiltonian varies at each step in the decomposition.
Furthermore, we carry through a complete demonstration of this scheme in  classical GPU simulator, and obtain key features of quantum $\mathbb{Z}_2$ lattice  gauge theory, including quantum phase transitions, topological properties, gauge invariance and duality. Hereby dubbed pseudoquantum simulation, classical demonstration of quantum simulation in  state-of-art fast computers  not only facilitates the development of schemes and algorithms of real quantum simulation, but also represents a new approach of practical computation.

\vspace{1cm}

{\em J. High Energ. Phys.} {\bf 2020}, 160 (2020). https://doi.org/10.1007/JHEP08(2020)160
}

\begin{document}

\maketitle

\flushbottom

\section{Introduction}

Quantum simulation and quantum computation can efficiently solve some problems that cannot be efficiently solved in classical computers~\cite{feynman,Lloyd1996,cirac,nori}, and is under extensive studies worldwide,  thanks to the rapid development of quantum science and technology. A controllable quantum  system, which may even be universal or programmable, simulates various quantum systems,  whose physical properties can be conveniently investigated with various parameter values.   Even   with only tens of qubits, far less than those   in full fault-tolerant quantum computing, and even in the presence of some noises, a quantum machine can perform some tasks  surpassing its classical counterparts,  exhibiting the so-called quantum supremacy~\cite{preskill,supremacy}. Many quantum simulations are such tasks. Hopefully they not only can solve specific problems, but also represent new scientific methods~\cite{cirac}.

Quantum simulations of important models in theoretical physics are enabled by some quantum algorithms, including the Trotter decomposition~\cite{Lloyd1996}, the sparse Hamiltonian quantum walk~\cite{Childs}, the dense Hamiltonian density matrix exponentiation~\cite{Lloyd2014,Rebentrost2018,Wossnig},  the adiabatic algorithm~\cite{adia,Hamma}, and so on. It is timely even to develop various quantum softwares~\cite{software}.

An important battlefield of quantum simulation appears to be lattice gauge theory (LGT)~\cite{reviews}, the major non-perpurbative approach to gauge theory. In particle physics, gauge theory is the framework of the Standard Model, describing both electroweak and strong interactions among elementary particles, and is also a guide beyond the Standard Model. A LGT is a gauge theory defined on a spacetime lattice in path integral formalism or on a space lattice in Hamiltonian formalism,  making the degrees of freedom countable and convenient for numerical calculations. Gauge theory is also important in condensed matter physics, where  the lattice can be a real structure. It is often an effective description of constraints in strongly correlated systems, and uses  emergent gauge fields to characterize topological orders, which exist  in fractional quantum Hall effect, spin liquids and possibly in high temperature superconductivity, as well as in topological quantum computing,  etc. Topological order represents a new and active paradigm beyond the traditional frameworks of  symmetry breaking and Fermi liquid theory.

Using Monte Carlo (MC) simulation, LGT has made great achievements~\cite{lgt,Kogut,lgtbooks}. But there are also difficulties, including  the lack of real-time dynamics because of the use of  Euclidean spacetime, and  the notorious Fermion sign problem~\cite{sign}, which was  shown to be  NP hard~\cite{troyer}. The cause of the  Fermion sign problem  is that in presence of Fermions, the Boltzmann weight  of an appropriately defined configuration, to which the quantum problem is mapped, may become negative, therefore  the  partition function  oscillates  violently, consequently  the  importance sampling in MC becomes invalid.   These difficulties are related to the unsolved problems in quantum chromodynamics, such as color confinement and phase diagram of quark-gluon plasma. The Fermion  sign problem exists in most of the MC-based methods such as quantum MC (QMC),  except   special algorithms for some specific models~\cite{yao}, and in some special issues to be told below.

As a new approach  avoiding Fermion sign problem  and an ideal avenue to study quantum phase transition (QPT) and real-time quantum dynamics, quantum  simulations  of LGTs are under study. They were first theoretically explored, mostly  but not exclusively, in cold atom platforms for the simulations of  U(N), SU(N)  and  $\mathbb{Z}_n$ LGTs~\cite{zoller,zohargroup,lewenstein,zohar2,
zohar3,lewenstein2}. Other proposals were also made~\cite{Byrnes2006,lamm}.  Among these schemes, some are analog, based on physical  mapping between the Hamiltonians of simulated and simulating systems, while others are digital, based on Trotter decomposition of the finite-time evolution to many small steps that are much easier to be implemented.  A quantum-classical algorithm was developed for two-site Schwinger model~\cite{Kolco}.

Experimentally, quantum simulations of (1+1)-dimensional quantum electrodynamics (QED)  or U(1) theory were performed using trapped ions~\cite{Martinez} and cold atoms~\cite{Kasper}. Initial attempts were also made in quantum simulation of  quantum  $\mathbb{Z}_2$ LGT using cold atoms~\cite{Schweizer}. A practical proposal using trapped ions was also made on quantum simulations of QED, Chern-Simons theory and $\mathbb{Z}_2$ theory~\cite{Davoudi}.

Quantum $\mathbb{Z}_2$ LGT is the simplest quantum LGT~\cite{Wegner,Kogut,fradkin,fradkinbook}. On one hand, $\mathbb{Z}_n$ theory   can be obtained as the discretization  of U(1) theory, suitable for quantum simulations~\cite{Ercolessi}. On the other hand, quantum $\mathbb{Z}_2$ LGT is also important in condensed matter physics~\cite{fradkinbook,Sachdev}.  $\mathbb{Z}_2$ toric code model~\cite{Kitaev}, which is important in topological quantum computing, can be regarded as a variant of quantum  $\mathbb{Z}_2$ LGT, and  has been experimentally demonstrated in a quantum simulation using nuclear magnetic resonance~\cite{Li,Luo}.

In two spatial dimensions, quantum $\mathbb{Z}_2$ LGT  is dual to quantum Ising model in a transverse field~\cite{Wegner,Kogut,fradkin,fradkinbook}, which is thus often invoked for QPT properties, especially the critical point~\cite{Rieger,Hamer,Blote2002,evenbly}. Besides, $\mathbb{Z}_n$ theory in one spatial dimension was studied by using density matrix renormalization group (DMRG)~\cite{Ercolessi}, and  was studied in two spatial dimensions  by using tensor network techniques, as the low energy limit of toric model in a magnetic field  under the constraint of gauge invariance~\cite{Vidal}. Direct DMRG study of quantum  $\mathbb{Z}_2$ LGT  is difficult  as realizing plaquette interactions consistently with the gauge symmetry is challenging. But such study is technically possible for Abelian and non-Abelian gauge theory in 2D by using symmetry-preserving tensor network techniques~\cite{Tagliacozzo2}. Interestingly, the duality between such a spin gauge theory and a generalized Ising model allows for scalable quantum simulation with Rydberg atoms~\cite{Celi2}.

Coupling of $\mathbb{Z}_2$ gauge field  with various kinds of matter  has  also  been studied, starting with Ising  matter~\cite{fradkin}. Fractionalization of electrons is obtained in   theories of strong correlations with   $\mathbb{Z}_2$ gauge fields, giving rise to the so-called orthogonal metals~\cite{senthil}. Interestingly, the issue of Fermions coupled with  $\mathbb{Z}_2$ gauge field is exactly one of the special issues  free of sign problem in MC-based methods~\cite{Trebst}, another issue being Fermions with an even number of flavors~\cite{Dagotto}. The cases with both of  these characteristics were studied by using QMC~\cite{Gazit}.  Coupling of Fermions with $\mathbb{Z}_2$ gauge field with Gauss law not imposed but emerged was also studied by using QMC~\cite{Assaad}. These QMC studies were all in two spatial dimensions.  A modified $\mathbb{Z}_2$ gauge theory coupled with Fermions was studied analytically  in one  and two spatial  dimensions~\cite{Prosko}.  One-dimensional  Bosons coupled with $\mathbb{Z}_2$  gauge field with Gauss law not imposed but emerged was studied also by using DMRG~\cite{Cuadra}. A quantum link model of QED was approached by using tensor network method~\cite{Felser}.

\begin{figure}[htb]
\centering
\subfigure[]{}
\includegraphics[scale=0.6]{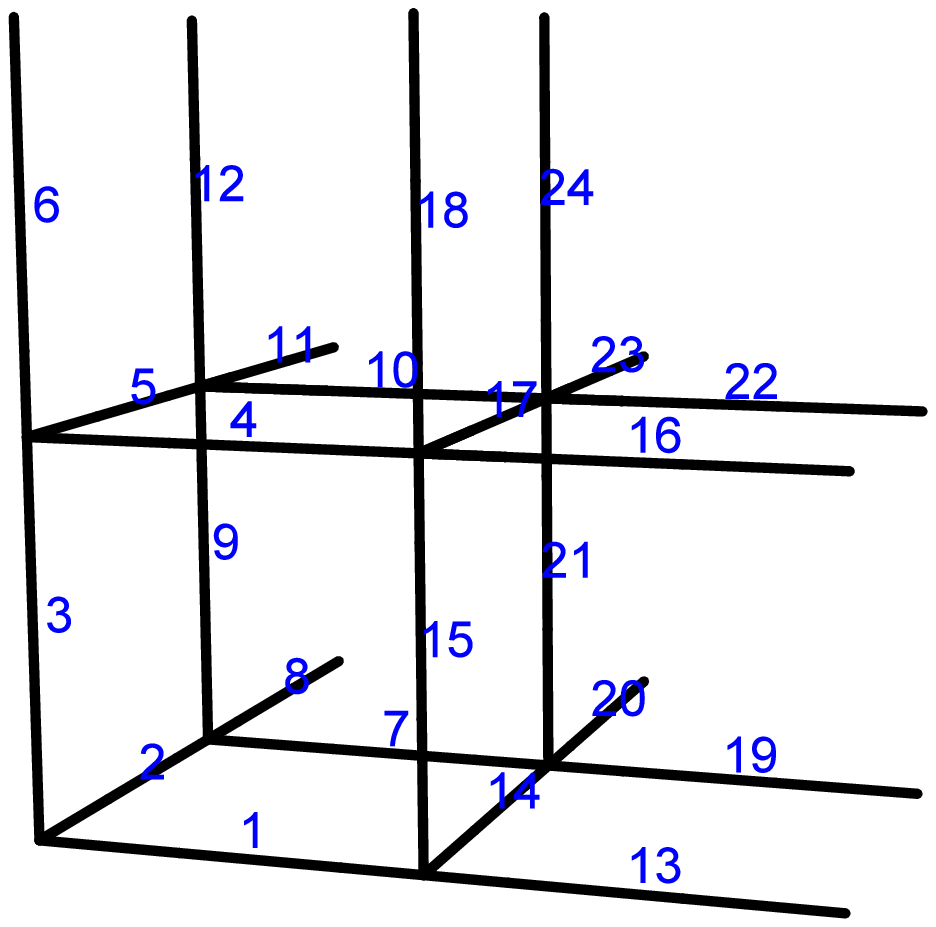}
\subfigure[]{}
\includegraphics[scale=0.6]{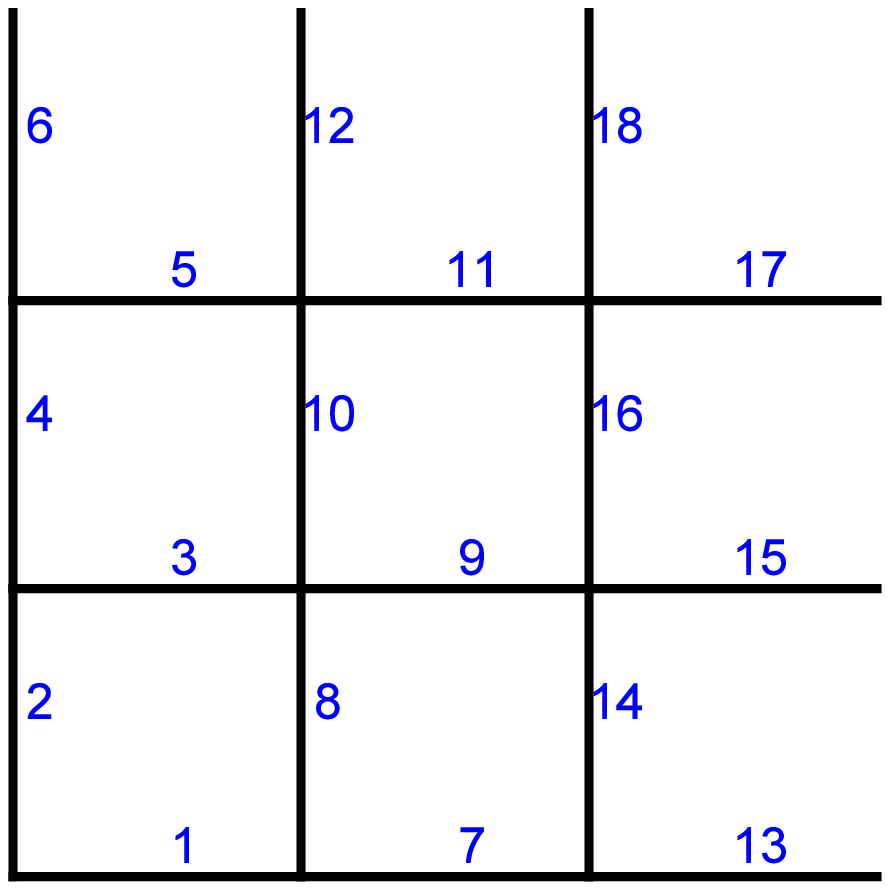}
\caption{(a) Three-dimensional $2\times2\times2$ lattice with periodic boundary condition.   (b)  two-dimensional   $3\times 3$ lattice with periodic boundary condition.  There is one qubit on each link,   with an ordering numbers indicated.  In (a), for example,  qubits numbered 1, 3, 4 and 15 form an elementary plaquette, while qubits numbered 4, 6, 1, and 18 form another, and there are $24$ qubits.  In (b), for example, qubits numbered 3, 4, 5 and 10 form an elementary plaquette, while qubits numbered 5, 6, 1, and 12 form another, and there are 18 qubits. }
\label{fig_lattice}
\end{figure}

In quantum  $\mathbb{Z}_2$ LGT, defined on a square lattice, each link is occupied by one qubit.  In the $2\times2\times2$ lattice, which is the smallest three-dimensional lattice, there are 24 links (Fig.~\ref{fig_lattice}).   In the $3\times3$ lattice, the second smallest two-dimensional lattice, there are 18 links  (Fig.~\ref{fig_lattice}).  In a quantum algorithm, additional qubits  may also be needed as ancillas  in simulating the adiabatic evolution, and in phase estimation simulating the  measurement, and so on.  Symmetries may reduce some degrees of freedom, but at the price of introducing nonlocal interactions, for example. Regarding the quantum adiabatic algorithm, tens of thousands steps are needed in  Trotter decomposition. Therefore, at present time, it seems  difficult for the experimental platforms to fully meet the  requirements on the qubit number, error rate and coherence time.

Therefore, it appears interesting to  use classical high-performance computing platforms to demonstrate  quantum simulation in general, and that of quantum  $\mathbb{Z}_2$ LGT in particular. We call it pseudoquantum simulation, which serves as a benchmark for real quantum simulation, and facilitates the development of quantum algorithm and  quantum softwares. Meanwhile, it is also a new  method of computation and simulation,  providing useful results on the computed problems.

Such pseudoquantum simulation is realistic if it is run on a fast enough classical computing platform. An example is the latest  graphics processing unit (GPU) parallel   computing architecture, which  greatly accelerates large-scale complex scientific computation. Indeed, a GPU simulator  called Quantum Exact Simulation Toolkit (QuEST) has been developed, as a new software platform simulating the quantum circuit \cite{quest}. It is based on the software platform and application programming interface CUDA created by Nvidia, which allows the development of parallel program using a CUDA-enabled GPU.  QuEST is designed as a C library, and allows quantum codes to be deployed in a variety of computing platforms. With high precision, it can simulate 29 to 31 qubits on a single Nvidia GPU card, as detailed below. Therefore it fits well the need of our  pseudoquantum simulation of quantum  $\mathbb{Z}_2$ LGT.

In this paper, we present a scheme of circuit-based digital quantum simulation  of (2+1)-dimensional and (3+1)-dimensional quantum  $\mathbb{Z}_2$ LGT. We use universal quantum circuit to implement the quantum adiabatic algorithm, and we generalize the Trotter decomposition and symmetrized Trotter  decomposition to the case of adiabatic varying the Hamiltonian in each step of decomposition. As the first work  and proof of principle, here we consider pure gauge theory, without coupling to Fermions yet. Moreover,  we perform a complete pseudoquantum simulation, that is,  we classically demonstrate the scheme of quantum simulation, by using QuEST  in Nvidia Tesla K40m and V100 GPU cards, which operate 1.682TFLOPS and 7.834TFLOPS, with double precisions, respectively. Our work demonstrates the advantages of  quantum simulation as well as the usefulness of pseudoquantum simulation.  Meanwhile we  obtain useful results regarding  various properties of   quantum  $\mathbb{Z}_2$ LGT.  It seems to contain the first numerical result on quantum  $\mathbb{Z}_2$ LGT in three spatial dimensions, including the  features of first-order QPT, as hinted by  the fact  that the thermal phase transition in the classical $\mathbb{Z}_2$ LGT  is first-order  in  4 spatial  dimensions~\cite{Creutz}.

The rest of the paper is organized as the following. In Sec.~\ref{scheme}, we introduce the quantum  $\mathbb{Z}_2$ LGT, the lattices we consider, the quantum adiabatic algorithms and their realization in terms of the quantum circuits, our generalization of the Trotter and symmetrized Trotter  decompositions, the simulation of measurement, as well as the hardware we use to demonstrate the scheme of quantum simulation. In Sec.~\ref{results}, we present the results from our GPU computation simulating the quantum simulation, including the expectations of Wegner-Wilson loop operators and of the Hamiltonian, which are used to determine the critical points and orders of the QPTs, and confirm the self-duality in 3+1 dimensions. We  also calculate the densities of states, which confirm  the gauge invariance,  the  self-duality in 3+1 dimensions and its absence in 2+1 dimensions. We will also present the evidences of topological nature of QPT. Finally, a summary is made in Sec.~\ref{summary}.

\section{ Schemes and algorithms of quantum  simulation and pseudoquantum simulation   \label{scheme} }

\subsection{Quantum $\mathbb{Z}_2$ lattice gauge theory }

$\mathbb{Z}_2$ LGT is defined on a lattice. It was first proposed  as a generalization of Ising model, elevating the global up-down symmetry to a local symmetry, but without spontaneous magnetization at the phase transition, which is characterized by onset of topological order rather than symmetry breaking. So there is no local order parameter~\cite{Wegner}.  On the other hand, with some differences in details,  $\mathbb{Z}_n$ LGT  can also be obtained by discretizing U(1) gauge theory, by defining the matter field on the sites of a lattice, and the $n$-valued gauge potential and thus electric field on the links between sites.

Here we focus on the quantum  $\mathbb{Z}_2$ LGT with the Hamiltonian~\cite{fradkinbook,Sachdev}
\begin{equation}
 H = Z + gX,
\end{equation}
where $g$ is the coupling constant,
\begin{equation}
X \equiv  -\sum_l { \sigma_l^x },\,\,\,
Z \equiv \sum_{\square} { Z_\square},\,\,\,Z_\square \equiv  -\prod_{l \in \square}{  \sigma_l^z },
\end{equation}
with  $Z_\square$
defined for each elementary plaquette  (the smallest square)  $\square$ (Fig.~\ref{fig_lattice}).   For  qubit $l$,  $\sigma^z_l|0\rangle_l = |0\rangle_l$, $\sigma^z_l|1\rangle_l = - |1\rangle_l$. The Hamiltonian of the classical $\mathbb{Z}_2$ theory is $Z$ only, with each operator $\sigma_l^z$ reduced to a classical variable.  The quantum nature of $H$ is due to the noncommutativity between $\sigma_l^x$  and $\sigma_l^z$, resulting the competition between $Z$ and $X$, in a way like that between energy and entropy in a thermal phase transition. The coupling constant $g$ is a control parameter, playing  a  role in  QPT similar to the role of temperature in thermal phase transition.

For convenience, in our numerics, the links and thus the qubits are numbered  as  in Fig.~\ref{fig_lattice}.    For three spatial dimensions d=3, we   use $ 2\times2\times2$ lattice with periodic boundary condition, where there are 24 links and 24 elementary plaquettes. We  use 25 qubits, one of which is the ancilla. For two spatial dimensions d=2,  we use $3 \times 3$  lattice with periodic boundary condition, where there are 18 links and 9 elementary plaquettes. We use 19 qubits, one of which is the  ancilla. We assume periodic boundary condition, that is, each lattice is  a torus.   Although the lattice sizes are very small,  the key features of the quantum $\mathbb{Z}_2$  LGT do appear. In general, for a $d$-dimensional square lattice with linear size $L$, the number of links is $N_l=dL^d$, the number of  plaquettes  is $N_p=N_l(d-1)/2$.

This theory possesses $\mathbb{Z}_2$ gauge invariance, similar to Gauss law,  dictating  that each eigenstate  $|\psi\rangle$ of  $H$ must satisfy
\begin{equation}
G_i  |\psi\rangle =  |\psi\rangle
\end{equation}
where $G_i \equiv \prod _{l \ni i} \sigma^x_l$ is the product of the  $\sigma^x_l$'s on all the links  ending at each lattice point $i$.

\begin{figure}[htb]
  \centering
  \subfigure[]{}
  \includegraphics[scale=0.3]{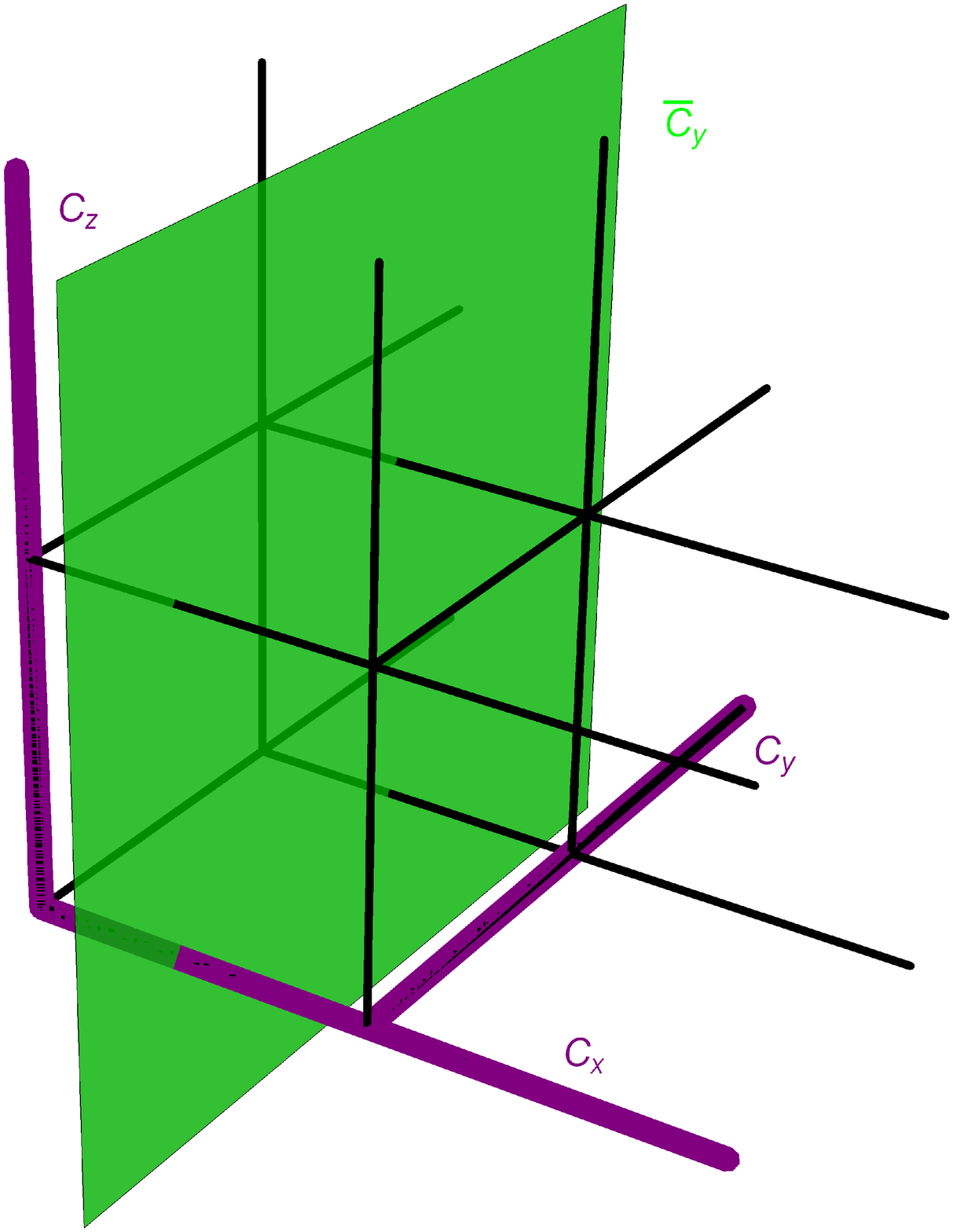}
  \subfigure[]{}
  \includegraphics[scale=0.4]{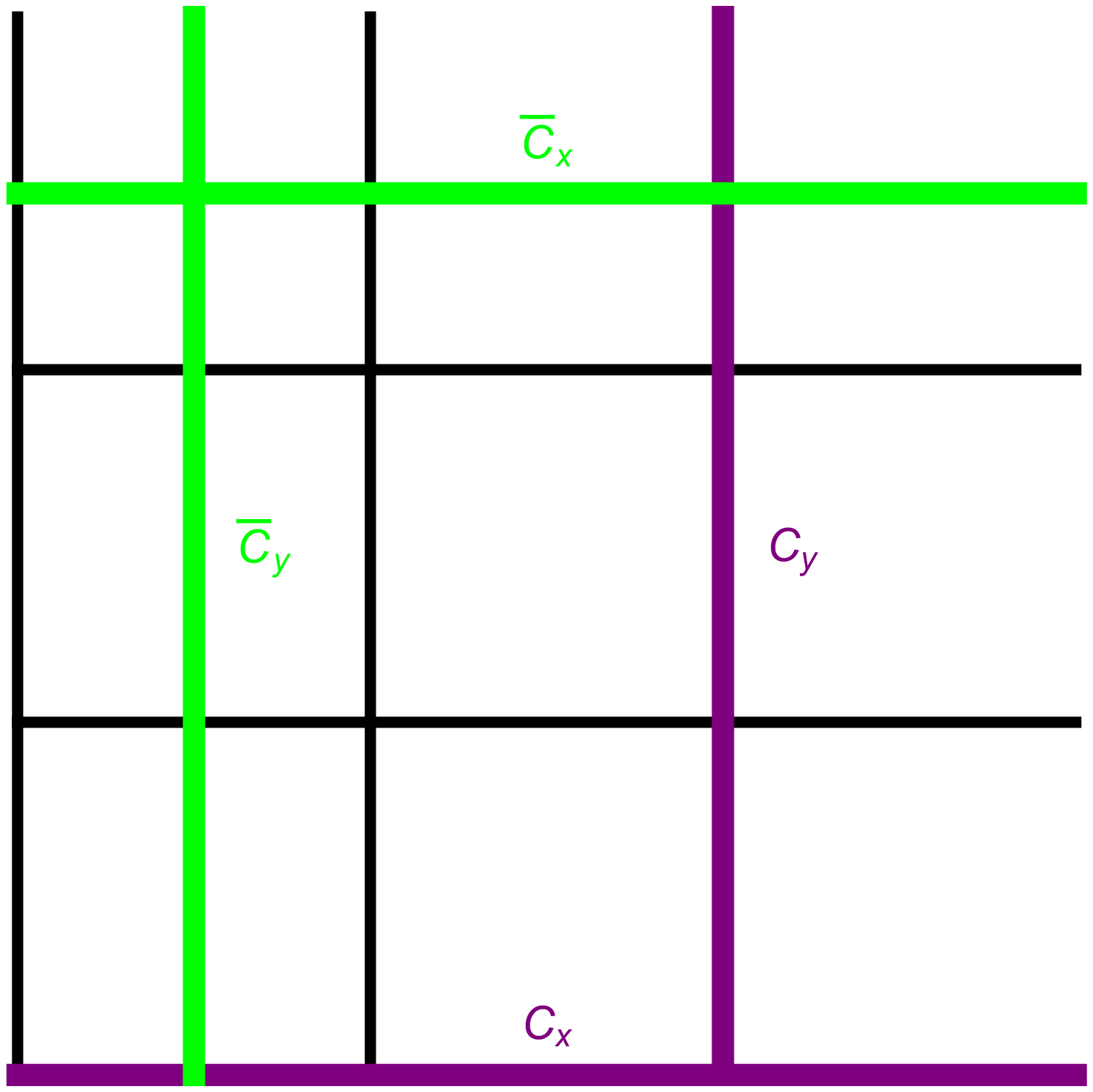}
  \caption{(a) d=3 Lattice, with  the indication of  non-contractible loops $C_{x}$, ${C}_{y}$ and ${C}_{z}$ on the direct lattice, and  non-contractible  $ \overline{C}_{y}$ on the  dual lattice. $C_{x}$, ${C}_{y}$ and ${C}_{z}$ are used in defining   specific Wegner-Wilson operators  $W_x$, $W_y$ and $W_z$.    $\overline{C}_{x}$ and $\overline{C}_{z}$ are similar to $ \overline{C}_{y}$. They are used in defining 't Hooft operators  $V_x$, $V_y$, $V_z$.       (b) d=2 Lattice, with  non-contractible loops  $C_{x}$ and $C_{y}$ on the direct lattice, used in defining specific Wegner-Wilson loop operators $W_x$ and $W_y$, as well as  non-contractible loops   $\overline{C}_{x}$ and  $\overline{C}_{y}$ in defining  't Hooft operators   $V_x$ and $V_y$. }
  \label{fig_torus_lattice}
\end{figure}

One gauge invariant operator is   Wegner-Wilson loop operator, which is defined as
\begin{equation}
W_{C} = \prod_{l\in  {C}} \sigma^z_l
\end{equation}
along a closed loop  ${C}$ on the direct lattice.  $G_i W_{C} G_i^{-1}=W_{C}$. A  Wegner-Wilson loop operator is  not necessarily  along a  non-contractible loop, for example,  $Z_{\square}$  is a Wegner-Wilson loop.  But  those along non-contractible loops play special roles. Generalizing  the well known    case in d=2,  here we define the special   Wegner-Wilson loop operators
\begin{equation}
W_\mu = \prod_{l\in  {C}_\mu} \sigma^z_l,
\end{equation}
along non-contractible loop  ${C}_\mu$,  $\mu=1,\cdots,d$  (Fig.~\ref{fig_torus_lattice}). Periodic boundary conditions means that topologically they  encircle a torus. As far as it encircles the lattice in $\mu$ direction, the details of  ${C}_\mu$ does not matter.

$\sigma^x$ and any product of  $\sigma^x$'s  are also gauge invariant.   $G_i\sigma^x G_i^{-1}=\sigma^x$. Generalizing  the well known case  in d=2, here we  make a definition of the so-called  't Hooft loop operator in  any dimention d,
\begin{equation}
V_{\mu} \equiv \prod_{l\in \overline{C}_{\mu}} {\sigma_l^x},
\end{equation}
which is a product of $\sigma_l^x$ pierced by a non-contractible $(d-1)$-dimensional surface   $\overline{C}_{\mu}$  on the dual lattice, $\mu=1,\cdots,d$   (Fig.~\ref{fig_torus_lattice}).  A   non-contractible $(d-1)$-dimensional surface  can be deformed by using $G_i$ operators, because of  gauge invariance.  $\overline{C}_{\mu}$  is  subscripted as $\mu$ as it can be deformed,  if needed, to be parallel to  ${C}_{\mu}$. Periodic boundary conditions means that topologically they  encircle a torus. In d=2, $\overline{C}_{\mu}$ is a loop.

$V_{\mu} $   commutes with  $H$, while $W_{\mu}$ does not unless $g=0$. For $\mu\neq \nu$,
\begin{equation}
   W_\mu V_\nu =-V_\nu W_\mu.
\end{equation}
Consequently  $W_\mu$   acting on an eigenstate of   $V_\nu$ ($\nu\neq \mu$)   yields another eigenstate of $V_\nu$, with  the eigenvalue of the opposite sign. Therefore,  from the state with $|V_{\nu}=1,\nu=1,\cdots,d\rangle$, using consecutive actions of $W_\mu$'s, one can generate   $|V_{\nu}=-1, V_{d-\nu}=1, \nu=1,\cdots m\rangle = \prod_{\nu=1}^m W_{\nu} |V_{\nu}=1,\nu=1,\cdots,d\rangle$, $m=1,\cdots,d$, hence $\sum_{m=1}^d C_n^m= 2^d-1$ states in total can be generated.    These $2^d$ common eigenstates $|V_1,\cdots,V_d\rangle$ of $V_\mu$ with eigenvalues  $\pm 1$, ($\mu=1,\cdots,d$), are also eigenstates of $H$. They are degenerate ground states when $g=0$. The degeneracy on a d-dimensional lattice with genus ${\cal G}$ is $2^{d{\cal G}}$. On the lattice considered here with period boundary condition, the degeneracy is $2^d$. The degeneracy changes  when $g \neq 0$, with  only $|V_{\nu}=1,\nu=1,\cdots,d\rangle$ remaining as the ground state. But they  represent different topological sectors, because $V_\mu$ is conserved  while each non-contractible $(d-1)$-dimensional surface can deform.

Depending on $g$, there are two different phases, deconfined phase at small  $g$  and confined phase  at large $g$, separated by a QPT at the critical point $g_c$, where there is a phase transition, which however cannot be characterized by a change of symmetry.   There is $\mathbb{Z}_2$ topological order in the deconfined  phase, while the confined  phase is trivial, and QPT in this theory is  topological~\cite{Wegner,Sachdev,fradkinbook}.

\subsection{Circuit-based quantum adiabatic algorithm \label{algo} }

Our digital scheme of quantum simulation of the  quantum  $\mathbb{Z}_2$ LGT is based on the quantum adiabatic algorithm, and we use   quantum circuits consisting of one-qubit and two-qubit quantum gates to implement it.  In  the adiabatic evolution, $g$ varies from $0$  to a large enough value $g_f$, passing  $g_c$. The slow  variation  allows  the system to adapt to the instantaneous ground state. According to the adiabatic theorem, the state of the system,  starting as a ground state of the initial Hamiltonian $H(g=0)=Z$, evolves as ground state of $H(g)$, and  ends up as the ground state of the final Hamiltonian $H(g_f)$.

In our simulation, each time $g$ is updated, the state evolves for a very short time, as realized by the quantum circuit under the present $g$ value, then $g$ is updated again, and the state evolves under the new value of $g$. The iteration continues until a final value $g_f$ of $g$. Therefore, the adiabatically varying Hamiltonian is  stepwise. We divide the evolution to $N_s$ steps, and each step is further divided to $n$ substeps, and $g$ varies at each substep. This  so-called ``substep''  really corresponds to the ``step'' in  Trotter  decomposition.   The reason of referring to the decomposition steps as substeps is that the evolution is paused after a period called step,  and calculations, or called pseudo-measurements, are done, afterwards the evolution is resumed. In real quantum simulation, the measurement  depends on the actual situation.

Therefore,  the stepwise Hamiltonian, for the $m$-th substep within  $k$-th step, is
\begin{equation}
H_{k,m}=Z+g_{k,m}X,
\end{equation}
where
\begin{equation}
g_{k,m}= (k-1)g_s+m\delta,
\end{equation}
$(m=1,\cdots,n)$,  $n$ is the total number of substeps in each step, which is freely set. $g_s$ is the increase of $g$ in each step, which lasts time $t_s$,  $\delta=g_s/n$ is the increase of $g$ in each substep.  The total  number of steps is $N_s=g_f/g_s$, and the total number of substeps is $nN_s$.

The evolution in  $k$-th   step  is \begin{equation}
\displaystyle\prod_{m=1}^{n} e^{-iH_{k,m}\frac{t_s}{n} },
\end{equation}
while the total evolution is \begin{equation}\displaystyle\prod_{k=1}^{N_s}\prod_{m=1}^{n} e^{-iH_{k,m}\frac{t_s}{n}}.
\end{equation}

The evolution in each substep of $t_s/n$,  under $H_{k,m}$, consisting  of  two noncommutative parts $Z$ and $g_{k,m}X$,   can be decomposed, in an asymmetric way, into  consecutive evolution of  $Z$ and $g_{k,m}X$,
\begin{equation}
   e^{ -i H_{k,m} \frac{t_s}{n}  }  \approx  e^{-i Z  \frac{t_s}{n} }e^{-ig_{k,m} X \frac{t_s}{n} }. \label{a}
\end{equation}
Therefore the evolution in $k$-th step is
\begin{equation}
   \prod_{m=1}^{n} \left( e^{-i Z  \frac{t_s}{n} }e^{-ig_{k,m} X \frac{t_s}{n} }\right).
\end{equation}
which generalizes Trotter decomposition. It reduces to Trotter decomposition if $H_{k,m}$ is independent of $m$.

From the  identity $e^{A+B}=e^{A}e^{B}e^{-\frac{1}{2}[A,B]+\cdots}$ for two operators $A$ and $B$, we have
$e^{-i(H_1+H_2)\tau}=e^{-iH_1\tau}e^{-iH_2\tau}
e^{\frac{\tau^2}{2}[H_1,H_2]+\cdots}$, thus
\begin{equation}
 ||e^{-i(H_1+H_2)\tau}-e^{-iH_1\tau}e^{-iH_2\tau}|| \approx  \frac{\tau^2}{2}||[H_1,H_2]||.
\end{equation}
Therefore, the error for one substep, in the asymmetric decomposition (\ref{a}), is
\begin{equation}
 \Delta^{asy}_{k,m}\equiv ||  e^{ -i H_{k,m} \frac{t_s}{n}  } - e^{-i Z  \frac{t_s}{n} }e^{-i g_{k,m}X \frac{t_s}{n} } ||
 \\ \approx 2g_{k,m}N_p\frac{t_s^2}{n^2},  \label{error1}
\end{equation}
where we have considered that each $Z_\square$ is noncommutative with 4 $\sigma^x$'s. It is calculated that
\begin{equation}
\sum_{k=1}^{N_s}\sum_{m=1}^{n} g_{k,m} = \frac{1}{2}(N_s^2n+N_s)g_s\approx N_s^2n g_s/2.
\end{equation}
Therefore the total error of the asymmetric decomposition is
\begin{equation}
\Delta^{asy}=\displaystyle\sum_{k=1}^{N_s}
\sum_{m=1}^{n}
\Delta^{asy}_{k,m}=
(N_s^2+\frac{N_s}{n})N_pg_s\frac{t_s^2}{n} \approx N_s^2N_pg_s\frac{t_s^2}{n}. \label{asymtotal}
\end{equation}

We have also used the symmetric    decomposition
\begin{equation}
  e^{ -i H_{k,m} \frac{t_s}{n}  }  \approx   e^{-i Z \frac{t_s}{2n}} e^{-ig_{k,m} X \frac{t_s}{n}} e^{-i Z  \frac{t_s}{2n}}.  \label{s}
\end{equation}
Therefore the evolution in $k$-th step is
\begin{equation}
   \prod_{m=1}^{n} \left(e^{-i Z \frac{t_s}{2n}} e^{-ig_{k,m} X \frac{t_s}{n}} e^{-i Z  \frac{t_s}{2n}}\right),
\end{equation}
which generalizes the symmetrized Trotter decompositon. It can be rewritten as
\begin{equation}
 e^{-i Z  \frac{t_s}{2n}} \prod_{m=1}^{n-1} \left( e^{-ig_{k,m}X \frac{t_s}{n}} e^{-i Z  \frac{t_s}{n}} \right)  e^{-ig_{k,n} X \frac{t_s}{n} }  e^{-i Z \frac{t_s}{2n}},
\end{equation}
which indicates a more convenient way of execution in our computation.

Using $\ln(e^{A/2}e^{B}e^{A/2})= A+B-([A,[A,B]]+2[B,[A,B]])/24+\cdots
$~\cite{Suzuki}, we   find
\begin{equation}
   e^{ -i H_{k,m} \frac{t_s}{n}  } -       e^{-i Z \frac{t_s}{2n}} e^{-ig_{k,m} X \frac{t_s}{n}} e^{-i Z  \frac{t_s}{2n}} \approx \frac{i}{24}\frac{t_s^3}{n^3}\left(g_{k,m}
[Z,[Z,X]]+2g_{k,m}^2[X,[Z,X]]\right), \label{error2}
\end{equation}
where    $[Z,[Z,X]]$ is of the order of $8(d-1)N_p$, while  $[X,[Z,X]]$ is of the order of $16N_p$, for the following reason.
Each $Z_\square$ is noncommutative with 4 $\sigma^x$'s, hence  $[Z_\square,X]$ is the sum of 4 products of one $\sigma^y$ and 3 $\sigma^z$'s. Each  $\sigma^y$ is noncommutative with the $2(d-1)$ $Z_\square$'s of the plaquettes sharing with the link $l$.  On the other hand, each  product of one $\sigma^y$ and 3 $\sigma^z$'s is noncommutative with 4 $\sigma^x$'s. Therefore  $[Z,[Z,X]]=O[8(d-1)N_p]$, while  $[X,[Z,X]]=O(16N_p)$. Another way of reasoning is the following. Each $\sigma^x$ is shared by $2(d-1)$ plaquettes, thus $[Z,\sigma^x_l]$ yields $2(d-1)$  products of $\sigma^y$ and 3 $\sigma^z$'s. Each  product  is   noncommutative with $Z_\square$'s of  the  $2(d-1)$ plaquettes, and with the 4 $\sigma^x$'s on the same plaquette. Consequently,   $[Z,[Z,X]]=O[4(d-1)^2N_l]$,   $[X,[Z,X]]=O[8(d-1)N_l]$. With $N_p=N_l(d-1)/2$, this is the same as above.

Therefore  the error, in the  symmetric decomposition (\ref{s}),    is
\begin{equation}
 \Delta^{sym}_{k,m} \equiv || e^{ -i H_{k,m} \frac{t_s}{n}  } -      e^{-i Z \frac{t_s}{2n}} e^{-ig_{k,m} X \frac{t_s}{n}} e^{-i Z  \frac{t_s}{2n}}  ||\approx  \frac{(d-1)g_{k,m}+4g_{k,m}^2}{3}N_p
 \frac{t_s^3}{n^3}. \label{symmerror}
\end{equation}

It  is calculated that
\begin{eqnarray}
\sum_{k=1}^{N_s}\sum_{m=1}^{n} g_{k,m}^2&=&\frac{(N_s-1)N_s(2N_s-1)n g_s^2}{6}+
\frac{N_sn(n+1)(2n+1)g_s^2}{6n^2}+\frac{N_s(
N_s-1)(n+1)g_s^2}{2} \nonumber \\ &
\approx & \frac{N_s^3n g_s^2}{3}.
\end{eqnarray}
Therefore the total error of the  symmetric decomposition is
\begin{equation}
\Delta^{sym}=\displaystyle\sum_{k=1}^{N_s}
\sum_{m=1}^{n} \Delta^{sym}_{k,m}
 =[\frac{(d-1)}{6}N_s^2g_s+\frac{4}{9}N_s^3g_s^2]
 N_p\frac{t_s^3}{n^2}\approx \frac{4}{9}N_s^3
 N_pg_s^2\frac{t_s^3}{n^2}. \label{total}
\end{equation}
The ratio of  the errors of the symmetric  and asymmetric decompositions is $\sim g_f ts/n$.   With $g_f=O(1)$,  the total error of the symmetric decomposition is less than the asymmetric one by a factor of $t_s/n$.
We have done our simulations   using both  decompositions. The results from the  symmetrized decomposition is  clearly better and thus  presented below.

Note that our decompositions are different from, and generalize, the usual Trotter decomposition and symmetrized Trotter decomposition, each of which repeats a constant evolution for a number of times.

We now discuss how to implement the evolution in each substep, or called each  decomposition step. First,  $e^{-i Z t_s/n}=
\prod_\square e^{-i Z_\square  t_s/n}$, hence the evolution of $Z$ can be realized by the consecutive evolution of all the plaquettes.

\begin{figure}
\centering	
\includegraphics[scale=0.7]{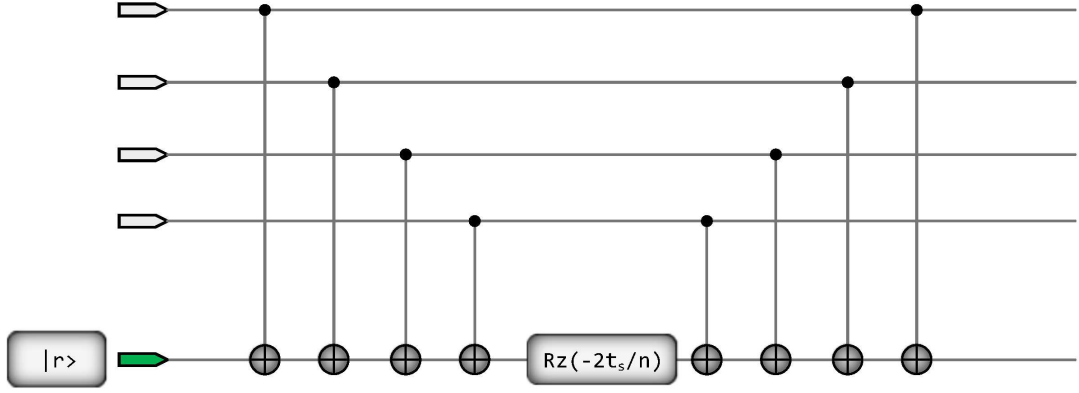}
\caption{
\label{fig_Z2}  The quantum circuit realizing   $e^{-i Z_\square t_s/n}$ or  $e^{i Z_\square t_s/n}$, depending on whether the  initial state of the ancila  $|r\rangle$ is set to be  $|0\rangle$ or $|1\rangle$.  }
\end{figure}

Evolution of $e^{-i Z_\square t_s/n}$, for each palquette $\square$,    is realized in terms of a  quantum circuit, where there is also an  ancilla, shown in Fig.~\ref{fig_Z2},
\begin{equation}
   e^{-i Z_\square \frac{t_s}{n}} = A^{-1}  R_z^a(-2\frac{t_s}{n}) A, \label{ara}
\end{equation}
with
\begin{equation}
A= \prod_{l \in \square} { CNOT_{l,a}},
\end{equation}
where $CNOT_{l,a}$ is a controlled-NOT gate controlled by the qubit $l$ and targeting on the ancilla $a$, $A^{-1}$ is the product of these CNOT gates in reversed order,  $R_z^a(\phi)\equiv e^{-i\sigma_z^a \phi/2}$ is single-qubit gate on ancilla  representing  rotation of angle  $\phi$ around  z-axis. Initially, the ancilla  is set to be  $ |r\rangle =\ket{0} $. Preceding  $R_z^a(-2t_s/n)$,   each $CNOT_{l,a}$   flips $|r\rangle$ if and only if the control qubit $l$ is $|1\rangle$. Therefore $R_z^a(-2t_s/n)$ acts as $e^{it_s/n}$ when there are even number of $|1\rangle$'s  on the plaquette, and acts  as $e^{-it_s/n}$ when there are odd number of $|1\rangle$'s on the plaquette. This is precisely the effect of $ e^{-i Z_\square t_s/n}$.  Afterwards,  the four CNOT gates after $R_z^a(-2t_s/n)$  return $|r \rangle$   to $\ket{0}$, which  can be used for the  next plaquette. So we only need one ancilla.  If we set $ |r\rangle  =\ket{1} $ initially, this circuit can be used to simulate the reversed   evolution $ e^{i Z_\square t_s/n}$, in other words $ e^{-i Z_\square t}$ for reversed time  $t=-t_s/n$.

It is straightforward to realize the evolution under  the other part $gX$ in the Hamiltonian,
\begin{equation}
   e^{-i gX \frac{t_s}{n}} = \prod_l{ e^{-i (-\sigma_l^x) g\frac{t_s}{n}} } =  \prod_l {  R_x^l(-2g\frac{t_s}{n})},
\end{equation}
where $R_x^l(\phi)$ on qubit $l$ represents rotation of angle $\phi$ around the x-axis. For $m$-th substep of $k$-th step, $g=g_{k,m}$.

Our realization of $e^{-i Z_\square t_s/n}$, as given in (\ref{ara}) and Fig.~\ref{fig_Z2}, is a direct application of the standard strategy for the evolution under an  interaction that is a tensor product of  $\sigma^z$ operators~\cite{nielsen}.  In some  previous  schemes of quantum simulation of LGTs, the four-body interactions are obtained
stroboscopically through a sequence of two-body interactions with ancillary degrees of freedom, and gauge invariance  in each step of Trotter decomposition is  emphasized~\cite{lewenstein,zohar3}.

Here, as $Z_\square$ and $\sigma^x_l$ are gauge invariant operators, the evolution operators are gauge invariant in every substep of the digital  decompositions.

\subsection{Preparation of the initial  ground state}

\begin{figure}
\centering
\includegraphics[scale=0.5]{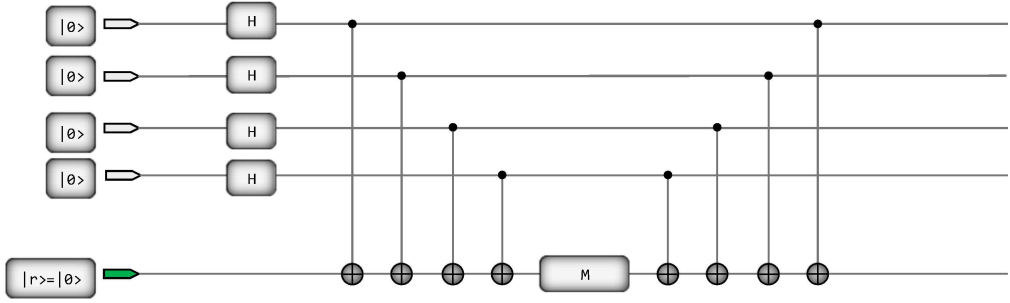}
\caption{The quantum circuit preparing  the initial  ground state of  ${ Z_\square}$ for one plaquette.
\label{fig_base} }
\end{figure}

The adiabatic quantum simulation  starts with $g=0$, i.e. $H=Z$, of which there are degenerate  ground states  satisfying
\begin{equation}
   Z_\square = -1, \forall   \square,
\end{equation}
which can be prepared by using  the quantum circuit   in Fig.~\ref{fig_base}.

Each qubit is initially in $|0\rangle$ and   is then transformed to be $(|0\rangle+|1\rangle)/\sqrt{2}$ by using a Hadamard gate $\mathbb{H}$. Therefore the state of the system is in the  equal superposition of all basis states,
\begin{equation}
|\psi_0\rangle =\frac{1}{\sqrt{2^{N_l}}} \sum_{i_1=0}^{1}\cdots\sum_{i_{N_l}=0}^{1}    |i_1\cdots i_{N_l}\rangle.
\end{equation}
Each qubit is not entangled any other qubit, each plaquette is also in the  equal superposition of all its basis states.

Then  the four CNOT gates between the four qubits of one plaquette and the ancilla initially in $|0\rangle_a$  produce the  state $\frac{1}{\sqrt{2}}(|r=0\rangle_a |  Z_\square = -1 \rangle + |r=1\rangle_a |  Z_\square = 1\rangle)$,  where  $|  Z_\square = \pm 1 \rangle$ are  states of all the qubits on the lattice satisfying  $Z_\square |  Z_\square = \pm 1 \rangle  = \pm |  Z_\square = \pm 1 \rangle$. The ancilla is entangled with the qubits on the lattice,  with   $\sigma_z^a \equiv 1-2r = -Z_\square $ in each branch. Then  the measurement operation $M$  on the ancilla  projects it to   $|r=0\rangle$,  thereby  selects the  state of qubits on the lattice to be  $|Z_\square = -1\rangle$.

Afterwards,  the ancilla  is returned to $|0\rangle_a$ by other four CNOT gates (Fig.~\ref{fig_base}). The same ancilla is ready to work on another plaquette, which may or may not share a qubit with a plaquette already worked on. Similar procedure  goes on, till all plaquettes have been worked on.  The state of the system  is  then an eigenstate of  $Z_\square$ for each $\square$,  with eigenvalue $-1$, and is  an equal superposition of all configurations satisfying $Z_\square=-1$ for each $\square$.

To summarize, we perform consecutive projections $ \prod_\square P_\square $, where $P_\square = |Z_\square=-1 \rangle\langle Z_\square=-1|$. Then the  state $ \prod_\square P_\square  |\psi_0\rangle$ is exactly a ground state of $Z$, and is the eigenstate of the 't Hooft operators  $V_{\mu}$ with eigenvalue $1$, for all $\mu$'s. It  satisfies the gauge invariance. This ground  state can adiabatically evolve  to the ground state for $g\neq 0$.

The evolution of each decomposition substep is also gauge invariant, so  the state preserves gauge invariance during the evolution. For a gauge invariant state $|\psi\rangle$, satisfying $G_i|\psi\rangle=|\psi\rangle$, and the evolution of a time period $\tau$ under a gauge invariant  operator $O$  satisfying $G_i O G_i^{-1}=O$,   one has $G_i(e^{-iO\tau}|\psi\rangle)=
G_i e^{-iO\tau}G_i^{-1}G_i|\psi\rangle
=e^{-iO\tau}|\psi\rangle$. Therefore, with the initial state gauge invariant, gauge invariance is always preserved in each evolution under $Z_\square$ or $\sigma^x_l$ in the digital decompositions.

The other degenerate ground states at $g=0$ can be obtained by using $W_\mu$'s as described above. This action changes the topological sector. When $g\neq 0$,  the state adiabatically evolves to the lowest energy state in this sector, which is not the ground state.

Alternatively, for  the adiabatic preparation of the ground states for various values of $g$, one can also start with the ground state  of $X$. For this approach, one had better redefine the Hamiltonian as $X+KZ$, with the coupling constant  $K$ corresponding  to $1/g$. The ground state of $X$  is with $\sigma^x=1$ for all qubits. With the increase of $K$, the ground state evolves from confined phase to deconfined phase. During the adiabatic evolution, the ground state remains in the topological sector of   $V_\mu=1$  for all $\mu$'s. To  enter other sectors  and  adiabatically approaches the other  degenerate ground states of $H(g=0)$, one can also use the actions of $W_\mu$'s.

In our demonstration, we use the first method, as it is easily simulated in our computing. Moreover,
QuEST provides a method function of controlling the ancilla to simulate the collapse to the destined state  in the GPU simulator.  In the real quantum simulation, ancilla measurements of $N_p$ times, each  conditioned on the result of the previous one, make the success rate only $2^{-N_p}$. Hence the second approach is preferred. We have actually also tried it in our demonstration, which yields result consistent with the first approach, so is omitted here, as the emphasis is on the deconfined phase.

\subsection{Measurement of physical quantities }

The energy $E$ in state $|\psi\rangle$ is the expectation value of $H$,
$\braket{H} = \braket{Z} + g\braket{X}$, where $\braket{Z}  =\braket{ \psi| Z |\psi } =  \sum_i { z_i P({z_i}) }$, $ \braket{X} = \braket{ \psi| X |\psi }= \sum_i { x_i P({x_i}) }$,   where $\{z_i\}$ and $\{x_i\}$  represent  the eigenvalues or measurement results of  $ Z$ and $X$,    respectively,   $P({z_i})$ and  $P({x_i})$ are the corresponding probability distributions,   called densities of states hereby. In our simulator, we obtain $E$ by summing  up $\braket{Z}$ and $g\braket{X}$. We also calculate the expectation values of Wegner-Wilson loop operators.

We use CUDA  parallel acceleration method to count the statistical summation of all basis vectors on GPU. We write our own codes for the calculation of the measurement results, which are not included in  QuEST.

The measurement is simulated at the end of each step, i.e. when $g=kg_s$, $(k=1,\cdots,N_s)$. All these quantities can be calculated in terms of the distribution $\{P({z_i})\}$ in the representation $\{\sigma^z_l\}$,
as  $\braket{W_C} =  \braket{ \psi | \prod_{l \in C}{\sigma_l^z} | \psi }$, $\braket{Z}  =  -  \sum_{\square}    \braket { \psi |   \prod_{l \in \square} {  \sigma_l^z } | \psi    }   $,     $\braket{X} =   -  \sum_{l}  { \braket { \psi | \mathbb{H}   \sigma_l^z  \mathbb{H}| \psi  }  }$, where $\mathbb{H}$ is Hadamard gate.

We mention that in real quantum simulation, $E$ can also be measured   by using quantum phase estimation. For an eigenstate $|u\rangle$ of a time-independent $H$, the evolution  for a time duration $t$ is  $ e^{-iHt}  \ket{u} = e^{-i\phi} \ket{u}$, where $\phi \equiv Et$.

$e^{-iHt}$ can be realized
in a  way similar to the decompositions described in Sec.~\ref{algo}, but now $g$ is fixed. That is,
\begin{equation}
 e^{ -i H t }  \approx ( e^{-i Z  \frac{t}{n} }e^{-ig  X \frac{t }{n} })^n,
\end{equation}
or
\begin{equation}
 e^{ -i H t }  \approx ( e^{-i Z  \frac{t}{2n} } e^{-ig  X \frac{t }{n} } e^{-i Z \frac{t}{2n} } )^n = e^{-i Z  \frac{t}{2n} }( e^{-ig  X \frac{t }{n} } e^{-i Z \frac{t}{n}} )^{n-1}e^{-ig  X \frac{t }{n} } e^{-i Z \frac{t}{2n}},
 \label{leap}
\end{equation}
where $n$ is the number of decomposition steps here. These are  Trotter decomposition and symmetrized Trotter decomposition.  The errors are just $n$ times those for $t/n$, given in Eq.~(\ref{error1}) and (\ref{error2}) with $g_{k,m}$ replaced as $g$,
 that is,   $$2g N_p \frac{t^2}{n^2}$$ and   $$|| \frac{i}{24}\frac{t^3}{n^2}\left(g
[Z,[Z,X]]+2g^2[X,[Z,X]]\right)|| \approx  \frac{(d-1)g+4g^2}{3}N_p
 \frac{t^3}{n^2},$$ respectively.

\begin{figure}
\centering	
\includegraphics[scale=0.5]{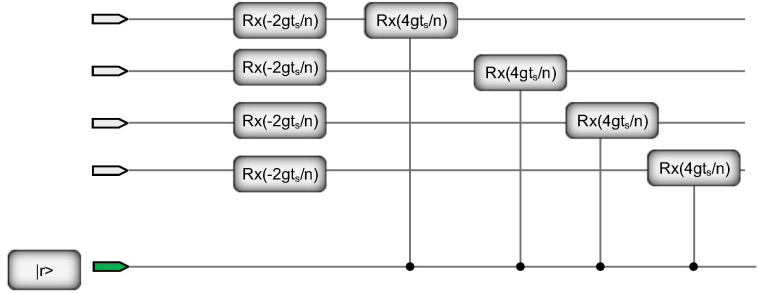}
\caption{
\label{fig_X}  The quantum circuit for   $ \prod_{l\in \square }{ e^{i \sigma_l^x gt_s/n} }$ or  $ \prod_{l\in \square }{ e^{-i \sigma_l^x gt_s/n} }$,  depending on whether the state of the ancila is $|0\rangle$ or $|1\rangle$.  }
\end{figure}

For the purpose of quantum phase estimation, we need   the conditional evolution controlled by an ancilla,
\begin{equation}
U(t)  = \ket{0}\bra{0} \otimes e^{-iHt} + \ket{1}\bra{1} \otimes e^{iHt},
\end{equation}
which can be realized in terms of controlled gates, and ancilla controlling the time direction.   For $Z$ part, the method is as shown in Fig.~\ref{fig_Z2}. For $X$ part, the method is as shown in Fig.~\ref{fig_X}.

Then $U(t)  \frac{1}{\sqrt{2}} (\ket{0}+\ket{1})\ket{u}
    =  \frac{1}{\sqrt{2}} ( e^{-i\phi}\ket{0}  + \ket{1} e^{i \phi} ) \ket{u}.$
Therefore, one can prepare  $U(2^k t)  \frac{1}{\sqrt{2}} (\ket{0}+\ket{1}) \ket{u}  = e^{-i2^k\phi} \frac{1}{\sqrt{2}}(\ket{0}+e^{i2^k\cdot 2\phi}\ket{1}) \ket{u}$, $(k =0,1,2,...)$, as required by the algorithm of quantum phase estimation. Subsequently,  the probability distribution of $2\phi$, and thus $E$,  can be obtained by using the standard procedure of  quantum  phase estimation.

\subsection{Computational hardware}

We use a Nvidia  Tesla K40m GPU card, which was used by QuEST team  in their simulation of 29 qubits with float decision~\cite{quest}, as well as a Nvidia Tesla V100-SXM2-32GB  GPU card. We estimated the maximal scales of quantum simulations that QuEST can simulate   under different precisions, as  listed in   Table.\ref{table_m40}. We use  double precisions.

\begin{table}
\begin{center}
\begin{tabular}{cccccccc}
\hline
GPU card & Simulation Precision&capacity (TFLOPS)& $N_{f}$ & $N_{qubit}$ & $M$  \\
\hline
K40m & float & 5.046 & 4 & 30 & 8.2GB  \\
K40m & double& 1.682 & 8 & 29 & 8.2GB  \\
V100 & float & 15.67 & 4 & 31 & 16.2GB  \\
V100 & double & 7.834& 8 & 30 & 16.2GB  \\
\hline
\end{tabular}
\end{center}
\caption{Maximal  scales that QuEST can perform with one Nvidia Tesla K40m-12GB or Nvidia Tesla V100-SXM2-32GB GPU card. $N_f$ is the byte number of  a floating point number, $N_{qubit}$ is the number of qubits, M is the memory requirement.
\label{table_m40} }
\end{table}

\subsection{Adiabaticity and parameter values}

We now estimate the total errors in the adiabatic process of the quantum simulation,  using
$\Delta^{asy}\approx N_s^2N_pg_s\frac{t_s^2}{n},$  $\Delta^{sym} \approx \frac{4}{9}N_s^3
 N_pg_s^2\frac{t_s^3}{n^2}$, given in Eqs. (\ref{asymtotal}) and ({\ref{total}).
We choose the  final value of $g$ to be $g_{f}=2$. For each step, the increase of $g$ is set to be $ g_{s}=0.001$. For each substep in the decomposition, the increase  is $g_s/n$.  Thus the  number of steps   is  $N_{s}= g_f/g_{s}=2000$, while the number of substeps is $2000n$. $n$ is different in different cases as described in the following.

First consider the asymmetric decomposition. For d=3 lattice, the number of plaquettes is $N_p=24$. The number of substeps in the decomposition is set to be  $n = 200$. The time for each step is chosen to be  $t_{s}=0.1$. The total evolution time is then $t_{f}=N_s t_s=200$. $t_s^3/n^2 = 0.25\times 10^{-7}$. The total error is about $2.1\times 10^{-3}$.  For d=2, the number of plaquettes is $N_p=9$. As the number of qubits are less than in d=3, we set the number of substeps   to be  $n = 5000$. We choose the time step to be  $t_{s}=0.2$. The total evolution time is then $t_{f}=N_st_s=400$.   Now  $t_s^3/n^2 \approx 0.32\times 10^{-9}$. The total error is about $1.1\times 10^{-5}$.

Now consider the asymmetric  decomposition. As the error is larger than the symmetric one by  one factor of $t_s/n$,  we use larger value of $n = 500$.  We use smaller value of  $t_{s}=0.01$ for  d=3,   and  $t_{s}=0.02$ for d=2.   Therefore,  $t_s^2/n=0.2\times 10^{-6}$ for  d=3, and  $t_s^2/n=0.8\times 10^{-7}$ for  d=2.  Therefore  the total error is $1.92\times 10^{-3}$   for  d=3,    $2.9 \times 10^{-3}$ for d=2.

These parameter values are chosen to allow the computation  to be completed in an acceptable time under adiabatic condition. The computation is proportional to $2^{N_q}$, where $N_q$ is the number of qubits, which is equal to $N_l+1$ in the present model. It is also proportional to  $N_s$ and $n$. For the parameter values given above, the time for the computation  based on asymmetric decomposition, run on a K40m server, is about 7 days for $d=3$ and 16 hours  for $d=2$, while the time for the computation  based on symmetric  decomposition, run on a V100 server, is about 28 hours for $d=3$ and 4 hours for $d=2$.  The accuracies are  all acceptable. The  difference between the computation times is due to the hardware difference rather than the decomposition methods.

Our simulation satisfies the adiabatic condition, which says that the variation of the Hamiltonian  should be  slower than the dynamical time scale, in other words,  the matrix element  of $\partial H/\partial t$ should be   smaller than the square of the  energy gap~\cite{Hamma,shi}.  In our simulation, the matrix element  of $\partial H/\partial t$   is of the order of $\partial g/\partial t = g_s/t_s$, which  varies from $0.005$ to $0.1$.

In the weak coupling limit $g\rightarrow 0$, the ground state is with all $Z_\square =-1$, while the first excited state is one with a pair of visons, that is, a pair of plaquettes with $Z_\square =1$, which can be created by flipping the qubits on the links pierced by a string on the dual lattice. Thus the gap is $4$. Its square   is a lot larger than $g_s/t_s$.  At $g=0$,  the ground state is $2^d$-fold degenerate, with different eigenvalues $\pm 1$ of the $d$ 't Hooft operators. They become nondegenerate when $g\neq 0$, and the one  with eigenvalues of all $V_\mu$'s being 1 becomes the unique ground state.  Hence there are  small energy splittings between the ground state and other eigenstates of $V_\mu$.  However,  $V_\mu$'s  are conserved because of gauge invariance, consequently  these states belong to different topological sectors. Once the initial state is prepared in one of the topological sectors, it remains there  when $g$ is varied. Consequently, the small splittings between the ground state and other $V_\mu$ eigenstates  do not matter.

In the strong-coupling limit $g\rightarrow \infty$, the ground state is with   $\sigma^x=1$ for all qubits. The gauge invariance dictates that in the first excited state, there is a plaquette with $\sigma^x_l=1$ on its four links. Hence the energy gap is $8g$, which is very  large. It is known that ground states and spectra in weak and strong coupling limits are all  stable up the QPT point~\cite{fradkinbook}.

For a finite size $L$, the adiabatic condition is satisfied  for general values of $g$, including the critical point as  the energy gap is of the order of $1/L$,  $L$ being the linear size of the system~\cite{Hamma}. Hence the gap is about $0.5$ for a general value of $g$.

\section{Results of pseudoquantum simulation  \label{results} }

Now we turn to the  results of our actual pseudoquantum  simulations of quantum $\mathbb{Z}_2$ LGT on  d=3 and on d=2 lattices. In each case, we first prepare the initial ground state at $g=0$, then execute the adiabatic algorithm by varying $g$ from $0$ to $2$ in substeps of $t_s/n$. After each step of  $t_s$,  several  quantities are calculated.

\subsection{Wegner-Wilson loops}

As $g$ increases, the ground state evolves, from the equal superposition of the configurations  with all plaquettes  in $Z_\square = -1$,  to be near the state with all qubits in $\sigma^x=1$. During this process, it undergoes a QPT, which can be characterized in terms  of the  Wegner-Wilson loop operator  $W_C$ defined along a loop $C$ on the direct lattice~\cite{fradkinbook,Sachdev}. In  the  confined phase  $g > g_c$,  $W_C$ obeys the area law $\braket{W_C} \propto g^{-A_C} = \exp(- A_C\ln g)$, where $A_C$ is the area enclosed by the contour $C$. In the deconfined phase at  $ g <g_c$, $W_C$ obeys the perimeter law $ \braket{W_C} = \exp(- B(g) P_C)$, where $P_C$ is the perimeter of the contour $C$, $B(g)$ is a smooth function of $g$ and vanishes as $g\rightarrow 0$.  At $g=0$,  all $Z_\square=-1$, the $\mathbb{Z}_2$ flux is expelled, $\braket{W_C} =1$. At small values of $g$, the  fluctuations  lead to the perimeter law. Our simulations confirm this picture.

In our simulation, for   d=3 and d=2 lattices respectively, we choose three contours denoted as  c1, c2, c3, as shown in  Fig.~\ref{fig_wloop_c}. On each lattice, the perimeter ratios are  1:1.5:2 while the  area ratios are 1:2:3.

\newcommand{\SCLwl}{0.30}
\newcommand{\SCLwlpdf}{0.20}
\begin{figure}[htb]
\centering
\subfigure[]{}
\includegraphics[scale=\SCLwl]{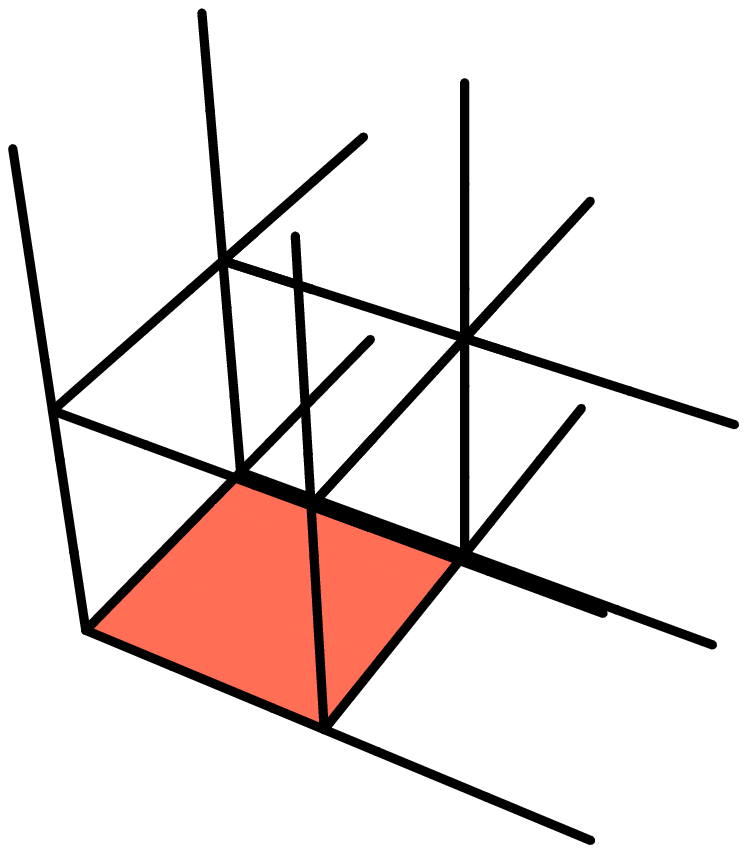}
\subfigure[]{}
\includegraphics[scale=\SCLwl]{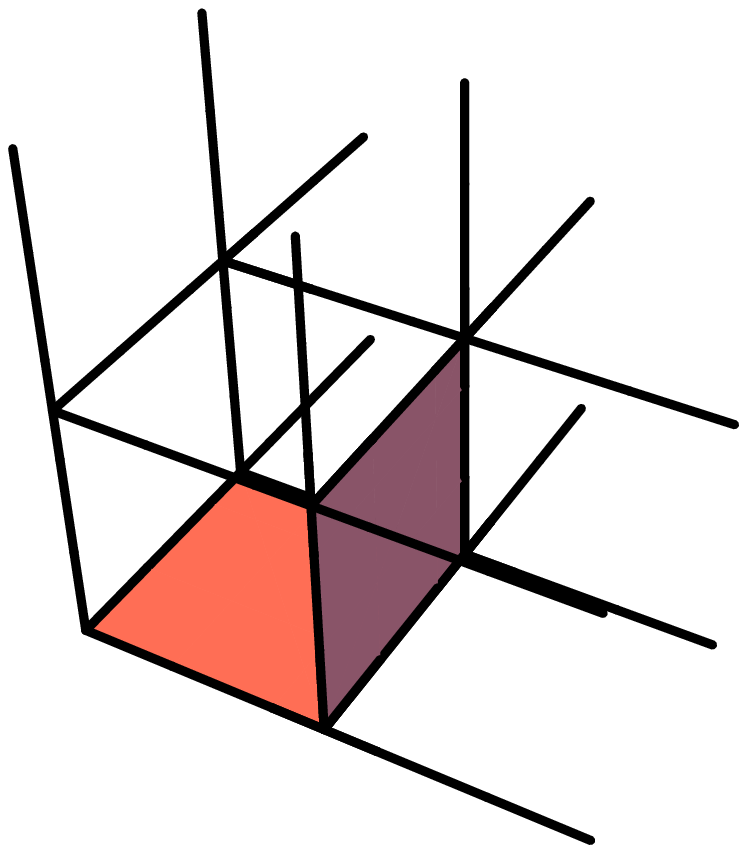}
\subfigure[]{}
\includegraphics[scale=\SCLwl]{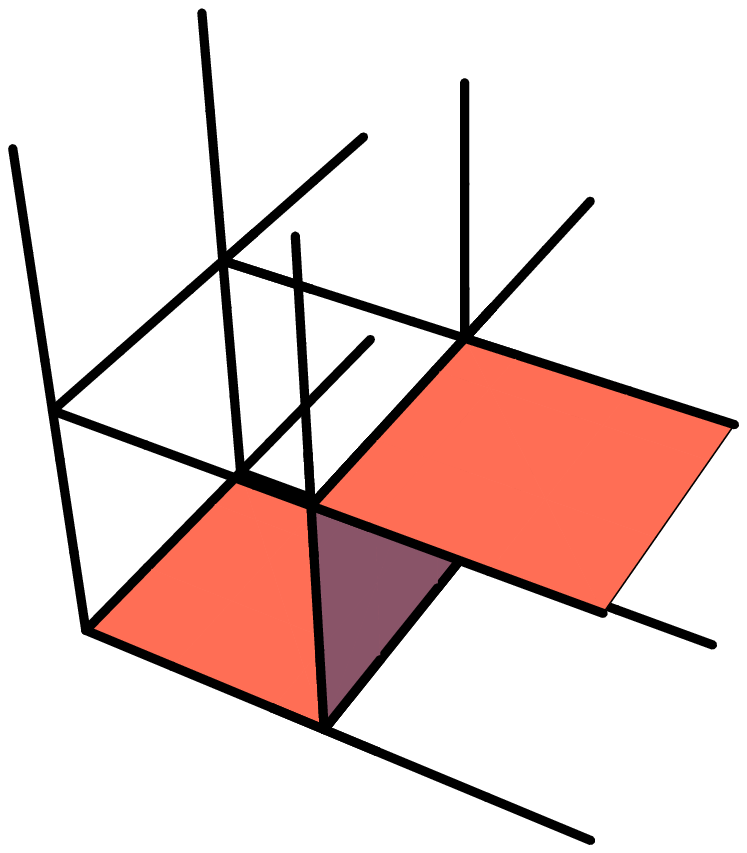}

\subfigure[]{}
\includegraphics[scale=\SCLwl]{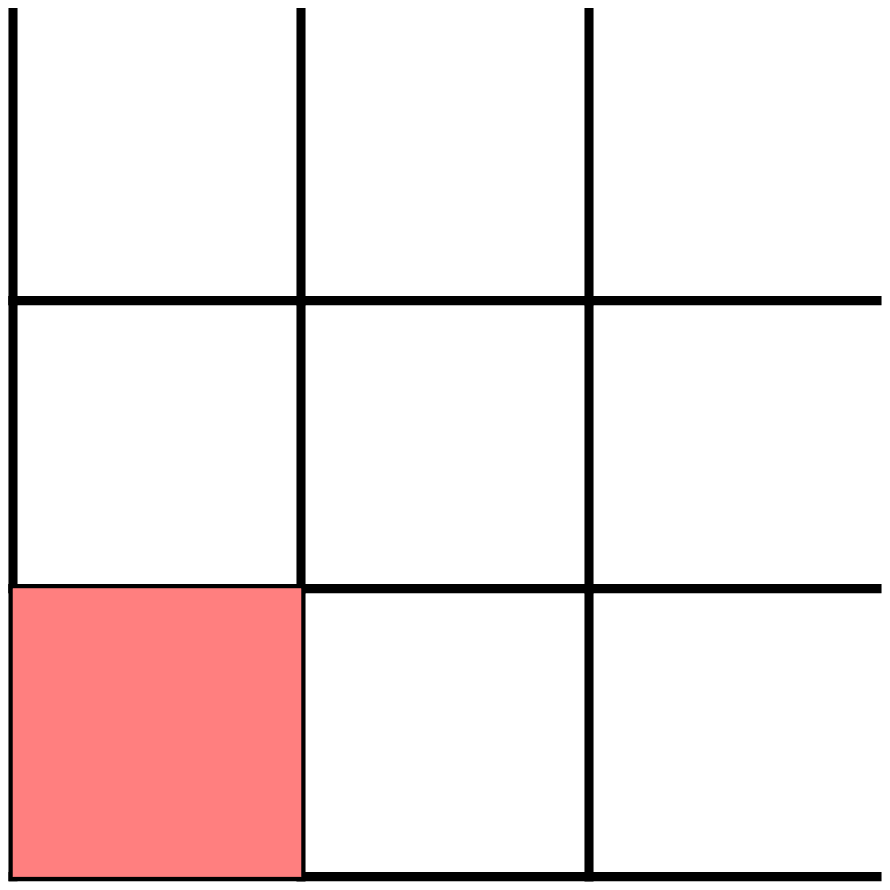}
\subfigure[]{}
\includegraphics[scale=\SCLwl]{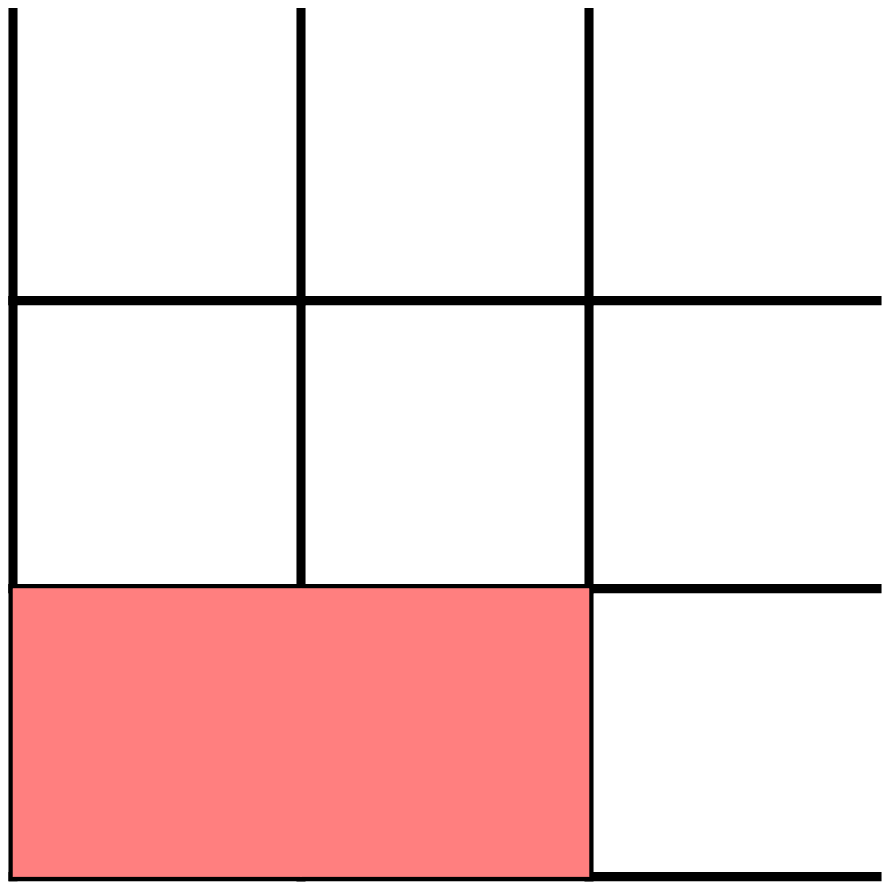}
\subfigure[]{}
\includegraphics[scale=\SCLwl]{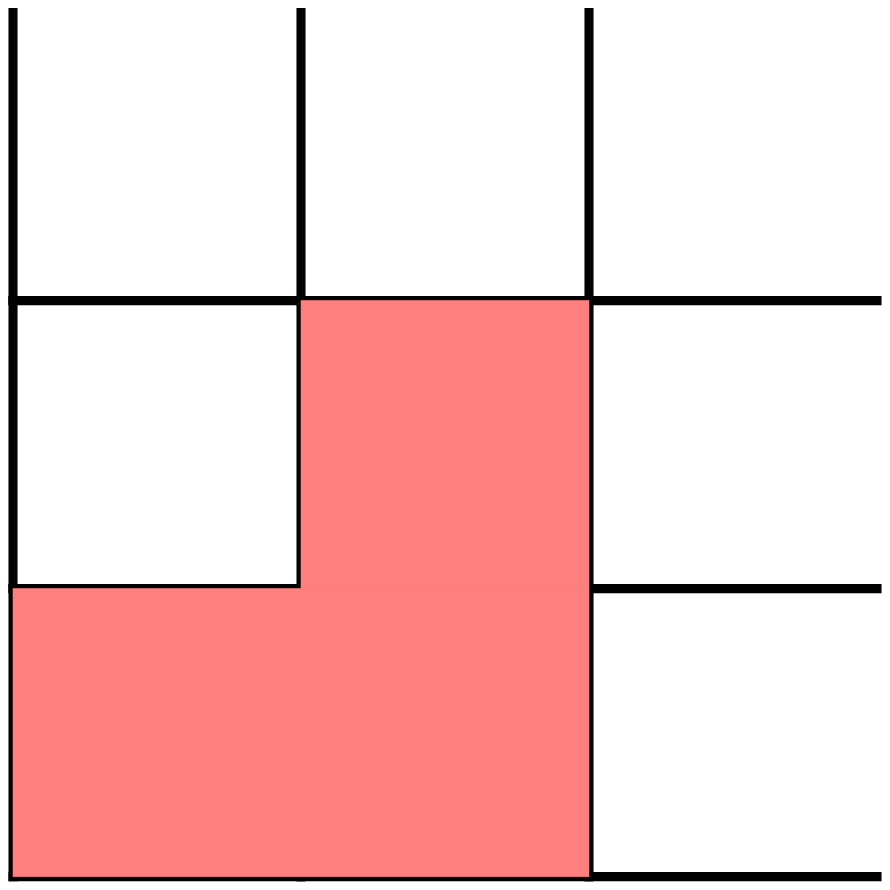}
\caption{(a-c) Wegner-Wilson loops c1, c2 and c3 on  d=3 lattice.    (d-f) Wegner-Wilson loops c1, c2 and c3 on  d=2 lattice. }
\label{fig_wloop_c}
\end{figure}

\begin{figure}[htb]
  \centering
  \subfigure[]{}
  \includegraphics[scale=\SCL]{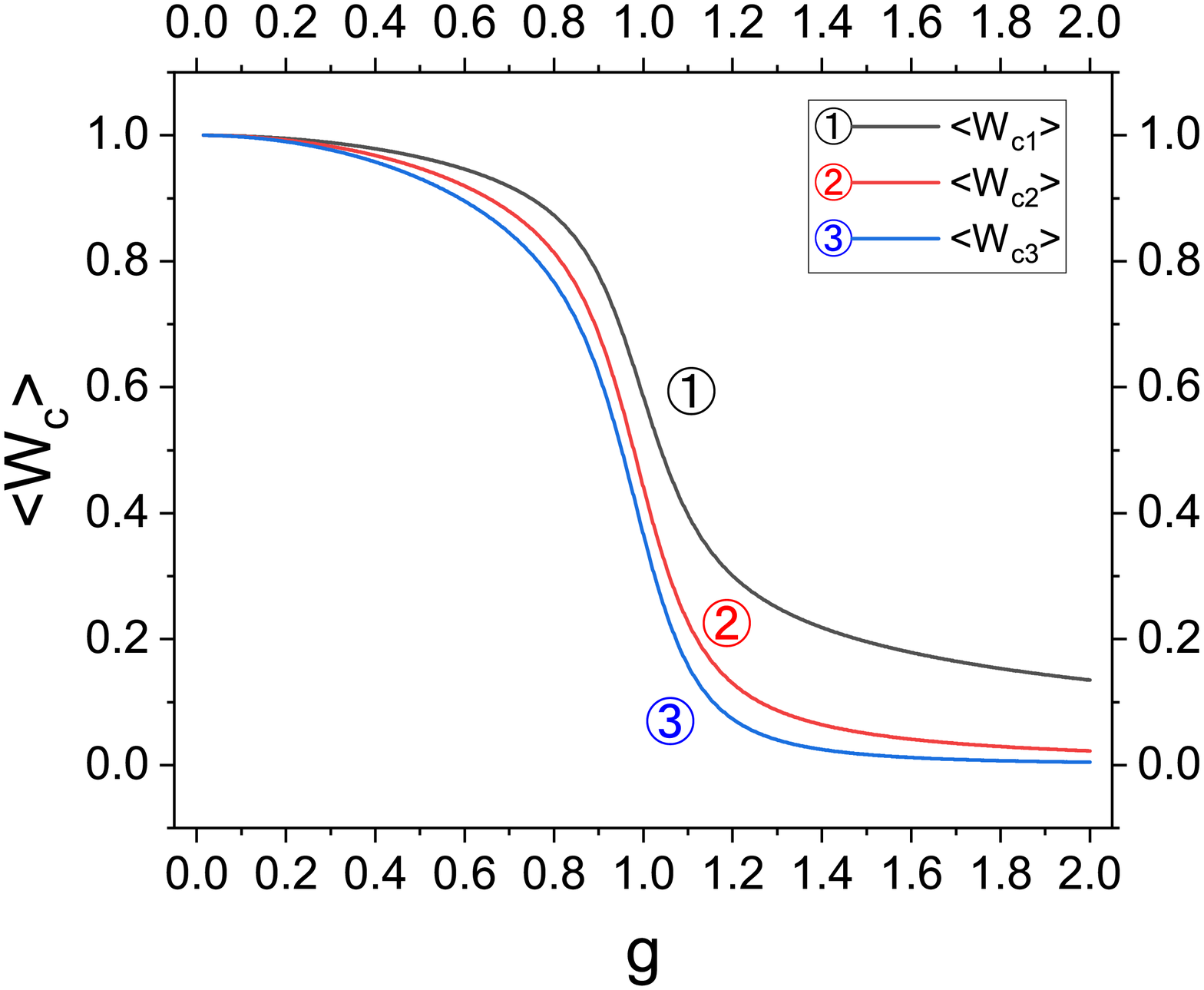}
  \hspace{0.01\linewidth}
  \subfigure[]{}
  \includegraphics[scale=\SCL]{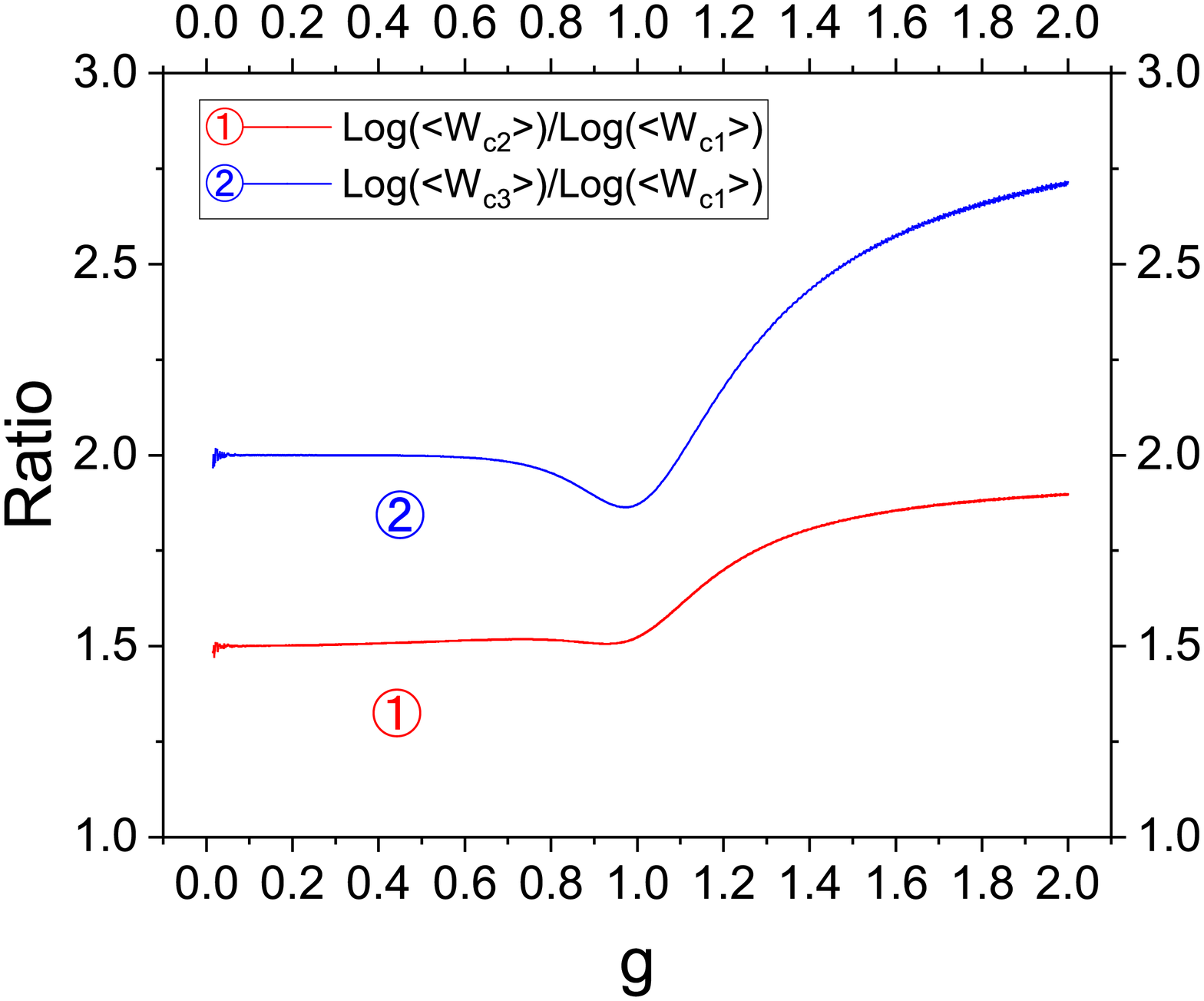}
  \hspace{0.01\linewidth}

  \subfigure[]{}
  \includegraphics[scale=\SCL]{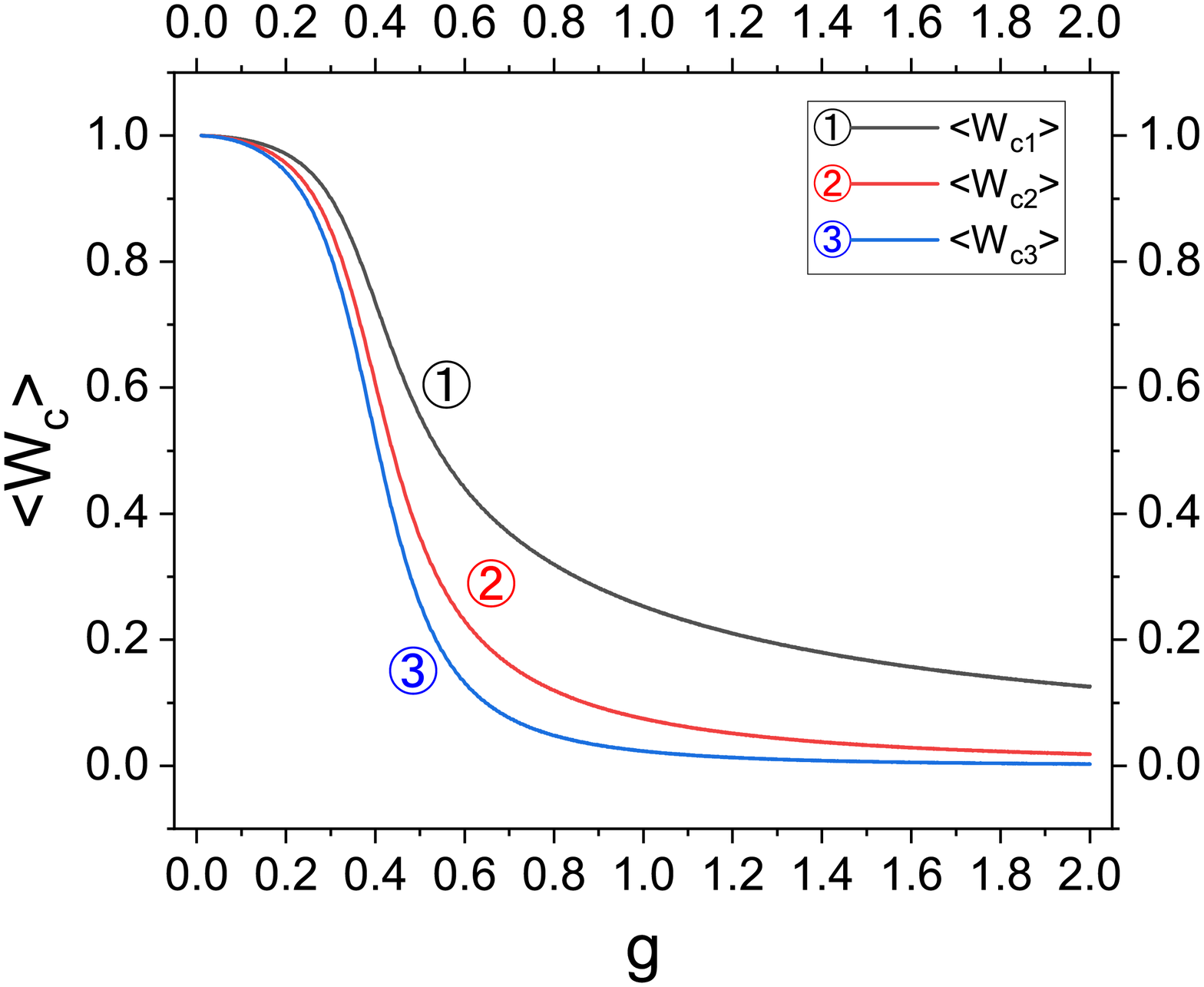}
  \hspace{0.01\linewidth}
  \subfigure[]{}
  \includegraphics[scale=\SCL]{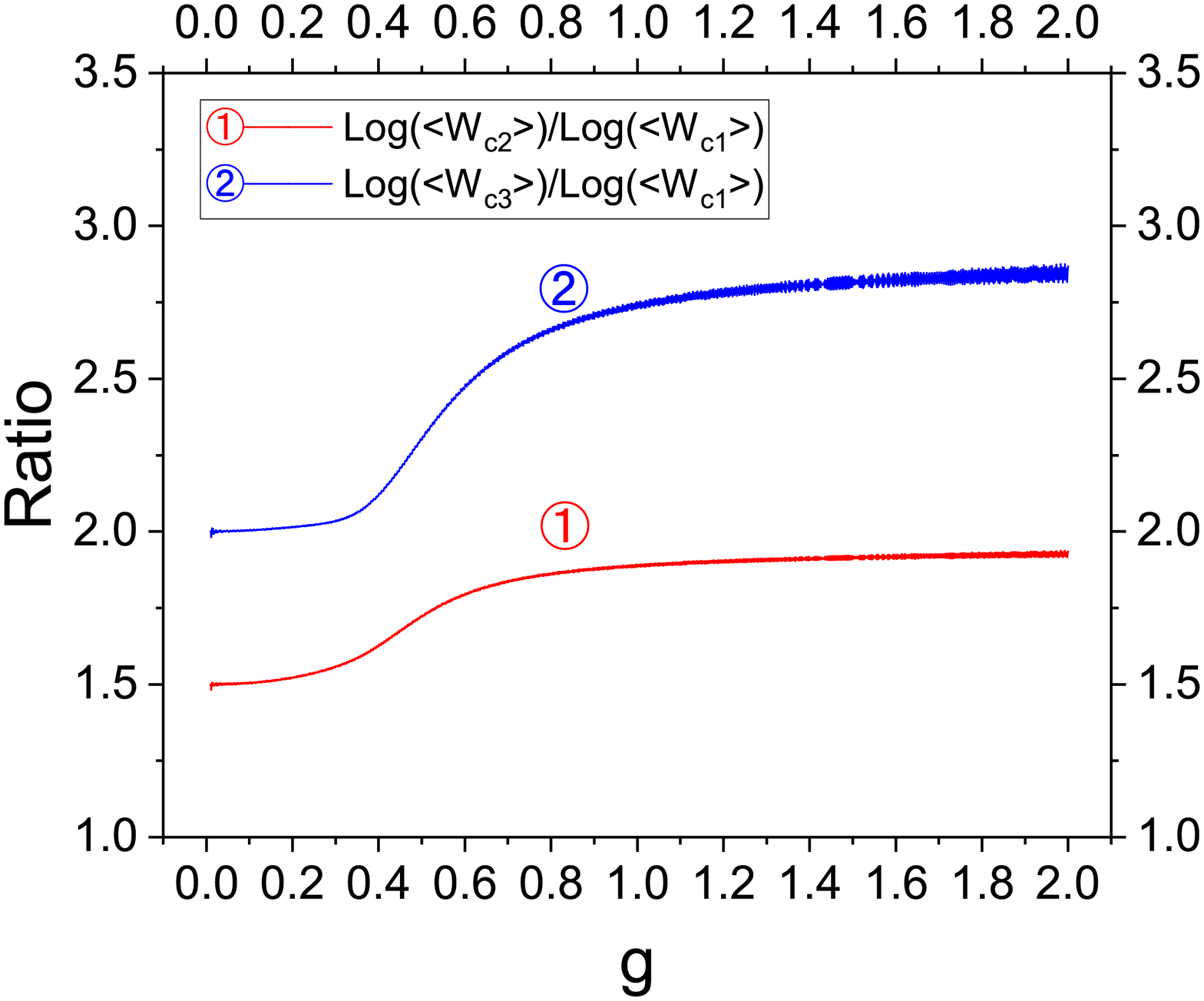}

  \caption{ Ground-state expectation values of  Wegner-Wilson loop operators   as functions of $g$, for d=3 and d=2 respectively.    (a) Results for d=3 $2\times 2\times 2$ lattice.  (b) Two ratios between logarithms of the expectation values in (a).   (c) Results for d=2 $3\times 3$ lattice. (d) Two ratios between logarithms of the expectation values in (c).   }
  \label{fig_wloop_results}
  \end{figure}

Features on Wegner-Wilson loops are shown in Fig.~\ref{fig_wloop_results}.  As $g \rightarrow 0$,   $\braket{W_C} \rightarrow 1$. With the increase of $g$,  $\braket{W_C}$  decrease relatively slowly when $g$ is small,  as indicated in subfigures (a) and (c), and  the ratios between $\log\braket{W_C}$'s    of different loops equal  the  ratios between the perimeters, as indicated in subfigures (b) and (d). When $g$ is relatively large,  $\braket{W_C}$'s decrease  as some powers of $g$,  and the powers are  shown to be areas, since  the ratios between $\log\braket{W_C}$'s for different loops equal  the  ratios between  areas, as can be seen in subfigures (b) and (d). Features in d=3 and d=2 are similar, except that  in d=3, there is a dip in the $g$-dependence of the $\log\braket{W_C}$ ratio, which will be discussed  below.

Previous tensor network calculation for d=2 gave   $\braket{W_C}$ as functions of perimeter and area for a small and a large values of coupling constant, respectively~\cite{Vidal}. Complementarily, here we give $\braket{W_C}$ as a  function  of $g$, satisfying the area and perimeter laws for three coutours in d=2 and d=3 respectively.

\subsection{Critical points  and duality}

In the adiabatic evolution, in steps of $t_s$, we pause the evolution, and calculate  $\langle Z \rangle$ and  $\langle X \rangle$, which are summed to give  $\langle H \rangle$, as shown in   Fig.~\ref{fig_e}, from which we also obtain the first and second  derivatives with respect to $g$, as shown in Fig.~\ref{fig_de}.
\begin{figure}[htb]
\centering
\subfigure[]{}
\includegraphics[scale=0.3]{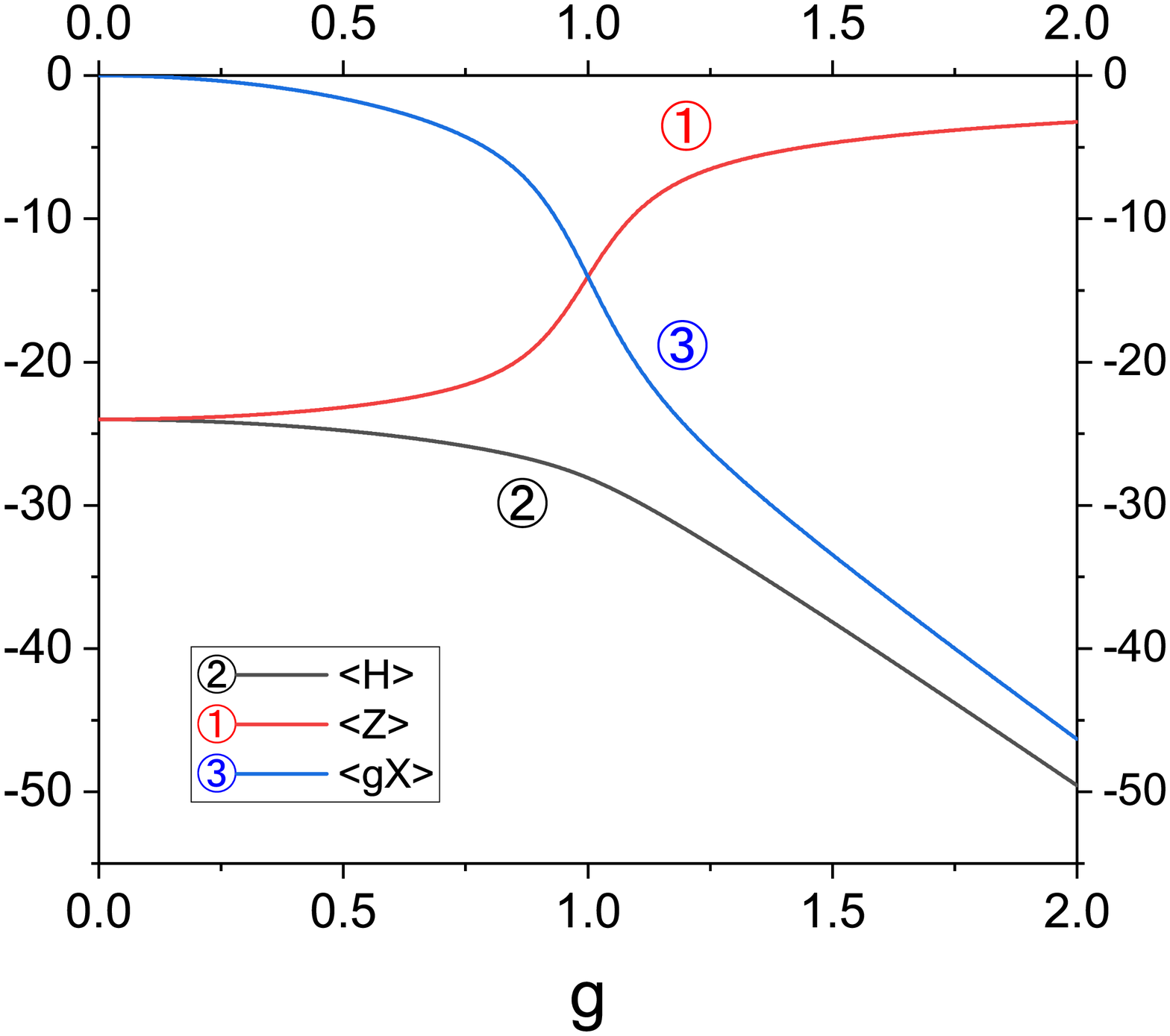}
\subfigure[]{}
\includegraphics[scale=0.3]{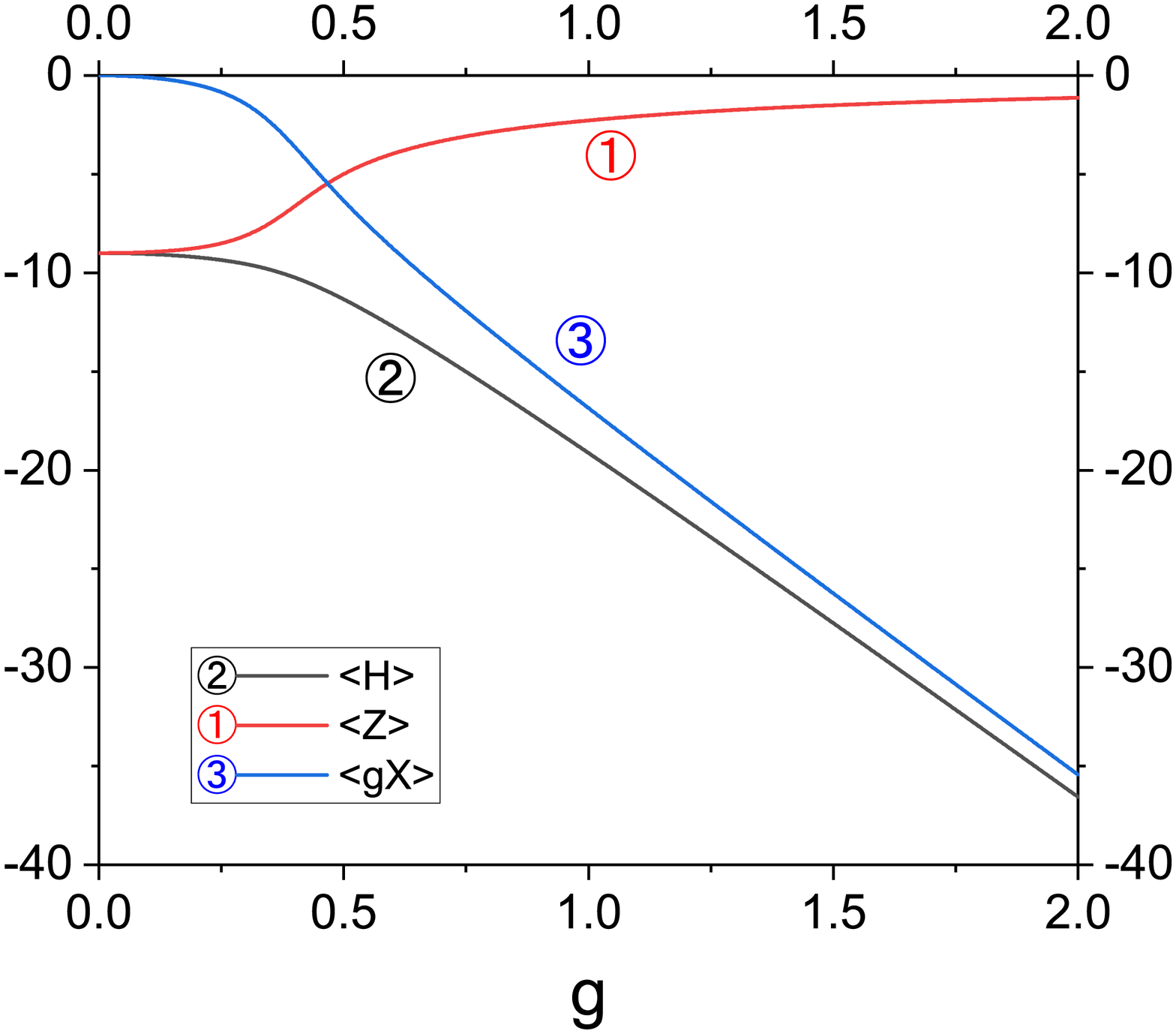}
\caption{$\langle Z \rangle$,   $g\langle X \rangle$ and  $\langle H \rangle$ as functions of $g$, which adiabatically   varies from  0 to 2. (a) Results on d=3 $2\times 2\times 2$ lattice.   (b) Results on d=2 $3 \times 3$ lattice.    }
\label{fig_e}
\end{figure}

\begin{figure}[htb]
\centering
\subfigure[]{}
\includegraphics[scale=0.3]{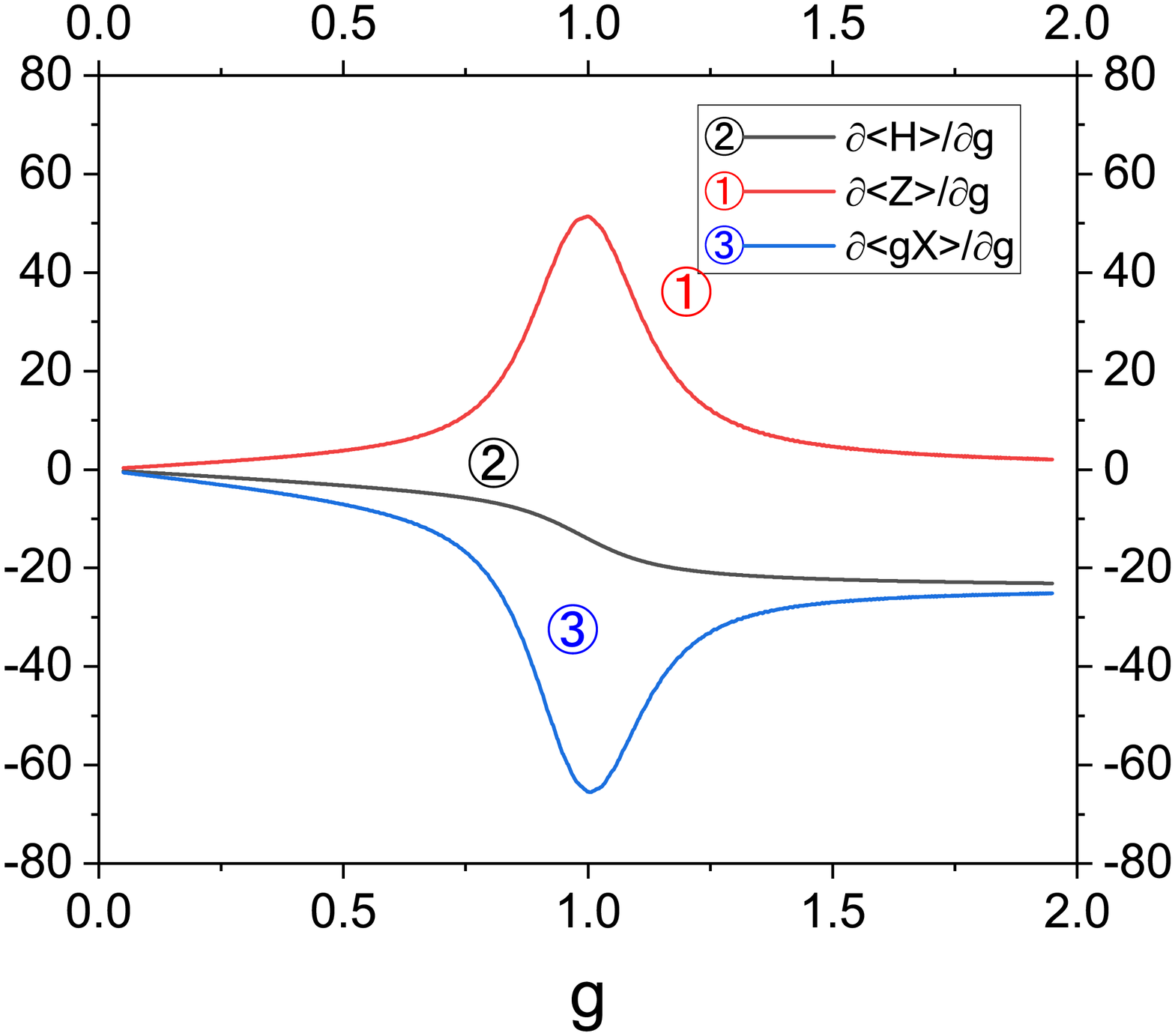}
\subfigure[]{}
\includegraphics[scale=0.3]{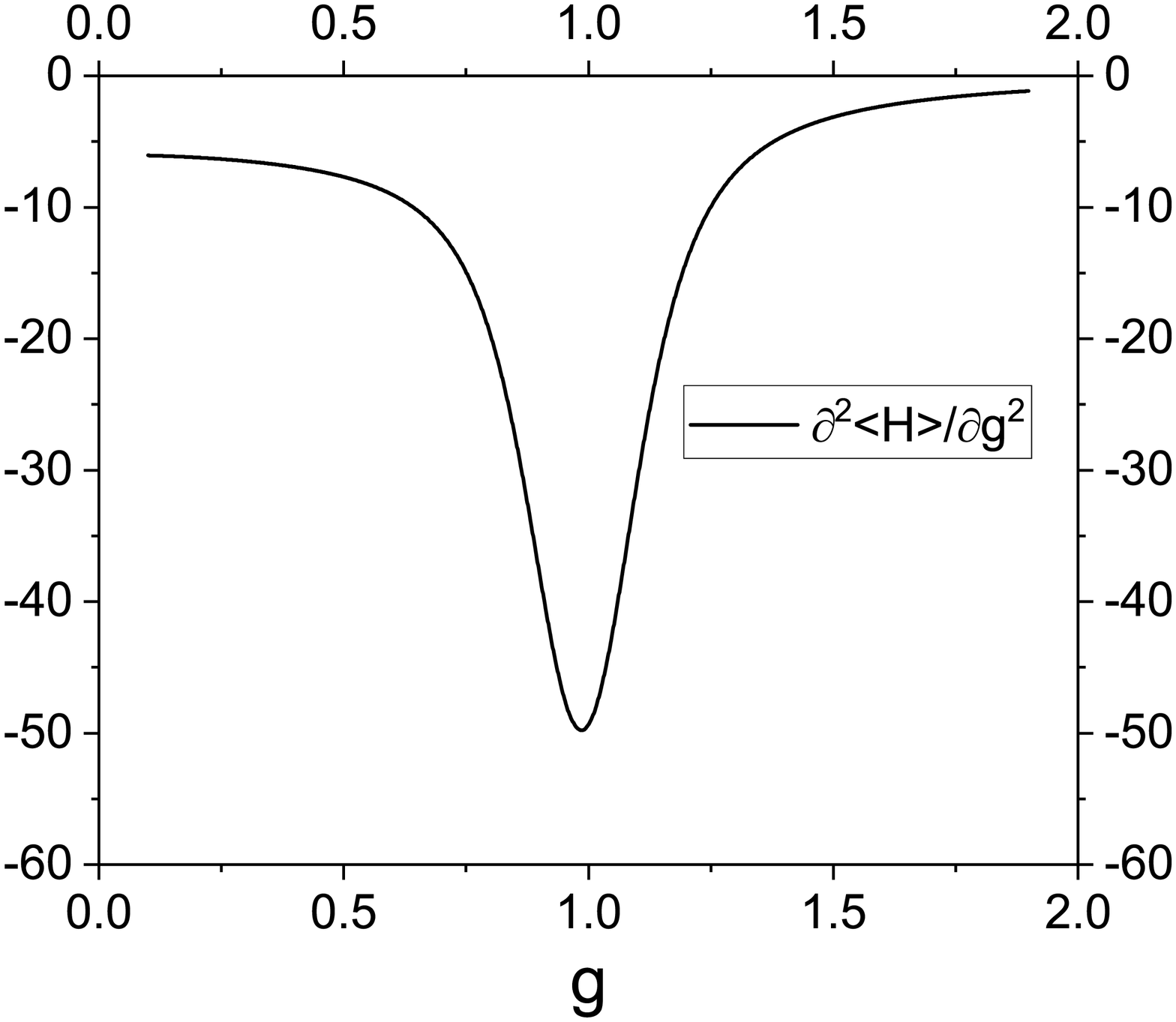}

\subfigure[]{}
\includegraphics[scale=0.3]{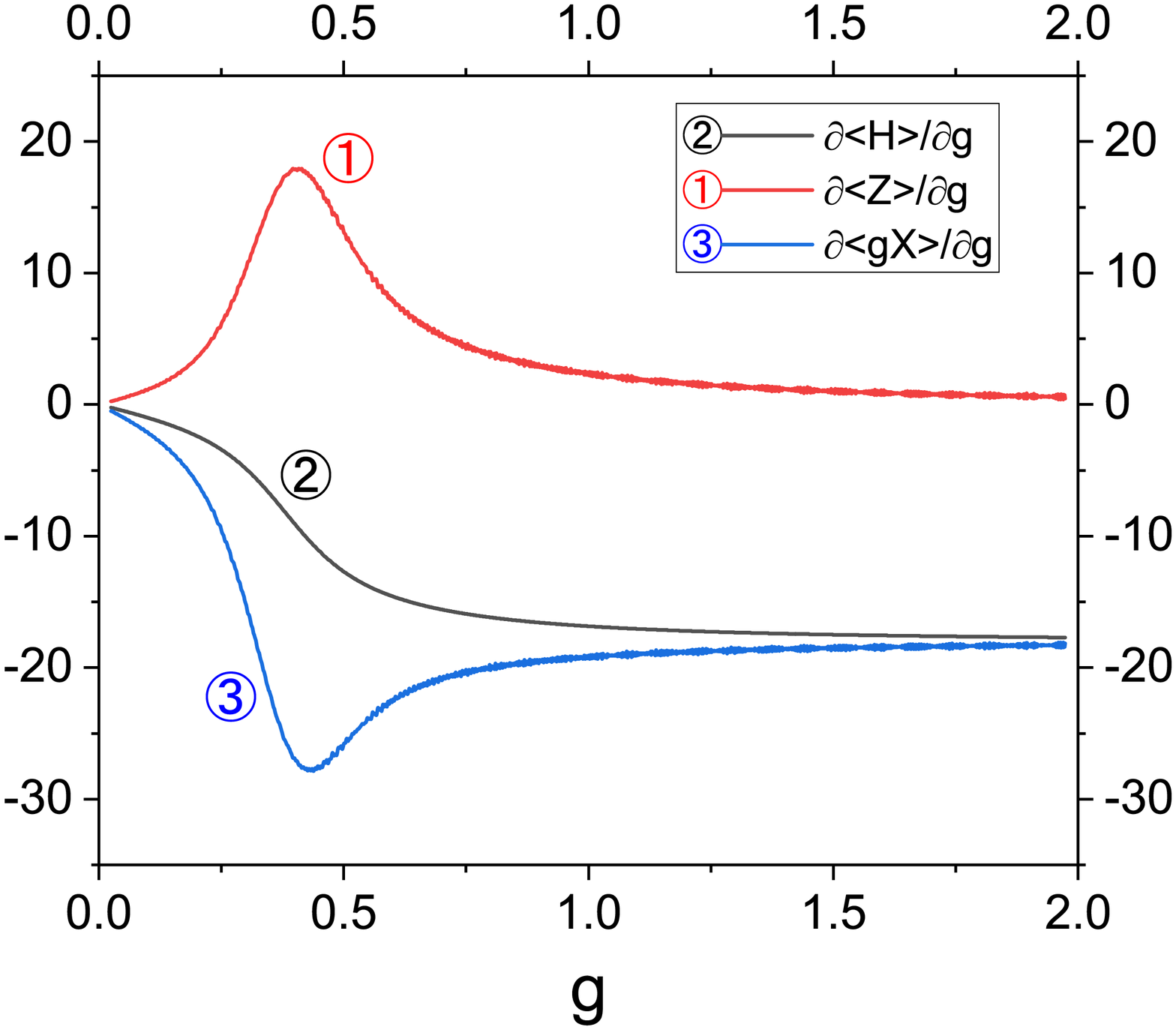}
\subfigure[]{}
\includegraphics[scale=0.3]{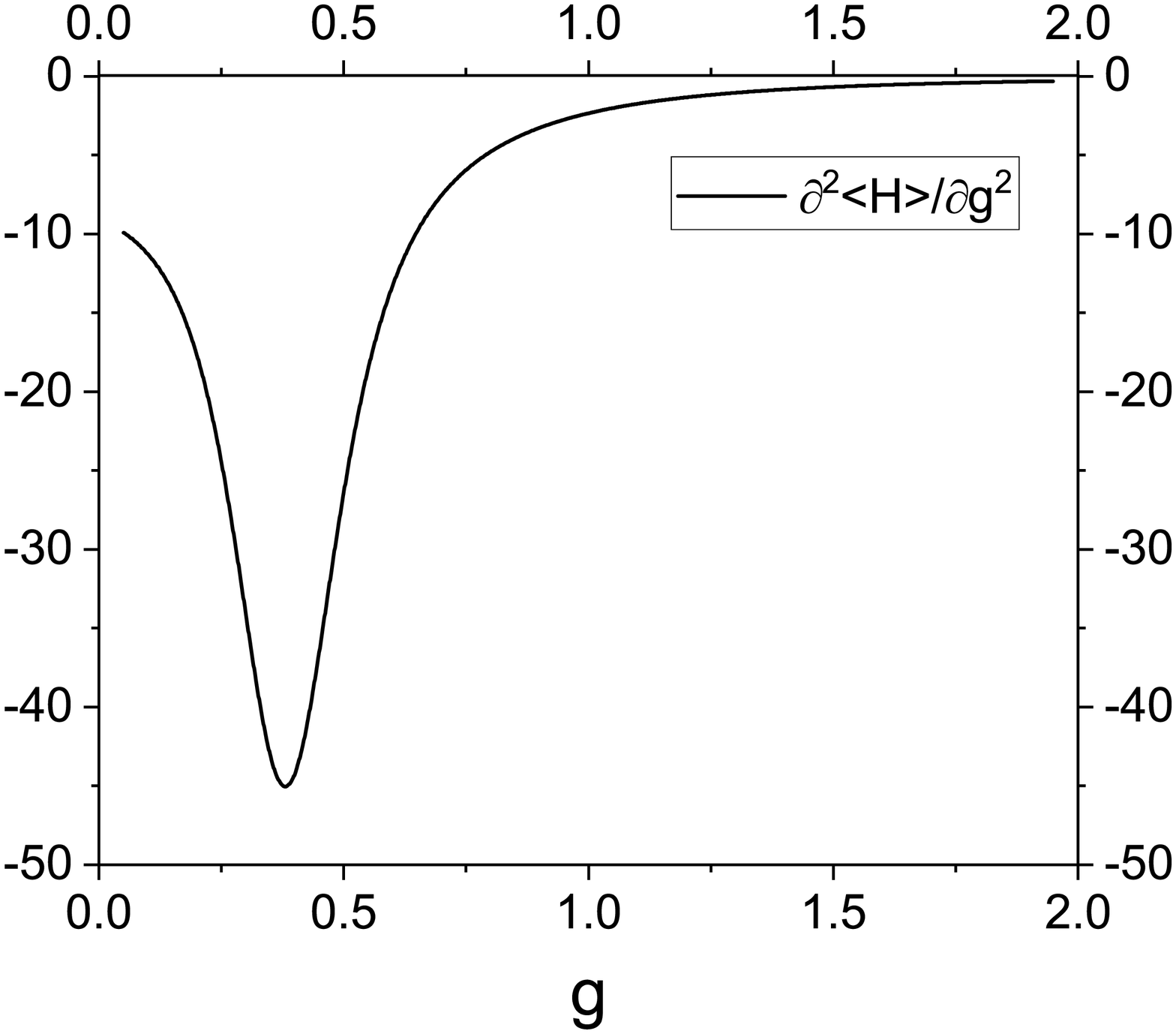}
\caption{ Derivatives with respect to $g$, as functions of  $g$. (a) First  derivatives  $\partial\langle Z \rangle /\partial g$,     $\partial\langle gX \rangle /\partial g$ and  $\partial\langle H \rangle/\partial g$ on d=3 $2\times 2\times 2$ lattice. (b) Second derivative  $\partial^2\langle H \rangle/\partial g^2$    on d=3 $2\times 2\times 2$ lattice, the lowest point  is at $g_c \approx 0.986$.   (c)  First  derivatives  $\partial\langle Z \rangle /\partial g$,     $\partial\langle gX \rangle /\partial g$ and  $\partial\langle H \rangle/\partial g$  on d=2 $3 \times 3$ lattice.     (d) Second derivative  $\partial^2\langle H \rangle/\partial g^2$    on    d=2 $3 \times 3$ lattice, the lowest point  is   at   $g_c \approx  0.380$.    }
\label{fig_de}
\end{figure}

Our simulations are on very small lattices, therefore, the singularities at the critical points of QPT are rounded out. Nevertheless one can observe  the critical points from the energy properties, where the transition is clearer  than   the transition  of the Wegner-Wilson loops from the perimeter law to the area law.

We have determined the critical points   $g_c$'s using two methods. First, from the lowest point in the valley   of the second derivative  $\partial^2 \langle H\rangle/\partial g^2$ for each lattice (Fig~\ref{fig_de}), it is observed  that
\begin{eqnarray}
g_c  \approx 0.986, \,\, d=3, \\
g_c  \approx 0.380,\,\, d=2,
\end{eqnarray}
which are not precise enough  without finite-size scaling.

\begin{figure}[htb]
\centering
\includegraphics[scale=0.3]{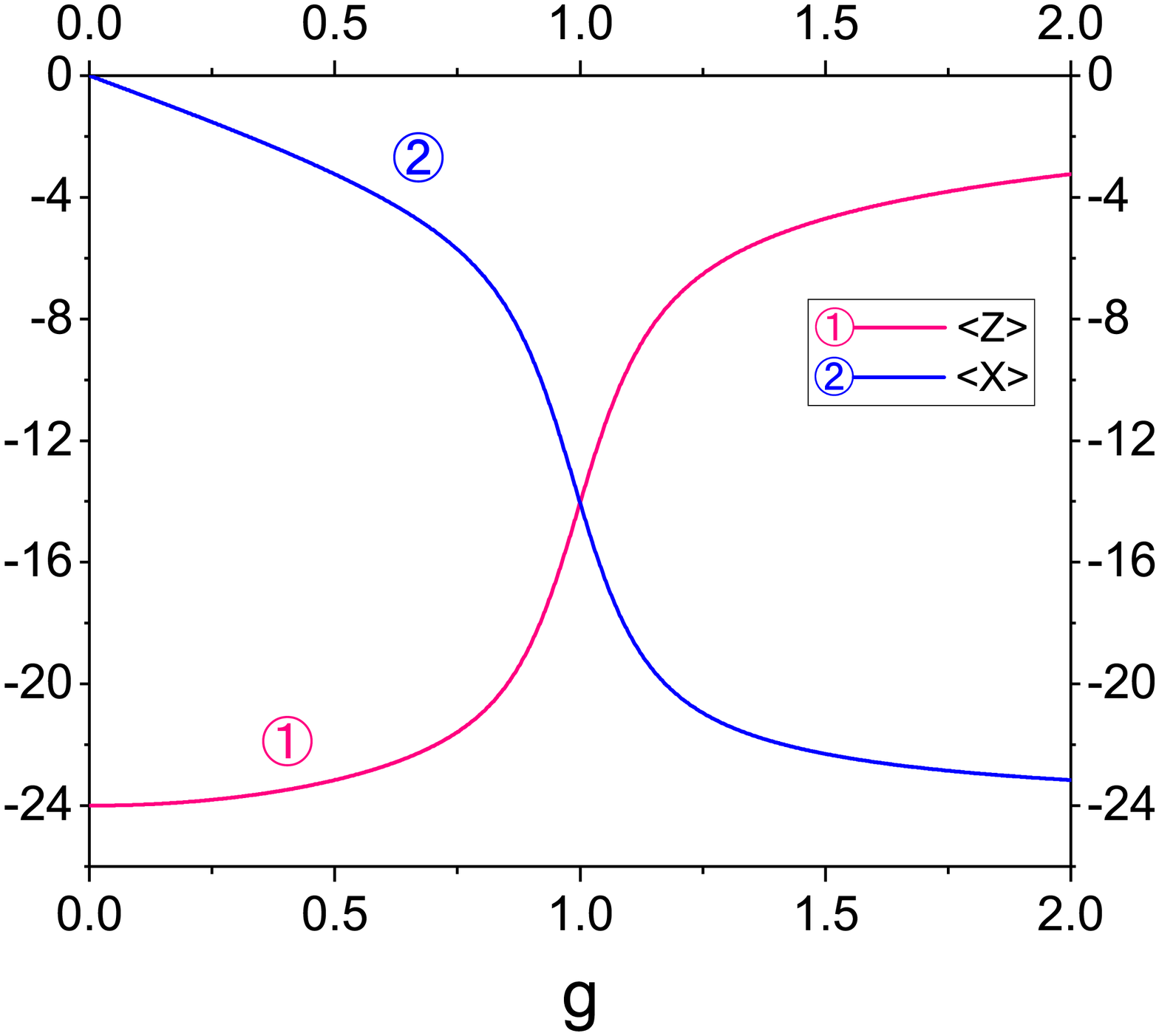}
\subfigure[]{}
\includegraphics[scale=0.3]{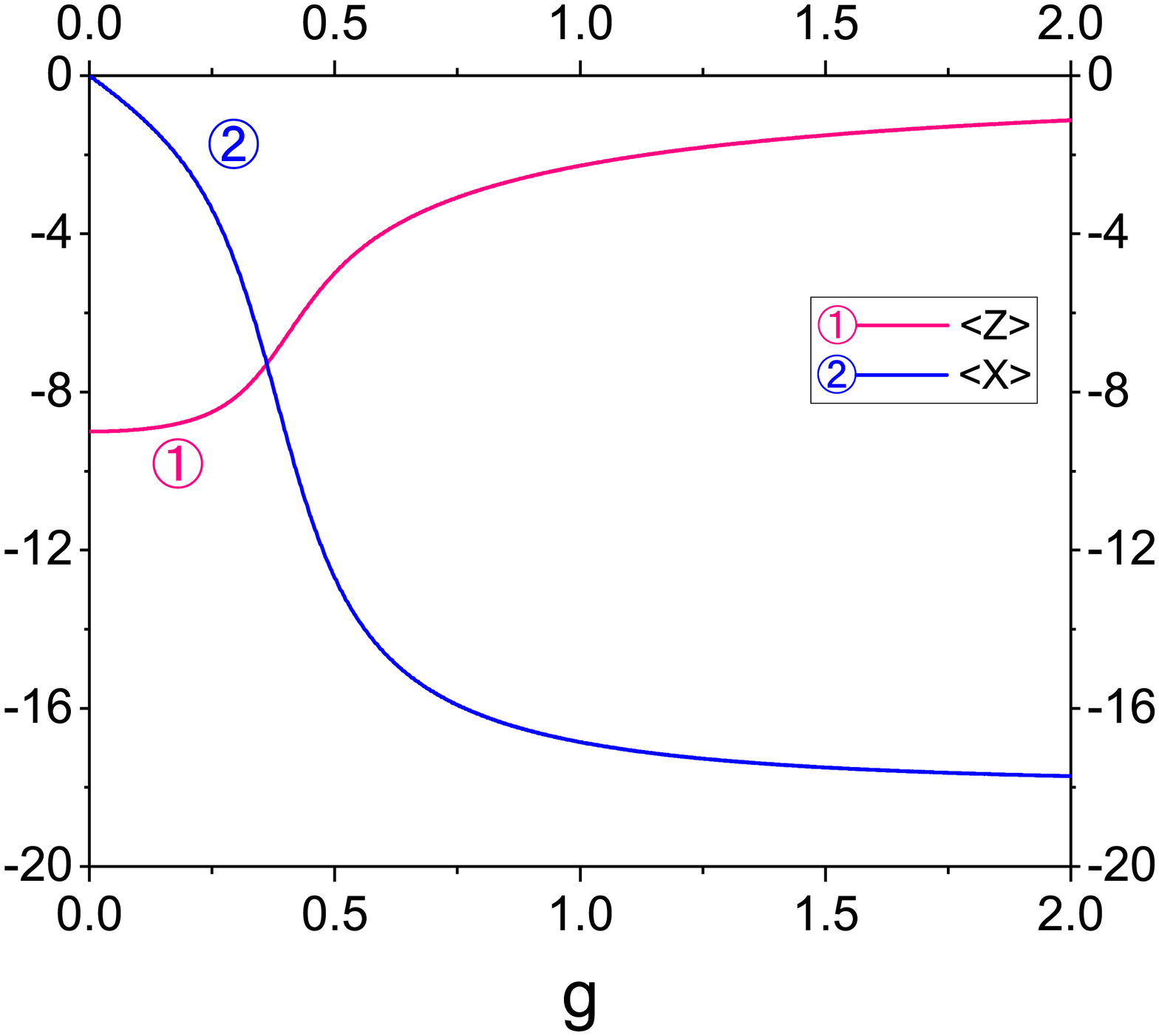}

\caption{ $\langle Z \rangle$ and  $\langle X \rangle$,  as functions of $g$. (a)  Results on  d=3 $2 \times 2  \times 2 $ lattice,  clearly displaying  self-duality and crossing at $g=1.0$.  (b)  Results on d=2 $3 \times 3$ lattice, showing the absence of self-duality.  }
\label{fig_duality}

\end{figure}

Under the duality transformation~\cite{fradkinbook},   $H(g)=g\tilde{H}(1/g)$, $|\psi(g)\rangle = |\tilde{\psi}(1/g)\rangle$,  where $\tilde{H}$ is the Hamiltonian defined on the dual lattice, $|\psi(g)\rangle$ and $ |\tilde{\psi}(g)\rangle$ are corresponding eigenstates of $H(g)$ and $\tilde{H}(1/g)$,  respectively. The occurrence of  QPT  doe not depend on the description in terms of $H(g)$ or  $\tilde{H}(\lambda=1/g)$, therefore the critical point $g_c$ of $H(g)$   is  related to the critical point $\lambda_c$ of $\tilde{H}(\lambda)$ as
\begin{equation}
g_c=1/\lambda_c.
\label{dualequal}
\end{equation}

In d=2, $\tilde{H}(1/g)$ is the Hamiltonian of transverse Ising model (TIM) with coupling constant $1/g$. A QMC calculation of TIM on $L=32$ lattice implies $g_c=1/3.044=0.3285$~\cite{Rieger}. A cluster Monte Carlo calculation implies $g_c=1/3.04428=0.3285$, while there were other calculations implying $g_c$ between $1/3.046=0.3283$  and $1/2.742=0.3647$~\cite{Blote2002}. An exact diagonalization calculation of TIM on $L=6$ lattice implies $g_c=0.32841$~\cite{Hamer}.
An entanglement renormalization calculation of TIM on $L=54$ lattice implies $g_c=1/3.075=0.3252$~\cite{evenbly}.

In the tensor network calculation on lattices with $L \leq 8$~\cite{Vidal}, $g_c$  is $0.3267$ from the energy gap in the same  topological sector of the ground state, and is $0.327$ from the string operator expectation, and is $0.3285$ from the overlap between the ground state and the lowest energy states in other topological sectors acted by Wegner-Wilson operators on the non-contractible loops.

In d=3,  $\tilde{H}(1/g)$ is the  quantum $\mathbb{Z}_2$  LGT with coupling constant $1/g$. This is self-duality. Hence $g_c=1/g_c$, consequently  $g_c= 1$ \cite{fradkin,Wegner}.

Moreover, in d=3, $Z = \tilde{X}$, $X = \tilde{Z}$, hence
\begin{equation}
\langle \psi(g)|Z| \psi(g)\rangle =
\langle \tilde{\psi}(1/g)| \tilde{X} | \tilde{\psi}(1/g)\rangle.
\end{equation}
On the other hand, as the expectation values,
\begin{equation}
\langle \tilde{\psi}(1/g)| \tilde{X} | \tilde{\psi}(1/g)\rangle  = \langle \psi(1/g)| X |\psi(1/g)\rangle.
\end{equation}
Therefore
\begin{equation}
\langle \psi(g)|Z| \psi(g)\rangle = \langle \psi(1/g)| X |\psi(1/g)\rangle. \label{dual}
\end{equation}
Setting $g=1$, we obtain
\begin{equation}
\langle \psi(1)|Z| \psi(1)\rangle = \langle \psi(1)| X |\psi(1)\rangle,
\end{equation}
which means that $\langle Z \rangle$ and  $\langle X \rangle$ cross  at   the critical point $g_c=1$.

In our simulation in $d=3$, as the second approach determining $g_c$,    we find   that the crossing point  of  $\langle Z \rangle$ and  $\langle X \rangle$  is right  at $g_c=1.0$ (Fig.~\ref{fig_duality}), the same as the theoretical result.

As shown in Fig.~\ref{fig_duality}, our simulation also demonstrates (\ref{dual}) for various values of $g$, further confirming self-duality  in d=3.    It also  clearly shows the  absence of self-duality  in d=2.

\subsection{Densities of states }

In the ground state $|\psi(g)\rangle$ for $g$ equal to multiplies of $g_s$, we  numerically calculate the density of   eigenstates $\{|z_i\rangle\}$ of $Z$ and density of eigenstates $\{|x_j\rangle\}$ of  and $X$, respectively.  $Z|z_i\rangle=z_i|z_i\rangle$,  $X|x_j\rangle=x_j|x_j\rangle$. From the decompositions  $|\psi(g)\rangle=\sum_i \alpha_i(g)|z_i\rangle  =\sum_j \beta_j(g)|x_j\rangle $, it is known that the densities of states are  $D(g,z_i)= |\alpha_i(g)|^2$ and $D(g,x_j)= |\beta_j(g)|^2$.

First consider  d=3 $2\times2\times2$ lattice,  with periodic boundary condition.   As shown in Fig.~\ref{fig_vison}, because of the geometric constraint, flipping one qubit between $\sigma_z$ eigenstates reverses the signs of the eigenvalues of $Z_\square$'s of 4 plaquettes sharing this qubit, thus reverses the eigenvalue of $Z$ by $8$. Flipping the qubits on two crossing or neighboring parallel links   reverses the signs of the eigenvalues of $Z_\square$'s of 6 plaquettes,   thus changes the eigenvalue of $Z$ by $12$. In general, $4+2n$  plaquettes can be flipped $(n=0, 1, 2...)$. Consequently,   the possible eigenvalues of $Z$  are $\pm 24$, $\pm 16$, $\pm12$, $\pm8$, $\pm4$, $0$. The prominent feature here is that $\pm 20$ are forbidden, because at least 4  plaquettes are  reversed.

\begin{figure}[htb]
\centering
\subfigure[]{}
\includegraphics[scale=\SCLwl]{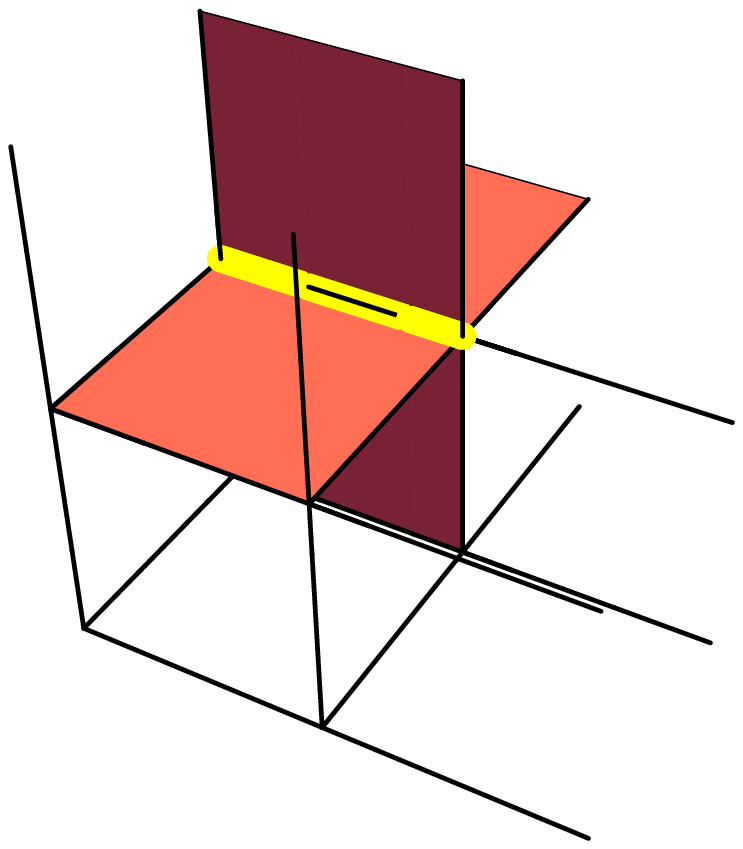}
\subfigure[]{}
\includegraphics[scale=\SCLwl]{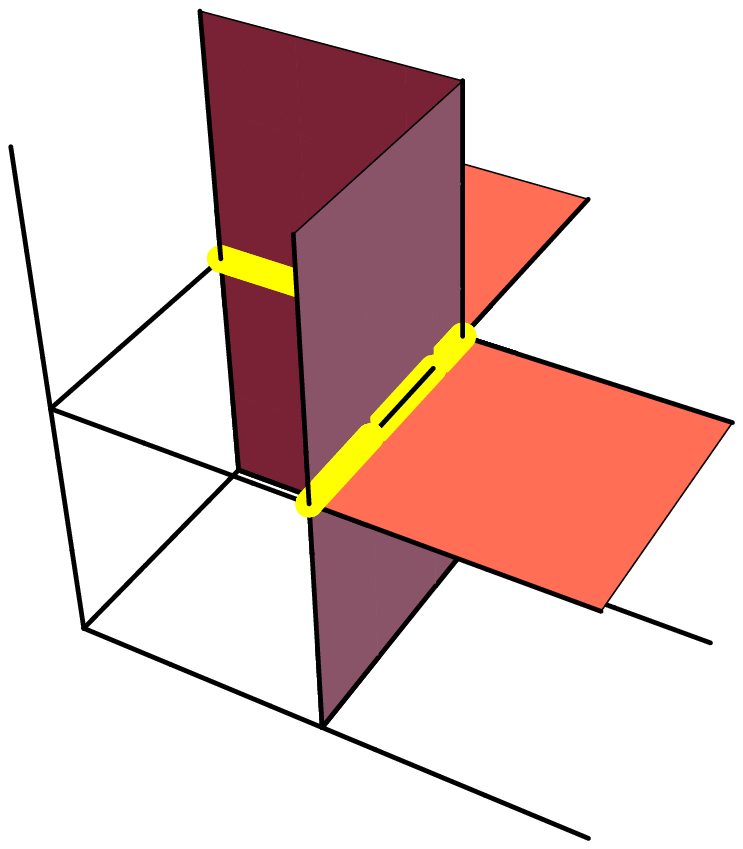}

\subfigure[]{}
\includegraphics[scale=\SCLwlpdf]{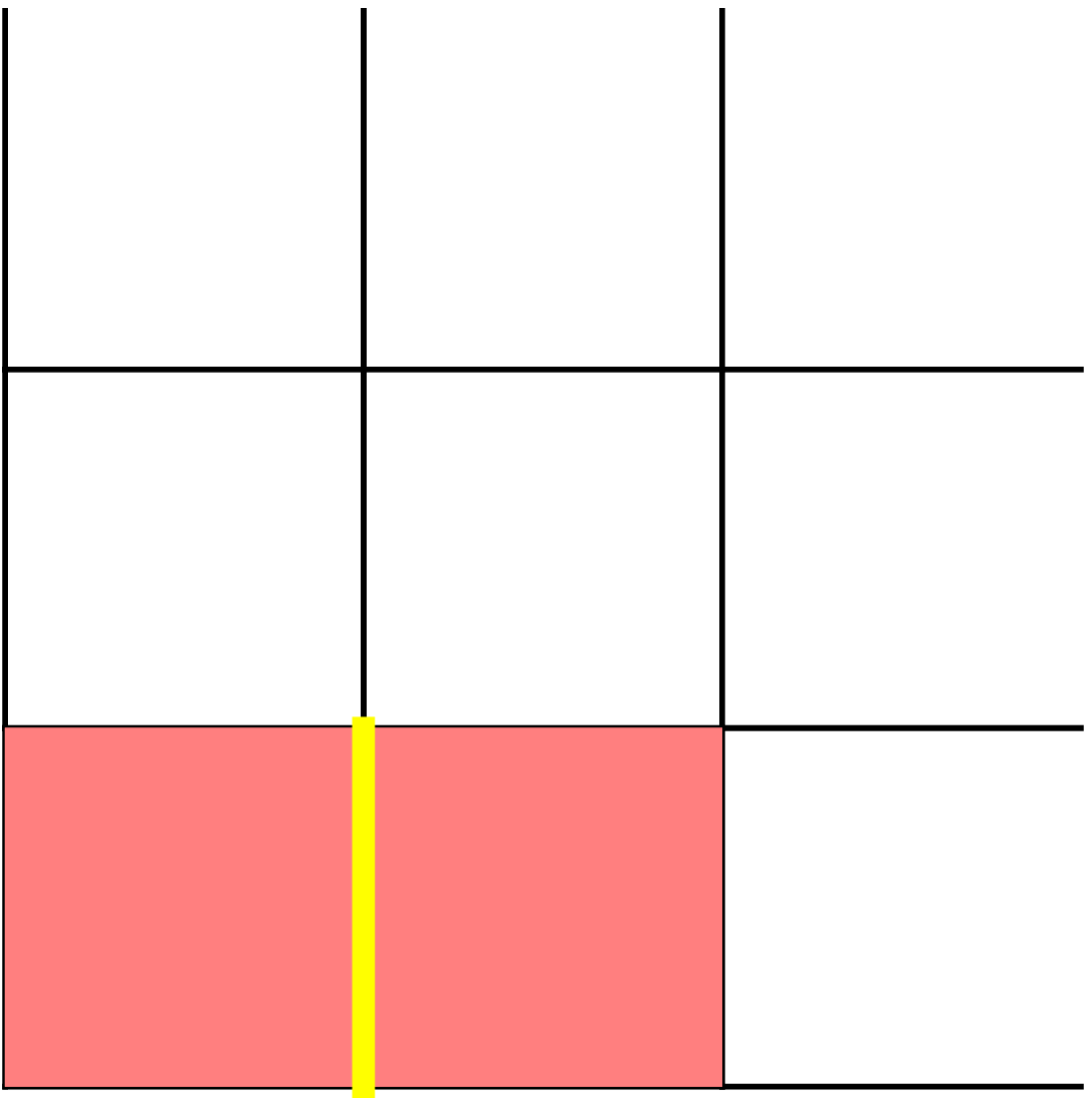}
\subfigure[]{}
\includegraphics[scale=\SCLwlpdf]{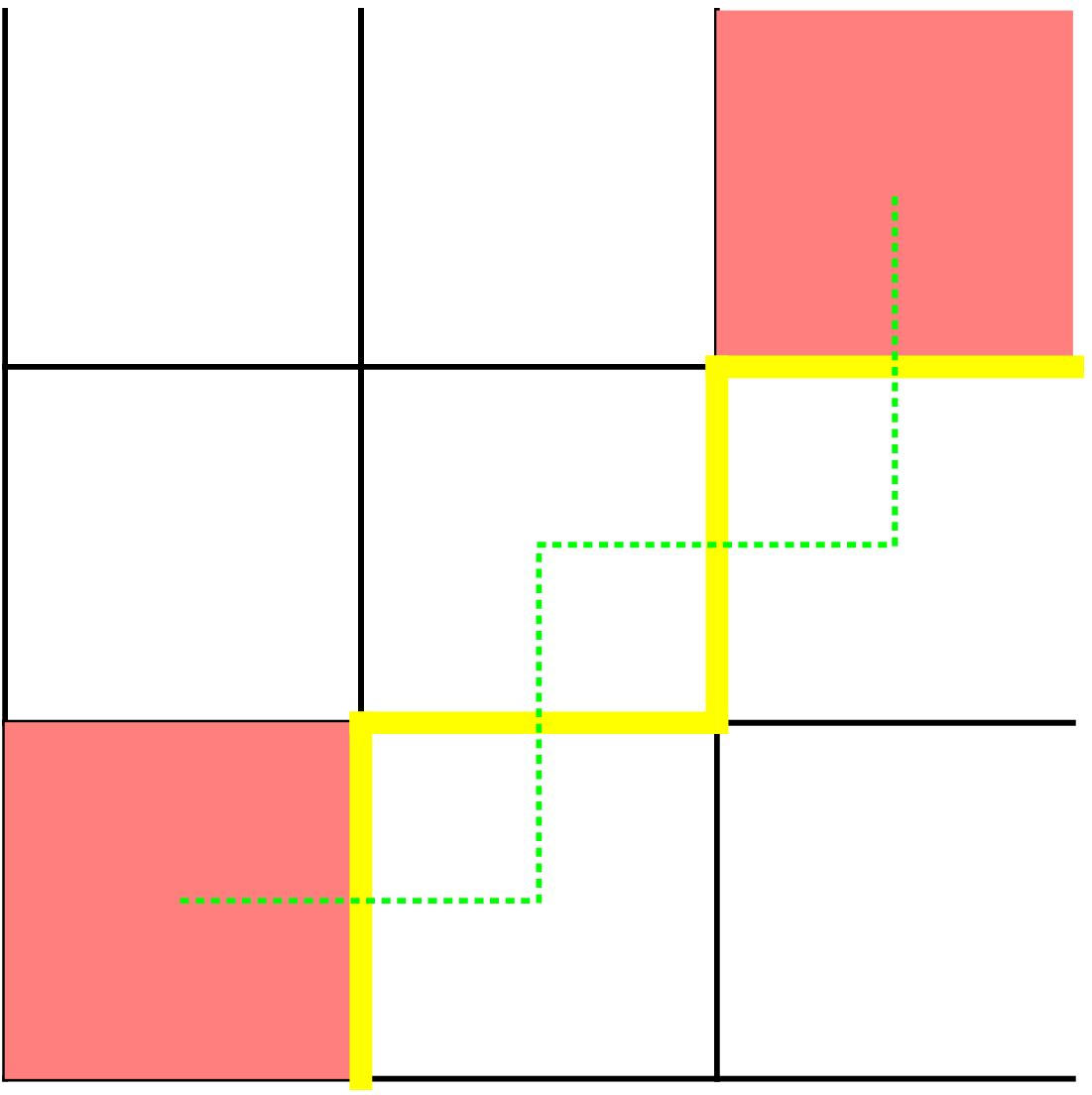}
\subfigure[]{}
\includegraphics[scale=\SCLwlpdf]{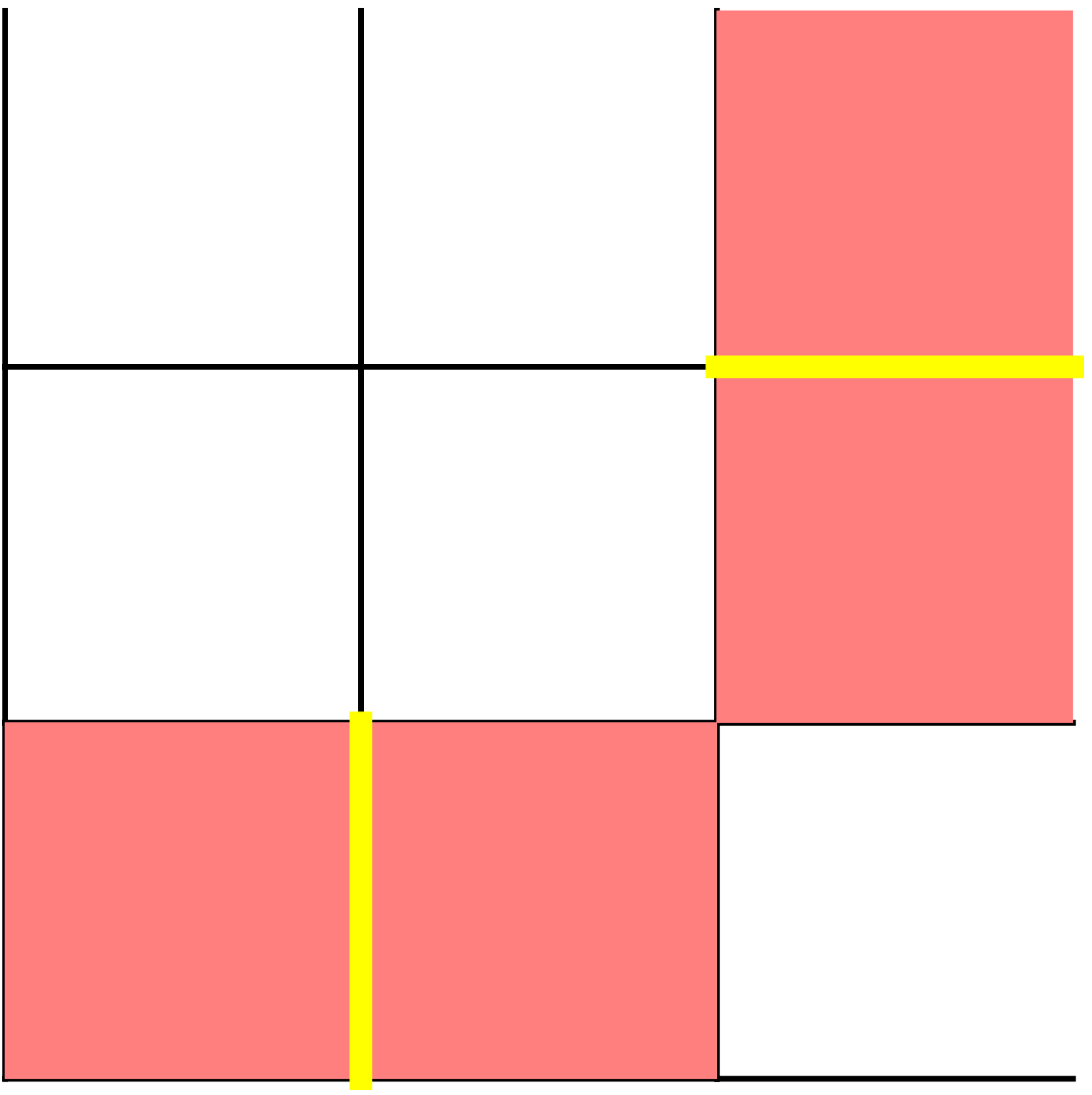}
\caption{Each yellow link represents flipping  the  eigenstate of  $\sigma_z$, each pink $\square$ means  $Z_\square =1$. (a) d=3 lattice with one qubit flipped,  thus 4  plaquettes reversed. (b) d=3 lattice with two neighboring qubits flipped,  thus  6  plaquettes reversed.  (c)  d=2 lattice with one qubit flipped, thus 2 plaquettes reversed. (d) d=2 lattice with  2 separated visons created by a string of $\sigma_x$ operators. (e) d=2 lattice with 2 qubits on different plaquettes flipped,  thus 4 plaquettes reversed.}
\label{fig_vison}
\end{figure}

As shown in Fig.~\ref{fig_dos} and Fig.~\ref{fig_dos_1st},  the above deduction is fully verified by our simulations, in which the possible eigenvalues $z_i$'s of $Z$ are indeed $-24$, $-16$, $-12$, $\cdots$, $12$, $16$, $24$.  Also note that in the ground state, those  $Z$ eigenstates with  large positive eigenvalues  are difficult to occupy, as can be seen in Fig.~\ref{fig_dos} and Fig.~\ref{fig_dos_1st}.

Interestingly, as can be seen in Fig.~\ref{fig_dos}   and Fig.~\ref{fig_dos_1st},  the possible eigenvalues of  $X$ are the same as those of $Z$, that is,  $-24$, $-16$, $-12$, $\cdots$, $12$, $16$, $24$, despite  flipping the $\sigma^x$ eigenstate of one qubit  changes the eigenvalue of $X$  only by $2$. This is a consequence  of gauge invariance, which dictates that   the qubits flipped in $\sigma^x$ eigenstates must be in closed loops, whose possible perimeters are 4, 6, 8, $\cdots$, or $4+2n$, ($n=0,1,\cdots$), thus $X$ can only be  changed by $8+4n$. This is valid in any $d \geq 2$. In d=3, the fact that $Z_\square$'s of at least 4 plaquettes are reversed corresponds to the fact that the perimeter of a loop is at least 4.

This also confirms the self-duality in d=3, which implies that the possible eigenvalues of $X$ must be the same as those of $Z$. Moreover, self-duality implies $D(g,z)= D(1/g,x=z)$, which is also clearly confirmed in our simulation (Fig.~\ref{fig_dos} and Fig.~\ref{fig_dos_1st}).

From DOS' of  $Z$  and   $X$, it is calculated that with the increase of $g$ from $0$,   the expectation value of $Z$ increases from    $-24$   towards $0$, and while  the expectation value of  $X$   decreases from $0$  to   $-24$ (Fig~\ref{fig_e}).

In d=2, $Z_\square$'s of  two plaquettes are reversed by flipping the $\sigma^z$  eigenstate of one qubit. More generally, it is  possible to create  2 visons by flipping $\sigma^z_l$'s of a string of qubits. Therefore the number of flipped  plaquettes is $2n$, $(n = 1, 2,\cdots)$,   as   shown in Fig.~\ref{fig_vison}. In d=2 $3\times3$ lattice with periodic boundary condition, the  possible eigenvalues of  $Z$ are $-9$, $-5$, $\cdots$, $3$, $7$.

In $d=2$,  the  possible eigenvalues of   $X$ are  $\pm 18$, $\pm 10$, $\pm 6$, $\pm 2$, because  the number of qubits flipped in $\sigma^x$ eigenstates must be  $4+2n$, that is, $X$ is changed by $8+4n$,  ($n=0,1,\cdots$), as said above for any $d \geq 2$.     In consistency  with the absence of self-duality, there is no identity $D_Z(z,g)= D_X(x=z,1/g)$, as can be seen in Fig.~\ref{fig_dos} and Fig.~\ref{fig_dos_1st}. As already shown in Fig.~\ref{fig_duality}, with the increase of $g$ from $0$,  the expectation value of $Z$  increases from  $-9$   and   towards $0$, while the expectation value of  $X$  decreases from  $0$    towards  $-18$.

\begin{figure}
\centering
\subfigure[]{}
\includegraphics[scale=0.27]{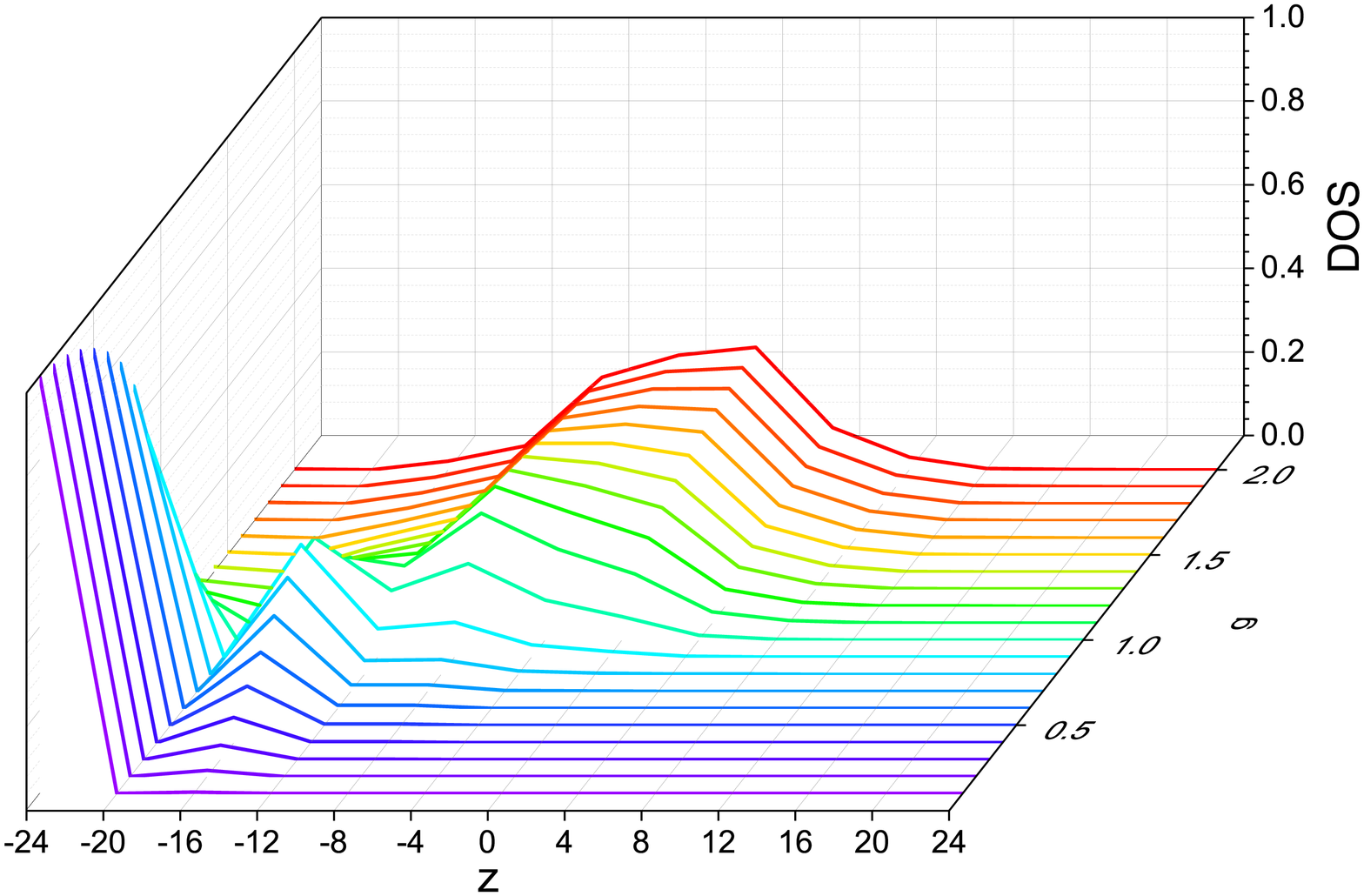}
\hspace{0.01\linewidth}
\subfigure[]{}
\includegraphics[scale=0.27]{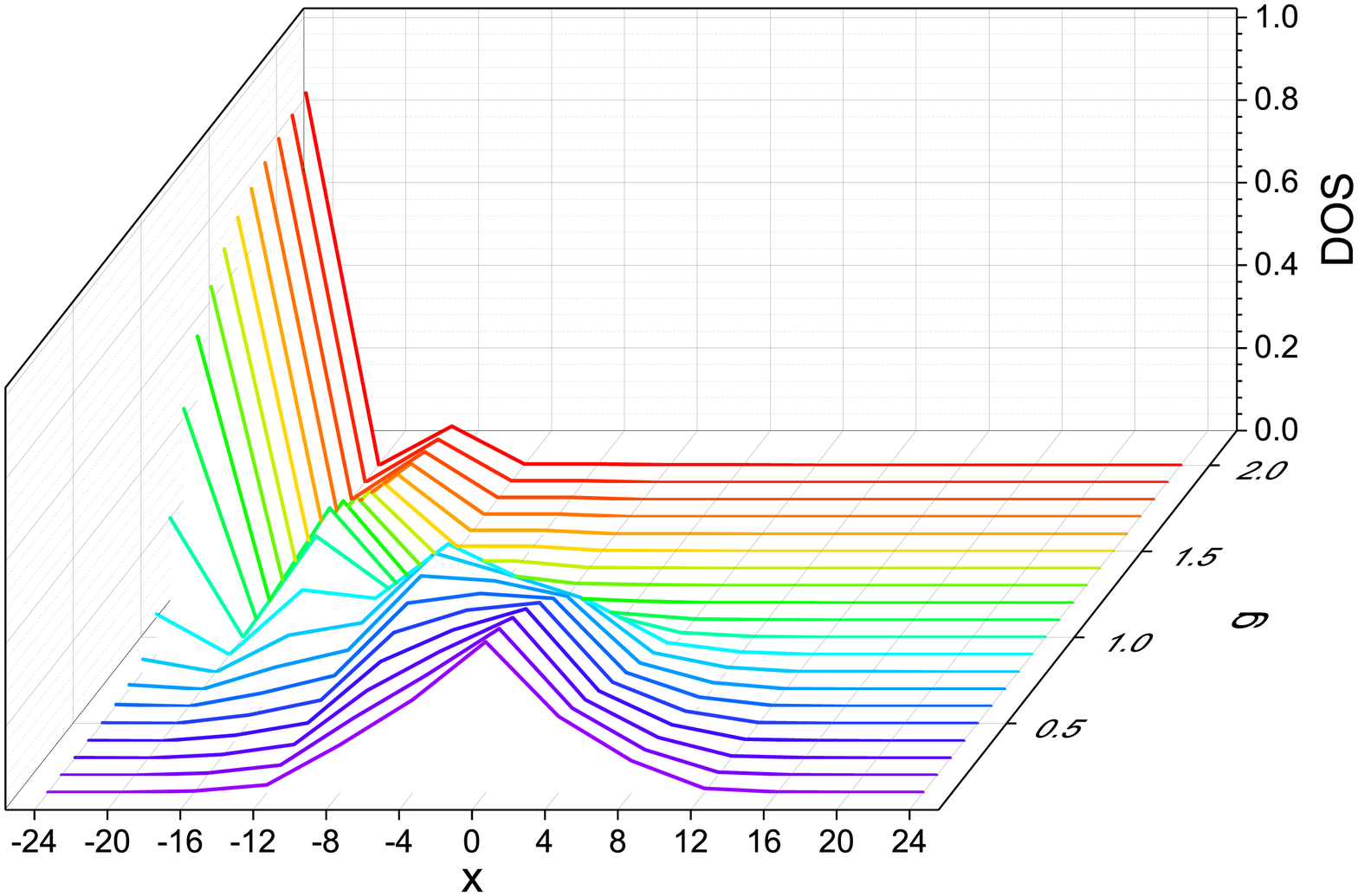}
\hspace{0.03\linewidth}

\subfigure[]{}
\includegraphics[scale=0.27]{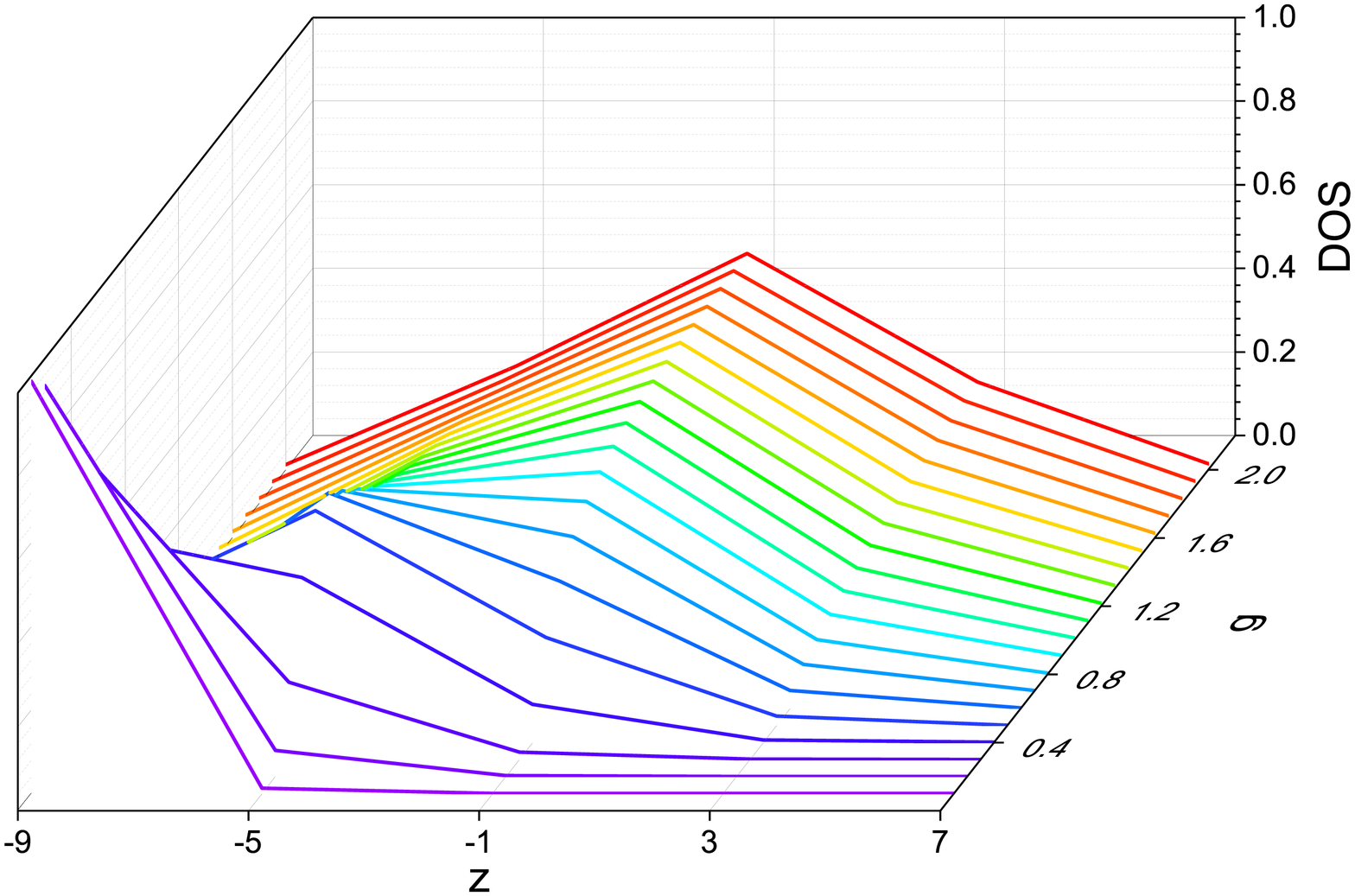}
\hspace{0.01\linewidth}
\subfigure[]{}
\includegraphics[scale=0.27]{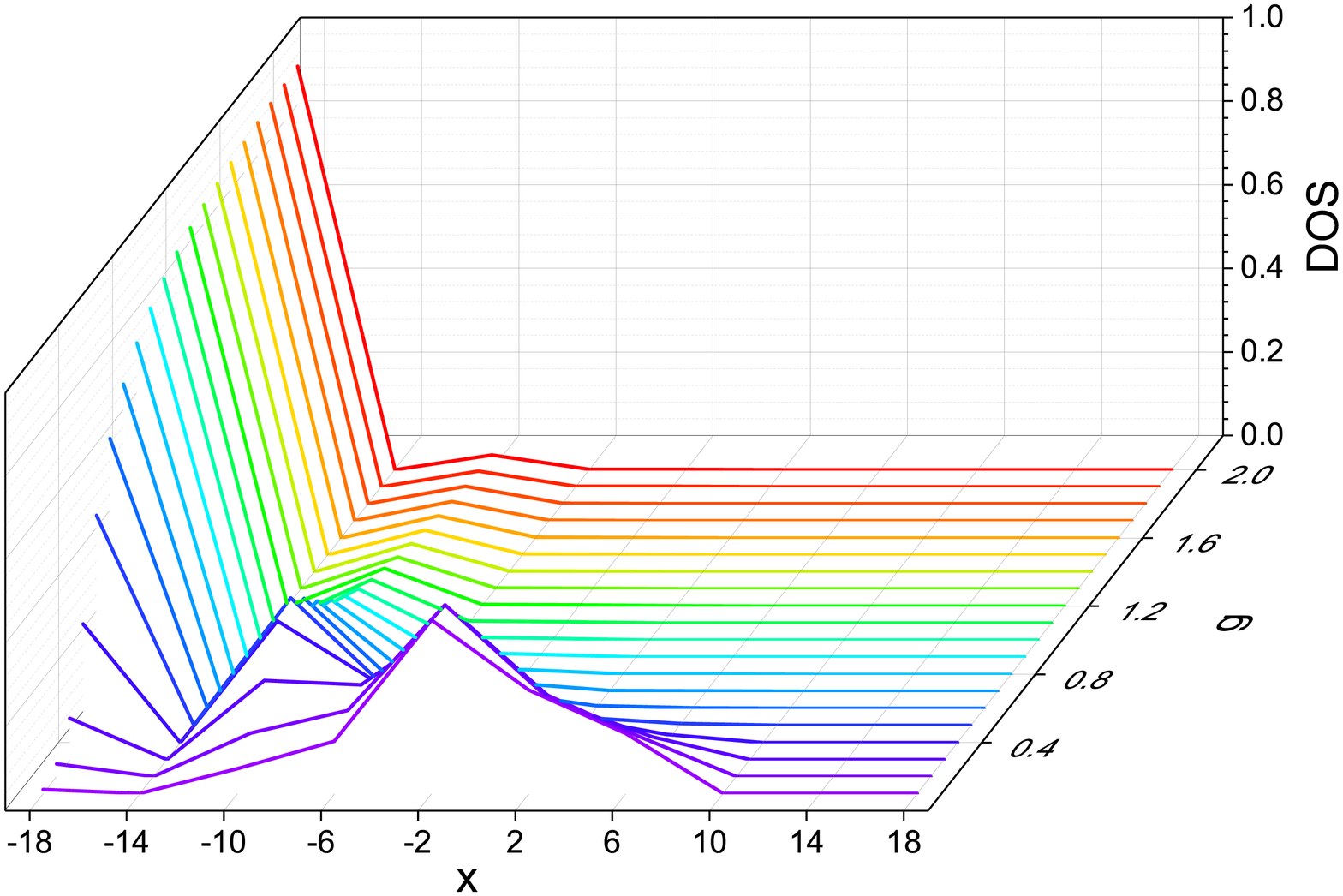}
\caption{Ground-state DOS of eigenstates of  $Z$ or  $X$  as  functions of the eigenvalues $z$ or $x$, and  $g$.  (a)  DOS of $Z$  eigenstates for   d=3 $2\times2\times2$  lattice. Note that it identically vanishes when   $z\neq -24+4n$, which are not considered in the plot. The prominent feature is that it also vanishes at $z=\pm 20$.   (b)   DOS of $X$ eigenstates for  d=3 $2\times2\times2$  lattice.  Note that it identically vanishes when   $x\neq -24+4n$, which are not considered in the plot. The prominent feature is that it also vanishes at $x=\pm 20$.  (c) DOS of $Z$ eigenstates for  d=2 $3\times3$  lattice.   Note that it identically  vanishes when  $z\neq -9+4n$, which are not considered in the plot.    (d) DOS of $X$ eigenstates for d=2 $3\times3$  lattice. It vanishes when   $x\neq -18+4n$, which are not considered in the plot. The prominent feature is that it also vanishes at $x=\pm 14$.   (a) and (b)  satisfy  $D_Z(z,g)= D_X(x=z,1/g)$, confirming  self-duality in d=3.  (c) and (d) confirm the absence of  self-duality in d=2.     }
\label{fig_dos}
\end{figure}

\begin{figure}[htb]
\centering
\subfigure[]{}
\includegraphics[scale=\SCL]{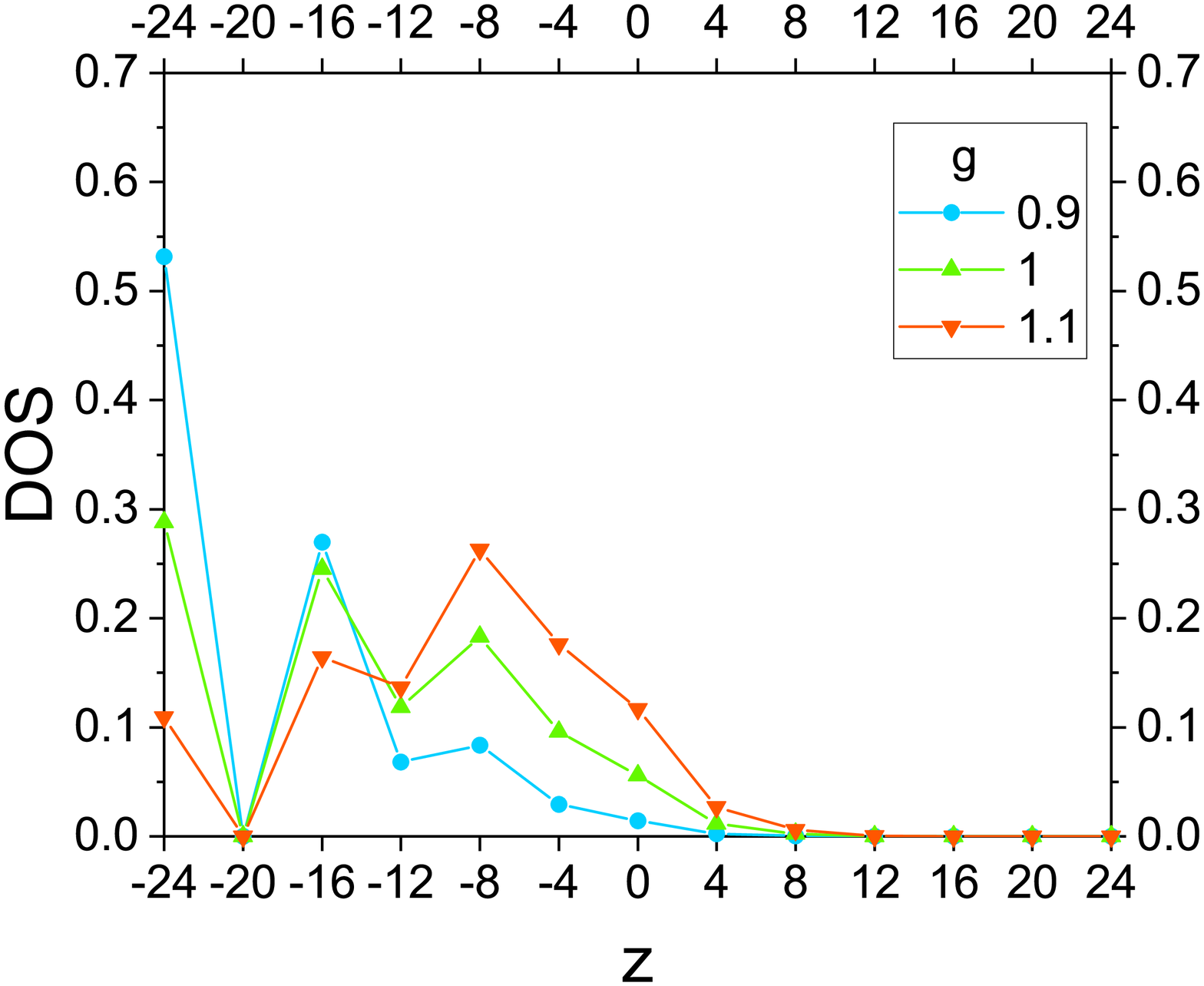}
\hspace{0.01\linewidth}
\subfigure[]{}
\includegraphics[scale=\SCL]{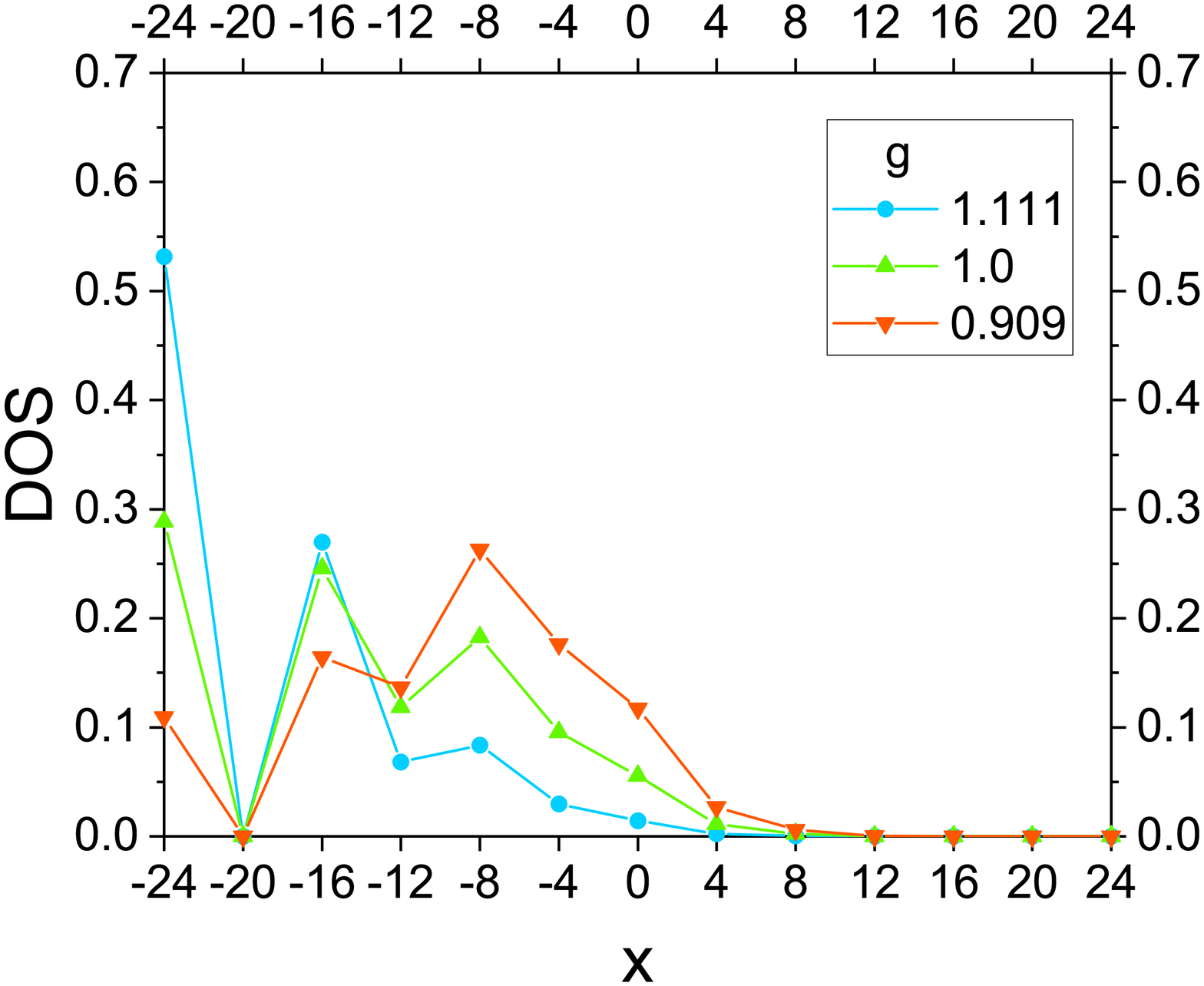}

\subfigure[]{}
\includegraphics[scale=\SCL]{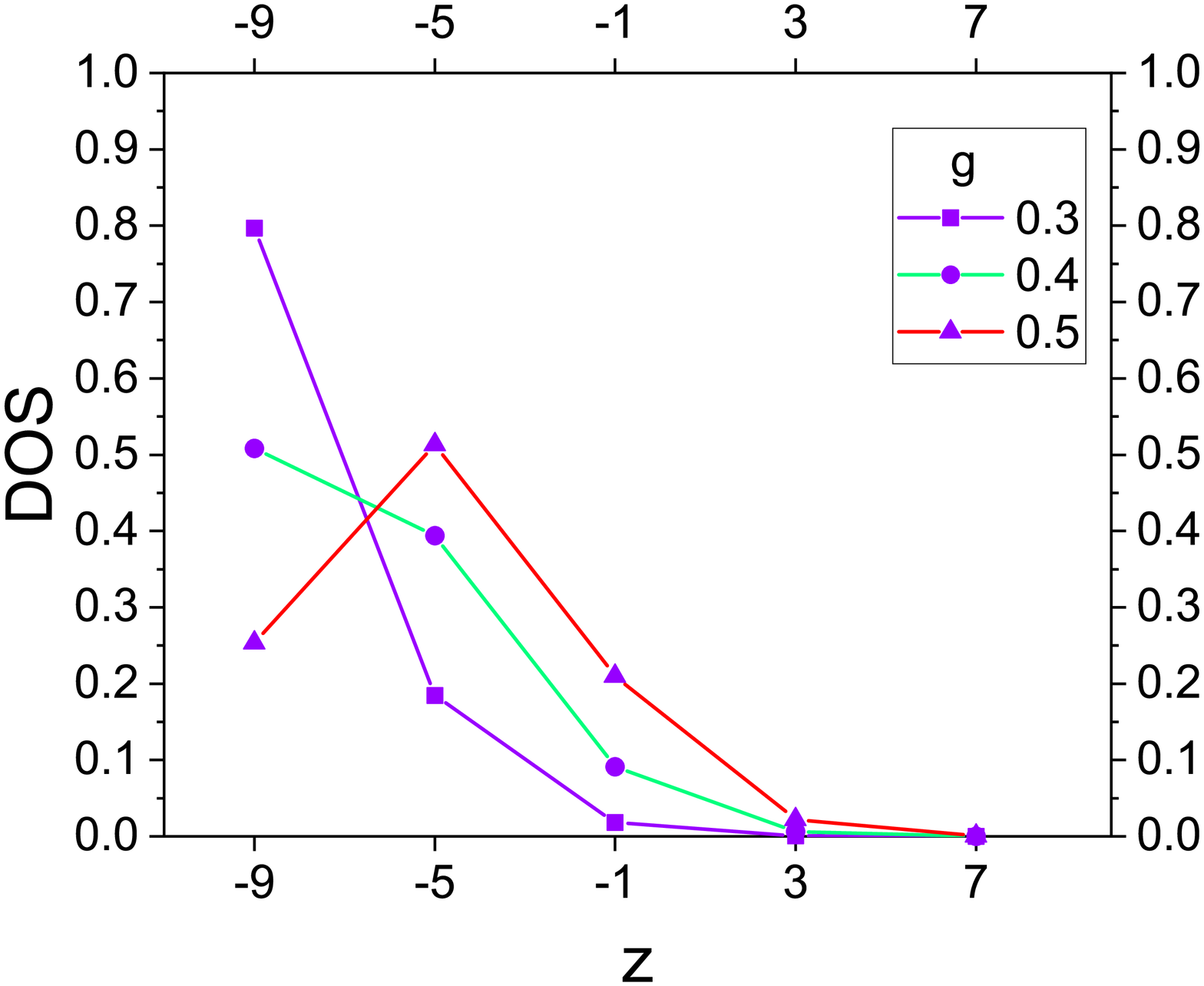}
\hspace{0.01\linewidth}
\subfigure[]{}
\includegraphics[scale=\SCL]{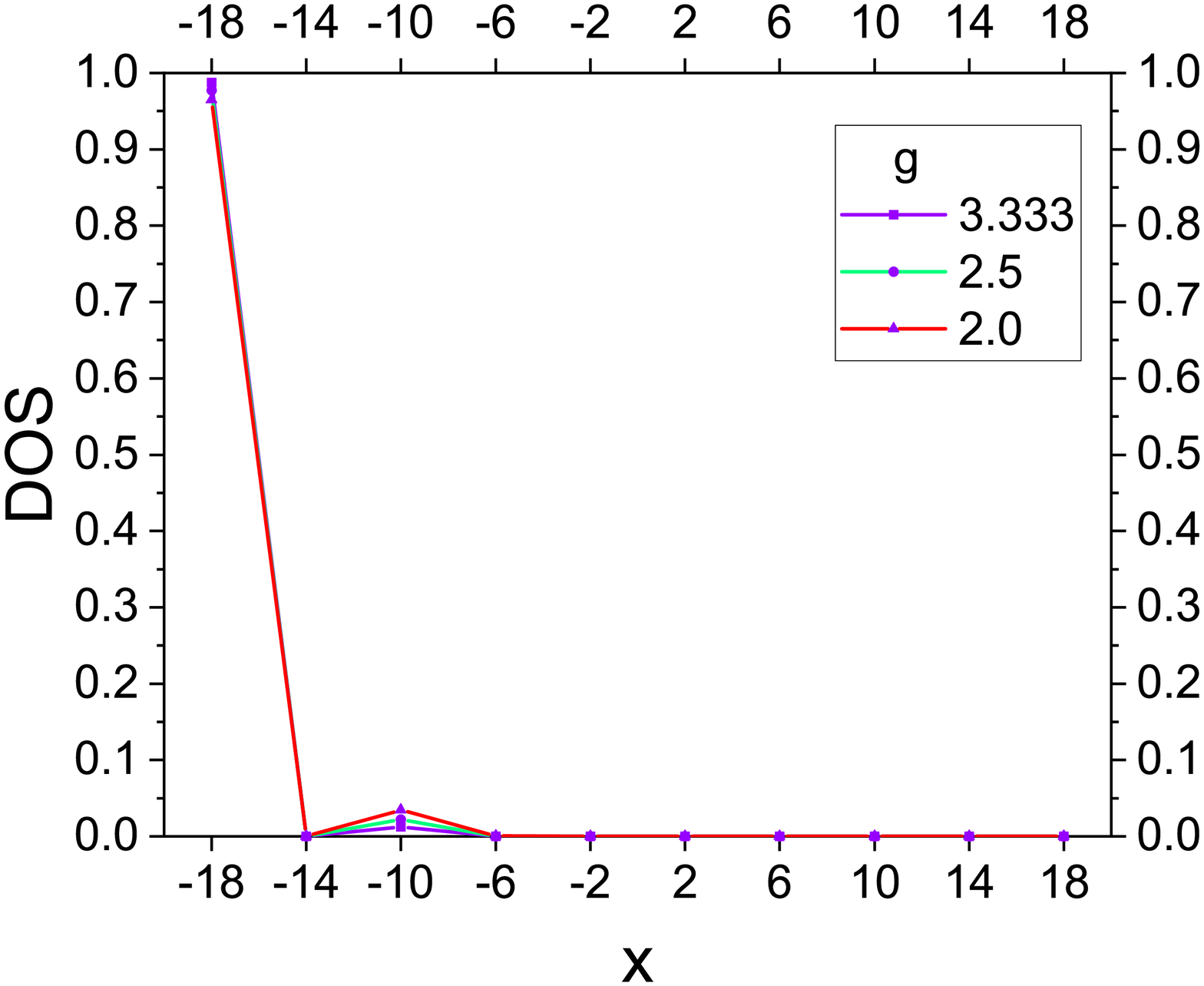}

\caption{ (a) Ground-state DOS of $Z$ eigenstates  as a function of $z$,  with  $g = 0.9$, $1.0$, $1.1$,  on   d=3 $2\times 2\times 2$ lattice. (b) Ground-state DOS of $X$ eigenstates as a function of $x$,  with  $g = 1.111$, $1.0$, $0.909$, which are the inverses of the values of $g$ in (a),  on  d=3 $2\times 2\times 2$ lattice. (c) Ground-state DOS of $Z$  as a function of $z$,  with  $g = 0.3$, $0.4$, $0.5$, on d=2 $3\times 3$ lattice.  (d) Ground-state DOS of $X$  as a function of $x$,  with  $g =2.5$, $2$, $3.333$,  which are the inverses of the values of $g$ in (c), on d=2 $3\times 3$ lattice.  To reach these values, we  specifically extend  the range of $g$ in the adiabatic evolution.  (a) and (b)  satisfy  $D_Z(z,g)= D_X(x=z,1/g)$, confirming  self-duality in d=3.   (c) and (d) confirm the absence of   self-duality in d=2.  }
\label{fig_dos_1st}
\end{figure}

\subsection{Orders of   quantum phase transitions  }

A first-order phase transition is one where there is a discontinuity of a first derivative of the free energy or energy,  at the critical value of the control parameter, in  contrast to a  second-order phase transition,    where the first derivatives are continuous   while there is a discontinuity of its first  derivatives at the critical point.

The thermal phase transition in the classical $\mathbb{Z}_2$ LGT  is first-order  in  4 spatial  dimensions, and  is  second-order  in 3   spatial  dimensions~\cite{Creutz}. This suggests that  for  quantum  $\mathbb{Z}_2$ LGT,  the QPT is first-order in   d=3 spatial dimensions, and is second-order in the d=2 spatial dimensions, because the  classical theory in D spatial dimensions   corresponds to  the   quantum theory  in d=D-1 spatial dimensions, in other words, D=d+1 spacetime dimensions, where 1 represents the time dimension.

Limited by the smallness of the lattice size, one cannot make conclusion about whether there exists discontinuity in the slope of $\langle H\rangle$  in  Fig.~\ref{fig_e} or $\partial\langle H\rangle/\partial g$ in Fig.~\ref{fig_de}.

\begin{figure}[htb]
\centering
\subfigure[]{}
\includegraphics[scale=0.3]{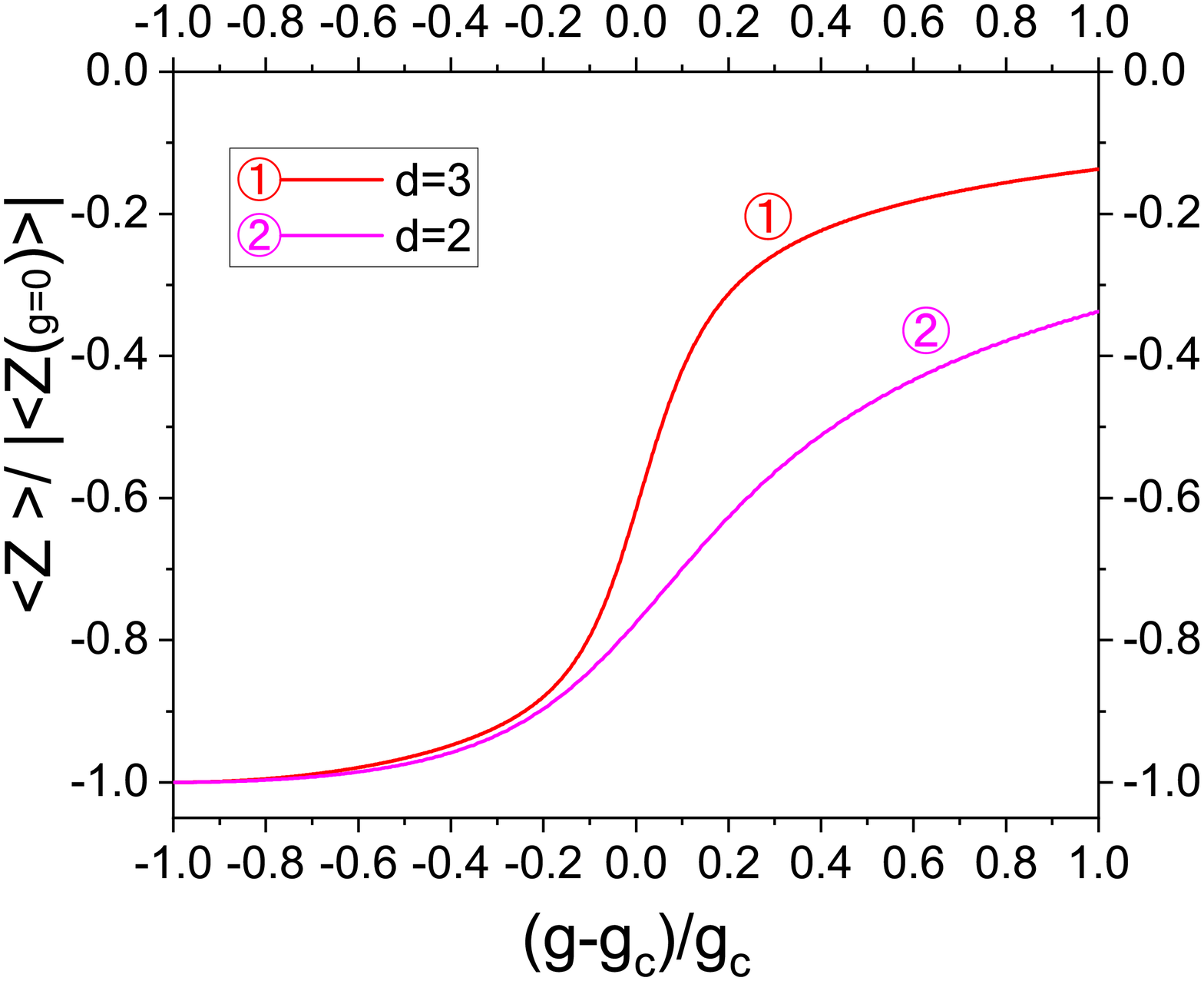}
\subfigure[]{}
\includegraphics[scale=0.3]{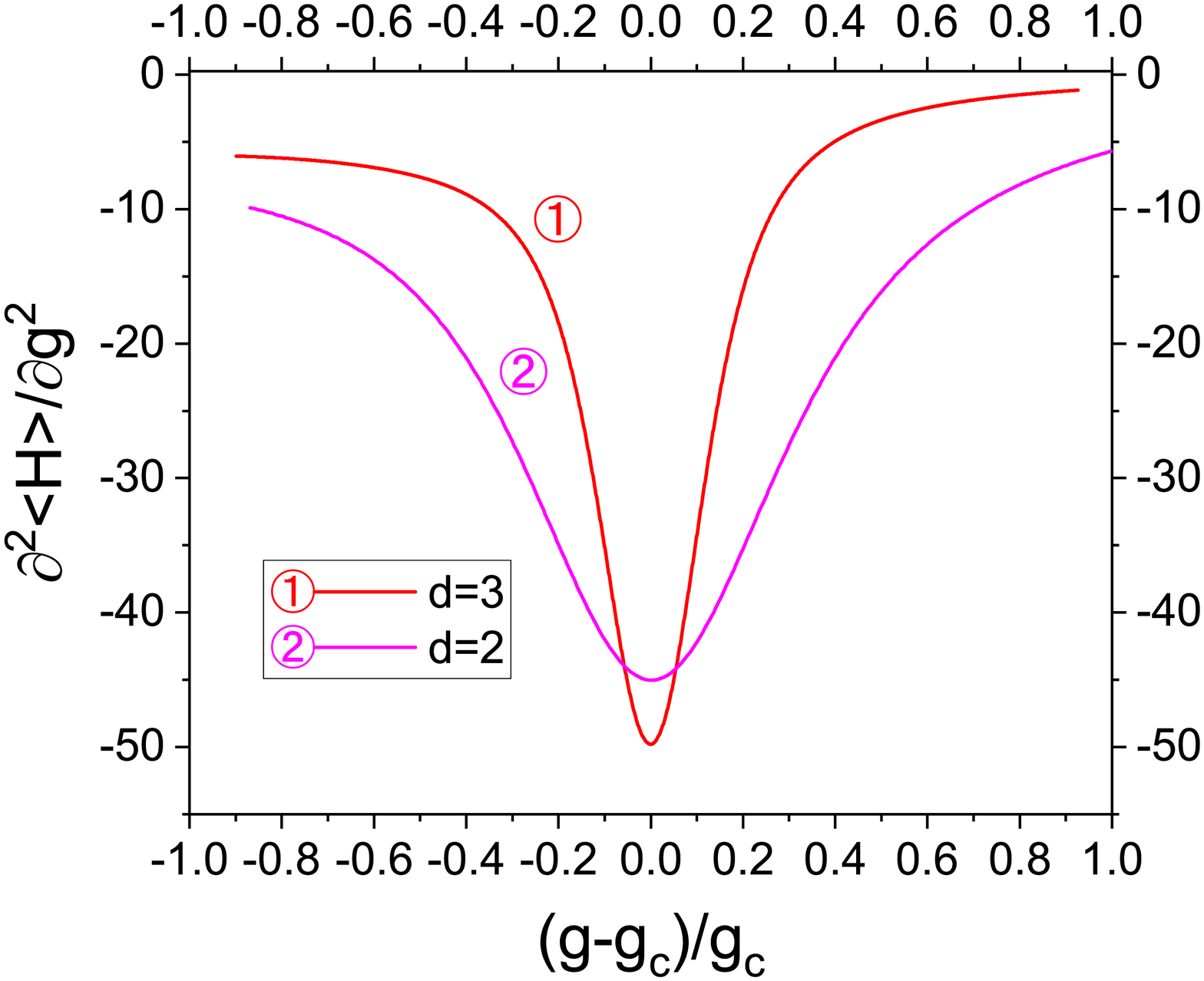}

\caption{ Comparison of results for d=3 $2\times 2\times 2$ and d=2 $3 \times 3$ lattices, as functions of $(g-g_c)/g_c$, where $g_c$ is the respective critical value in each dimensionality.  (a)  $\langle Z\rangle$, relative to the absolute value at $g=0$.  (b) Second derivative of the energy with respect to $g$.    It can be seen clearly that in d=3, the change of $\langle Z\rangle $ at QPT is much steeper, and  the valley of second derivative of the energy is much sharper.}
\label{fig_de2}
\end{figure}

Therefore, we make a comparison between d=3 and d=2,  by putting together $\langle Z\rangle$ and  second order derivatives  $\partial^2\langle H\rangle/\partial g^2$ for d=2 and d=3, as functions of $(g-g_c)/g_c$, which is  dimensionless   (Fig.~\ref{fig_de2}). Then it  can be seen clearly that in d=3, the change of $\langle Z\rangle $ at QPT is much steeper and the valley is much sharper.  This  would be consistent with  the claim  that  in  d=3, QPT is first-order, even though the discontinuity and the singularity are rounded out,   while in d=2, QPT is second-order.

Now we examine the properties of the ground states. As can be seen in  Fig.~\ref{fig_dos_1st}, in  d=3,  distributions of  DOS of $Z$ appear  significantly different before and after QPT. For example, when $g=0.9 < g_c$, the peaks are at  $z=-24$ and  $z=-16$; when $g=1.1>g_c$, the peak is at  $z=-8$.  When $g=g_c=1$, there are peaks   at $z=-24$ and  $z=-16$, i.e. the locations of peaks  when $g < g_c$, and at  $z=-8$, i.e. the location of  peak   when $g > g_c$. Same feature exists in DOS of X, by replacing $g$ as $1/g$ because of self-duality.

This feature indicates a quantum version of phase coexistence at the critical point, a hallmark of first-order phase transition.  When $g<g_c$, the ground state slowly varies with $g$, as a same phase. When $g>g_c$, the ground state is dramatically different from that for $g<g_c$, as indicated by their distinct DOS distributions. The ground state for  $g>g_c$    also slowly varies with $g$, as a same phase  within this regime. When $g=g_c$, the ground state is roughly a nearly-equal superposition of the two states. This feature is absent in d=2, also seen in Fig.~\ref{fig_dos_1st}.

Therefore, our simulations support the implication from 4-dimensional classical $\mathbb{Z}_2$ LGT that  QPT in  quantum  $\mathbb{Z}_2$ LGT  is first-order in d=3, while it is second-order in d=2.

Now we go back to the  expectations of  Wegner-Wilson operators of different loops (Fig~\ref{fig_wloop_results}). In d=3, there is a dip in the ratio of the logarithms, which is absent in d=2. We have examined that the dip is right  at  $g_c$ determined from the lowest point of the $\partial^2\langle H\rangle/\partial g^2$  in  Fig.~\ref{fig_de2}. It is not excluded that the dip is a finite-size effect. It is also  possible that the dip is a signature of first-order QPT at $g=g_c$, as  a consequence of the quantum phase coexistence.  The matrix element of the Wegner-Wilson loop operator  between the two states representing the two phases possibly decrease the expectation in their superposition representing the phase coexistence. The dip in the ratio between c3 and c1 is deeper than that in the ratio between  c2 and c1, in consistency with property that the larger the loop, the stronger the effect.  This conjecture is supported by the absence of such a dip in d=2, where QPT is second-order,  and also by the existence of a similar dip  in a quantity studied in a  quantum adiabatic algorithm for the exact cover problem, which was regarded as a criterion for first-order QPT~\cite{Young}. Nevertheless, no definite conclusion can be drawn yet.

\subsection{Topology}

The phase transition in the classical $\mathbb{Z}_2$ LGT is between states that cannot be distinguished in  symmetry~\cite{Wegner,Kogut}. It is now called topological phase  transition~\cite{Sachdev}. Our simulation verifies that QPT  in quantum  $\mathbb{Z}_2$ LGT is also topological.

First, the absence of symmetry breaking  is directly verified in  $\langle \sigma^z_l\rangle $ of qubit $l$, as a function of $g$. As can be seen in Fig. \ref{fig_one}, $\langle \sigma^z_l\rangle$ remains consistent with $0$ for all values of $g$.   For this qubit $l$, we have also studied the expectation values of $x_l \equiv  -\sigma_l^x$,  $z_l = \frac{1}{4} \sum_{\square \ni l } {Z_{\square} }$, which is the average of $Z_{\square}$'s of the plaquettes sharing the qubit $l$ normalized by the number of plaquett per link,   and  $h_l = z_l+x_l$.

\begin{figure}[htb]
\centering
\subfigure[]{}
\includegraphics[scale=\SCL]{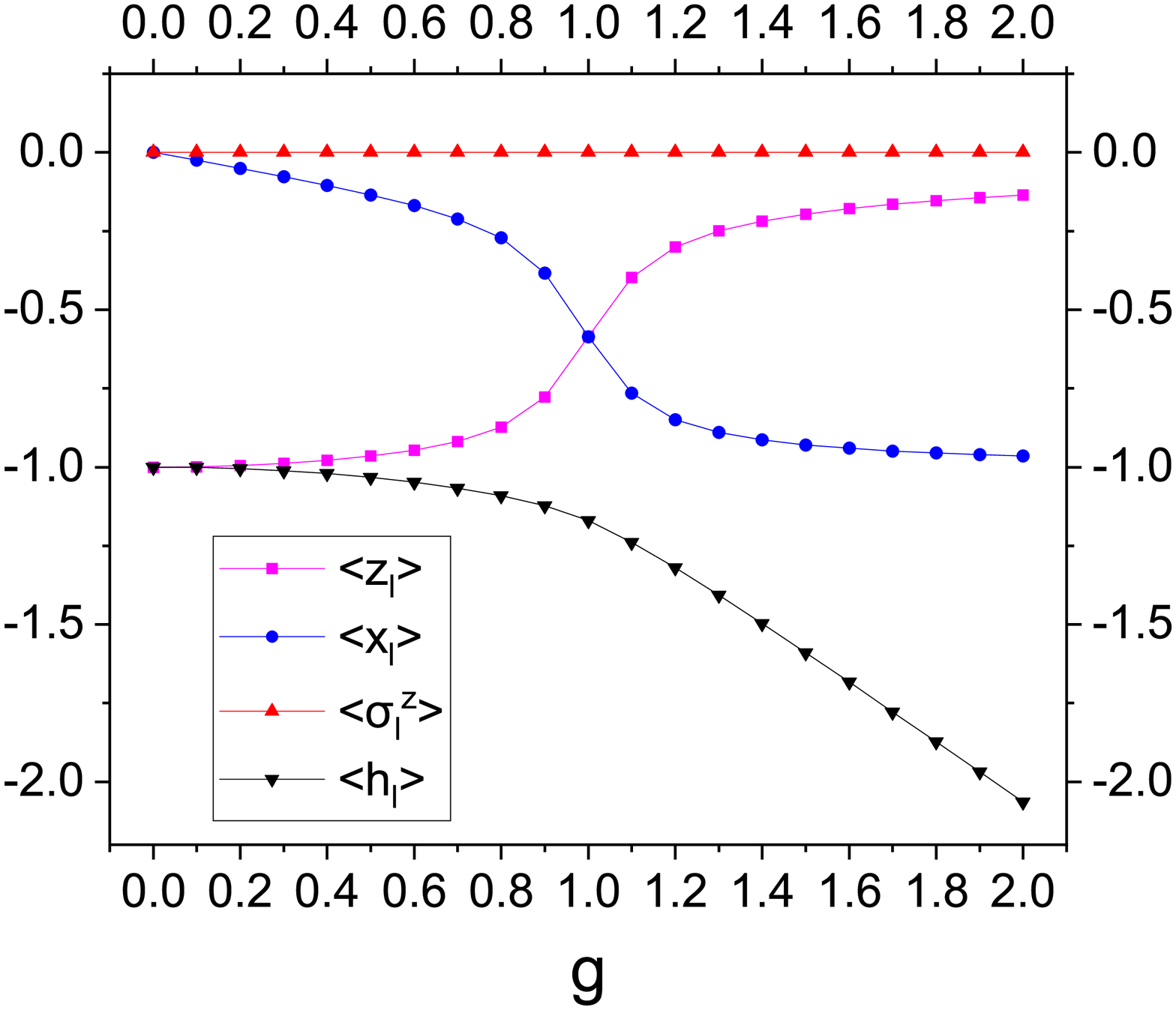}
\subfigure[]{}
\includegraphics[scale=\SCL]{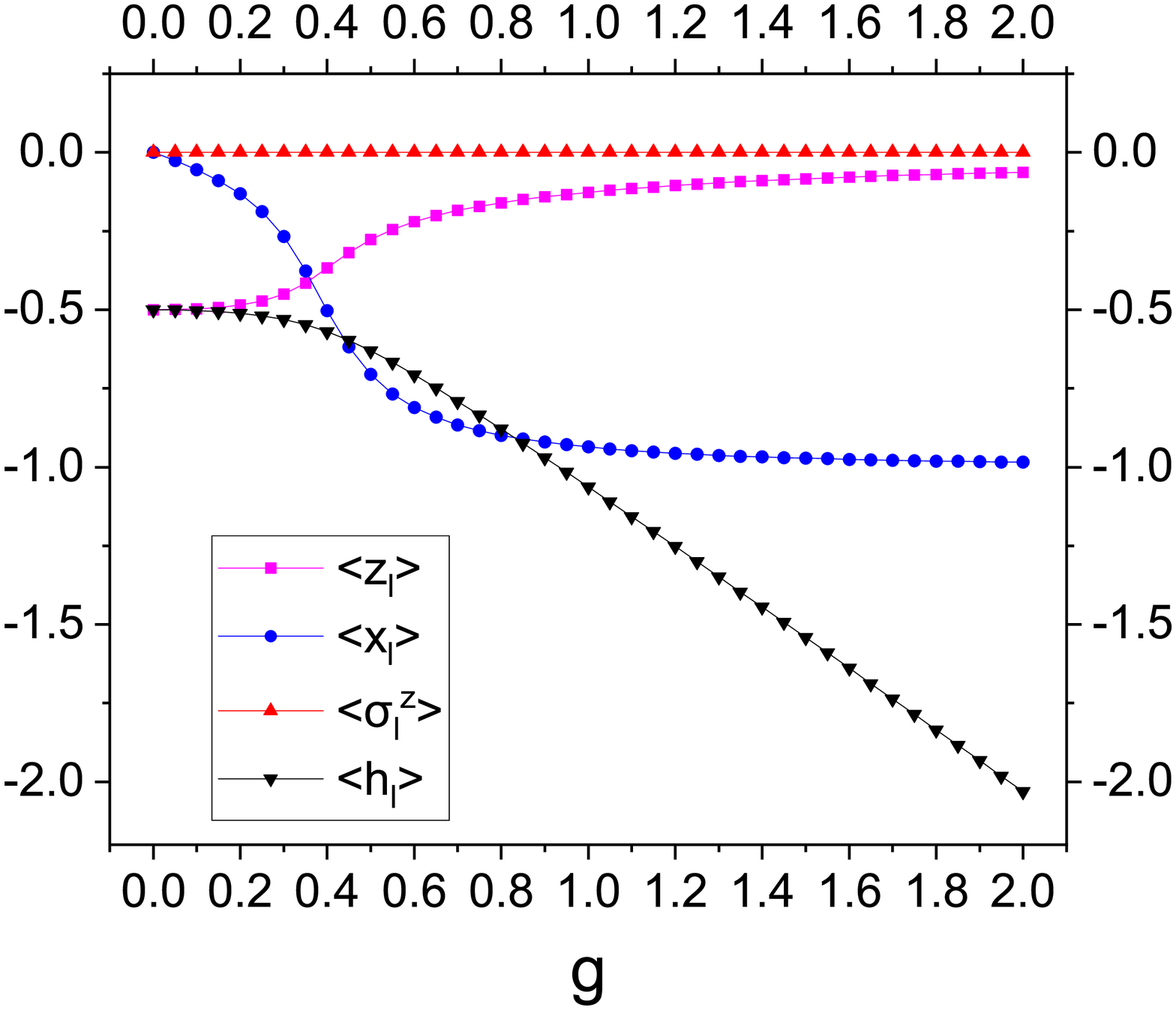}
\caption{ $\langle \sigma^z_l\rangle$, $\langle z_l\rangle$, $\langle x_l\rangle$ and $\langle h_l\rangle$ of one qubit $l$,  as functions of $ g$. (a) d=3 $2\times2\times 2$ lattice.     (b) d=2 $3\times 3$ lattice. }
\label{fig_one}
\end{figure}

Second, the above  results on the DOS of $X$  indirectly verifies the existence of  visons  (Fig.~\ref{fig_dos}).   Two separated visons can only be annihilated by a nonlocal operator or by contacting  each other.

Third, we have studied $g$-dependent energy splittings between the ground state and  the other $2^d-1$ common eigenstates of the $d$ 't Hooft loop operators  $V_{\mu}$'s, $\mu=1,\cdots,d$, (Fig.\ref{fig_torus_lattice}).  The presence of these $d$ lowest energy states with exponentially small energy splittings,  scaled as $g^{L+1}$ and vanishing with  $L$,  is a   defining characteristics of $\mathbb{Z}_2$ topological order~\cite{Sachdev}.

In d=3,  at $g=0$,  there are   $8$ degenerate ground states with $V_x=\pm1$, $V_y=\pm1$ and $V_z=\pm1$. For a generic value of $g$, there are 4 energy levels of $E(V_x,V_y,V_z)$, $E_1=E(1,1,1)$, $E_2=E(-1,1,1)=E(1,-1,1)=E(1,1,-1)$, $E_3=E(-1,-1,1)=E(1,-1,-1)=E(-1,1,-1)$, $E_4=E(-1,-1,-1)$.   The remaining degeneracy is due to rotational symmetry.

At $g=0$, we prepare the ground state  $(V_x, V_y,V_z)=(1,1,1)$ and  other three ones  $(1,-1,1)$, $(-1,-1,1)$ and $(-1,-1,-1)$,  using  the corresponding $W_\mu$ operators, defined along  non-contractible loops on the direct  lattice.  Other four  ground states are not studied,  because theoretically it is known that even when $g\neq 0$, each of them remain degenerate with one of the states studied.   Then we execute adiabatic algorithm to obtain the dependence of the energies and the splittings on $g$  before and after the QPT, as  shown in Fig. \ref{fig_torus}.

In d=2, at $g = 0$, we first prepare the ground state with  $V_x=V_y=1$, and then use $W_\mu$ operators to obtain  the other 3 degenerate ground state  with  $V_x=\pm1$ and $V_y=\pm1$. The  quantum adiabatic algorithm is executed  on each of them. They become nondegenerate when $g\neq 0$. In our simulation, we calculate the four energies   $E(V_x,V_y)$, which are $E_1=E(1,,1)$, $E_2=E(-1,1)$ and   $E_3=E(1,-1)$, $E_4=E(-1,-1)$.  The simulation  confirms that   $E_1$ is the lowest while $E_2=E_3$. As shown in the inset of  in (d) of Fig.~\ref{fig_torus},   the splitting $ E_{i1} \equiv E_i-E_1$  can be fitted by least-square method as    \begin{equation}
\begin{array}{rl}
E_{21}^{\frac{1}{L+1}}=E_{31}^{\frac{1}{L+1}}
&= 1.88719g+0.02516, \\
E_{41}^{\frac{1}{L+1}}
&= 2.08307g+0.0441229,
\end{array}
\end{equation}
with $L=3$, which  verify
\begin{equation}
    E_{i1}   \propto g^{L+1},  \label{gap}
\end{equation}
with $L=3$ fixed.

Previous tensor network calculation found that the splittings between the four eigenstates of $V_x$ and $V_y$ exponentially decay with  $L$ for given  small $g$~\cite{Vidal}.  This is consistent with and complements our result, as   $g^{L+1}=ge^{-L/\xi}$, with $\xi \equiv -1/\ln g$. Our result reveals the $g$ dependence but does not give  the $L$ dependence, while their result gives the $L$ dependence but does not give the  $g$ dependence.

\begin{figure}[htb]
\centering
\subfigure[]{}
\includegraphics[scale=0.3]{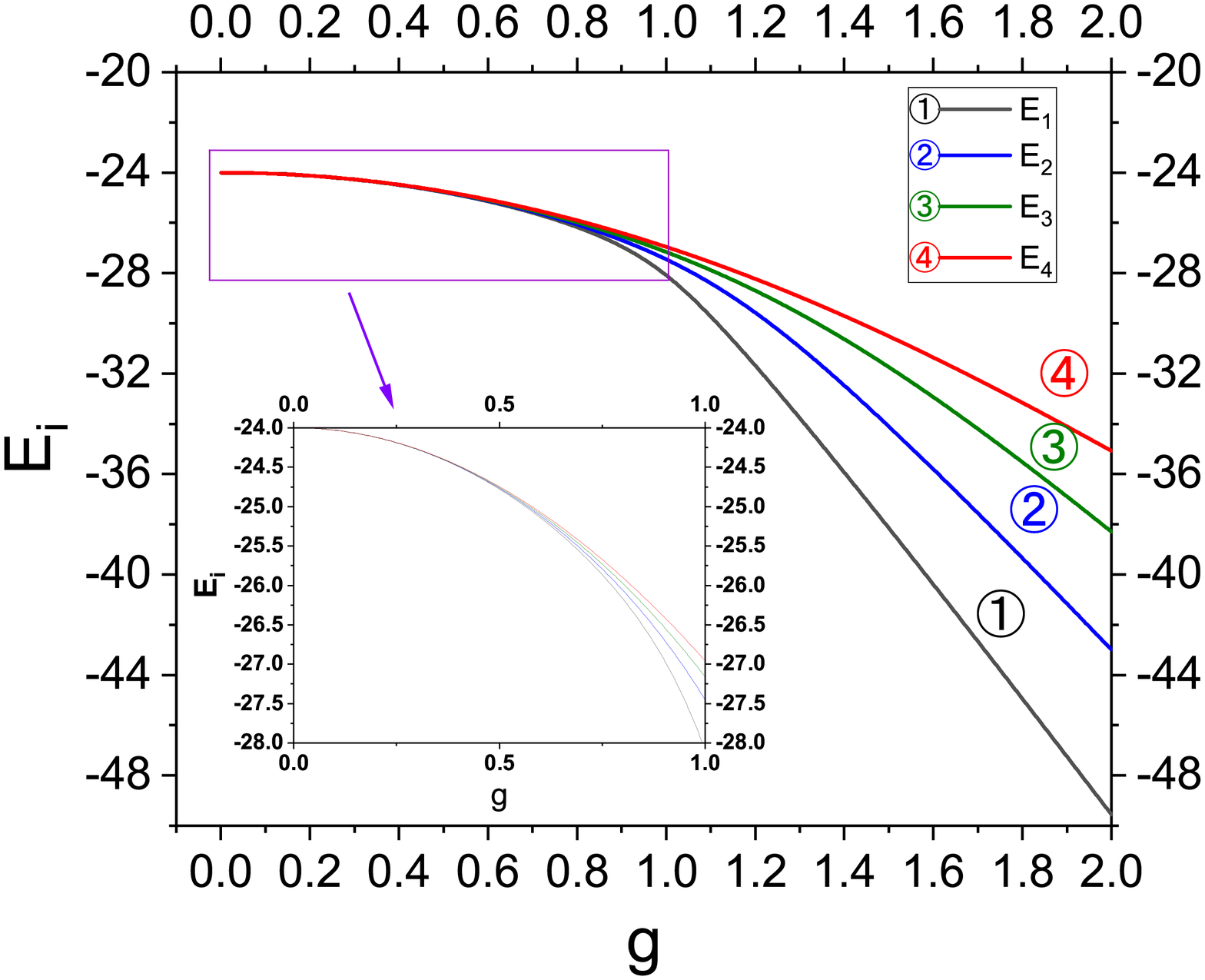}
\subfigure[]{}
\includegraphics[scale=0.3]{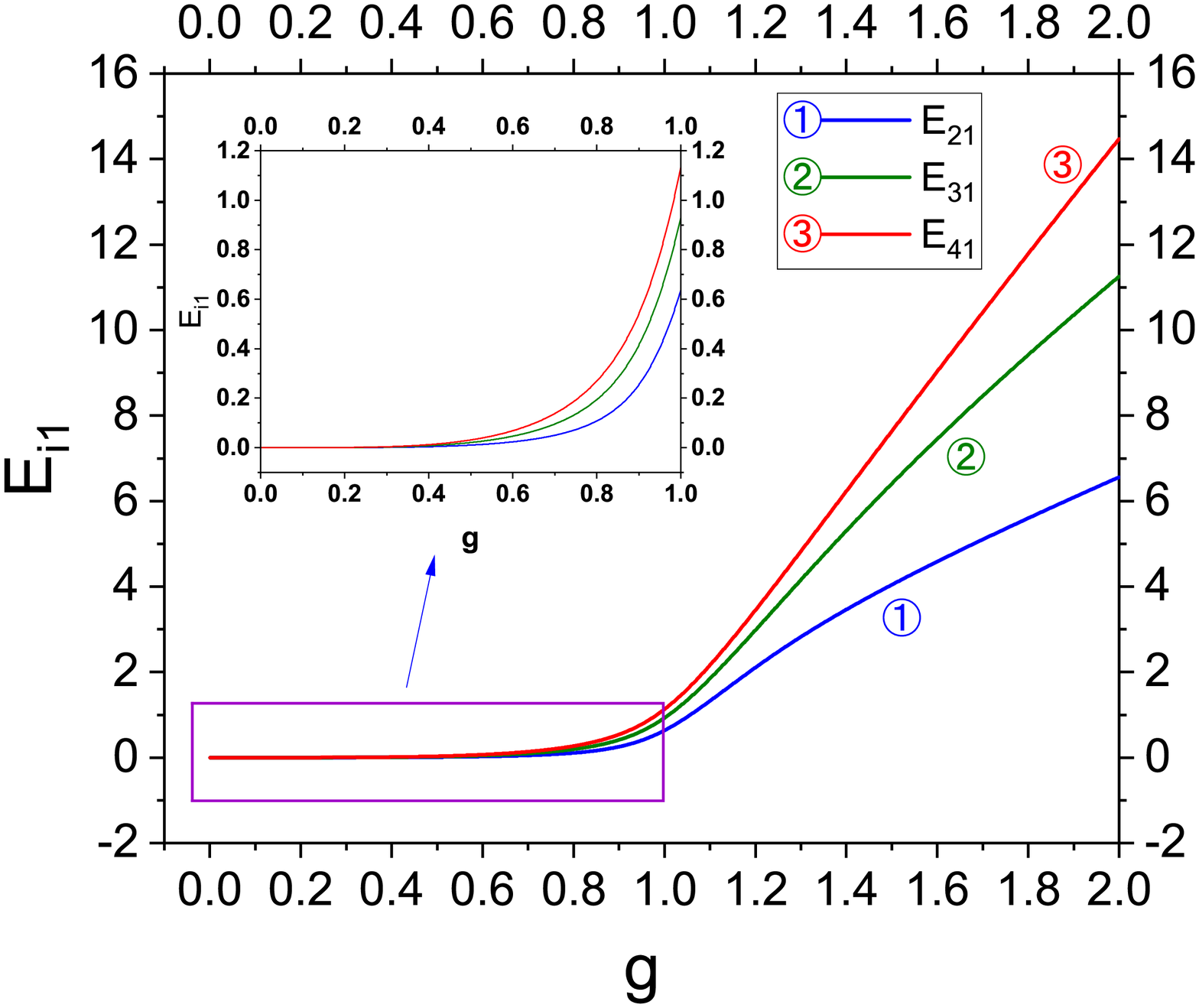}

\subfigure[]{}
\includegraphics[scale=0.29]{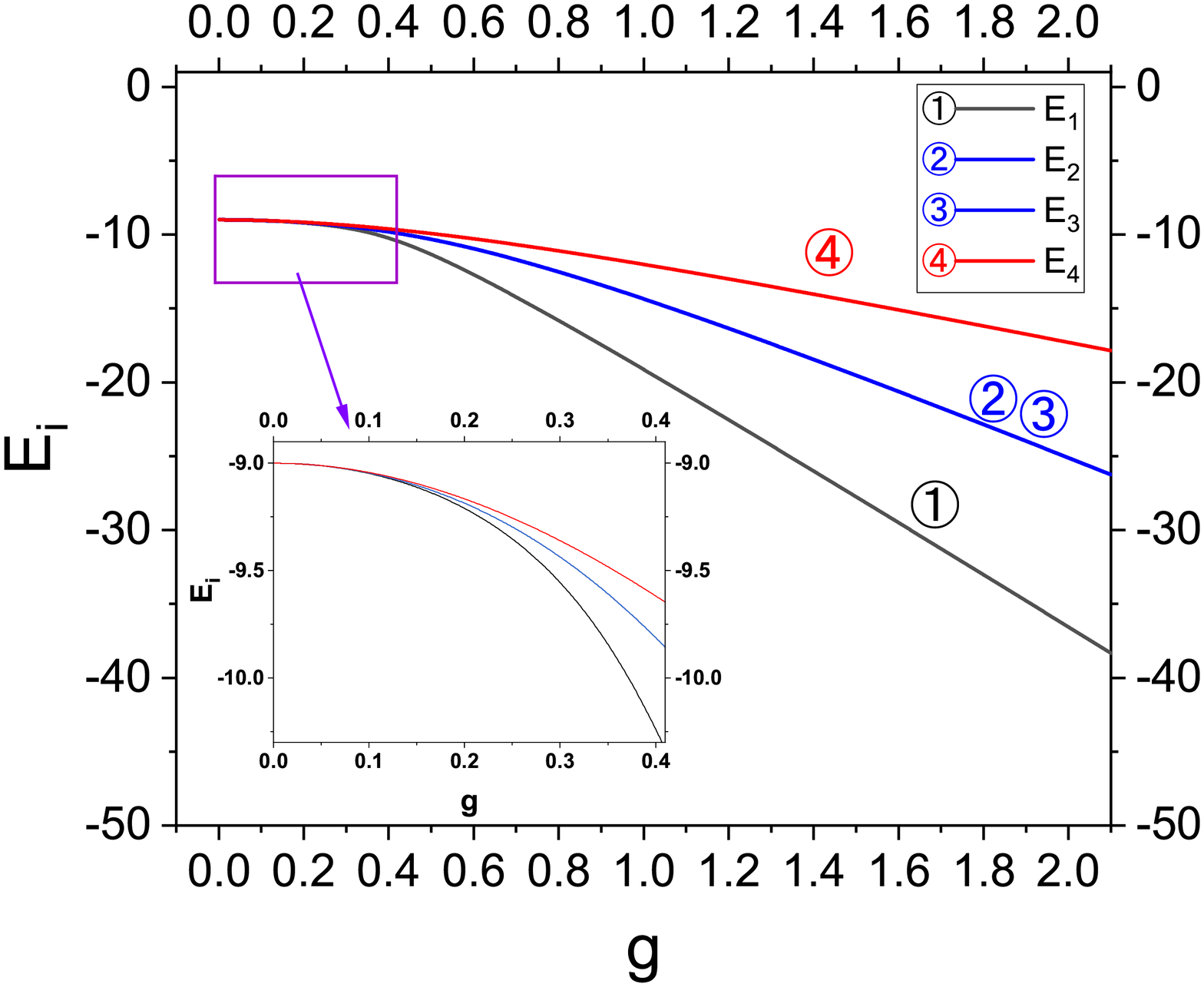}
\subfigure[]{}
\includegraphics[scale=0.29]{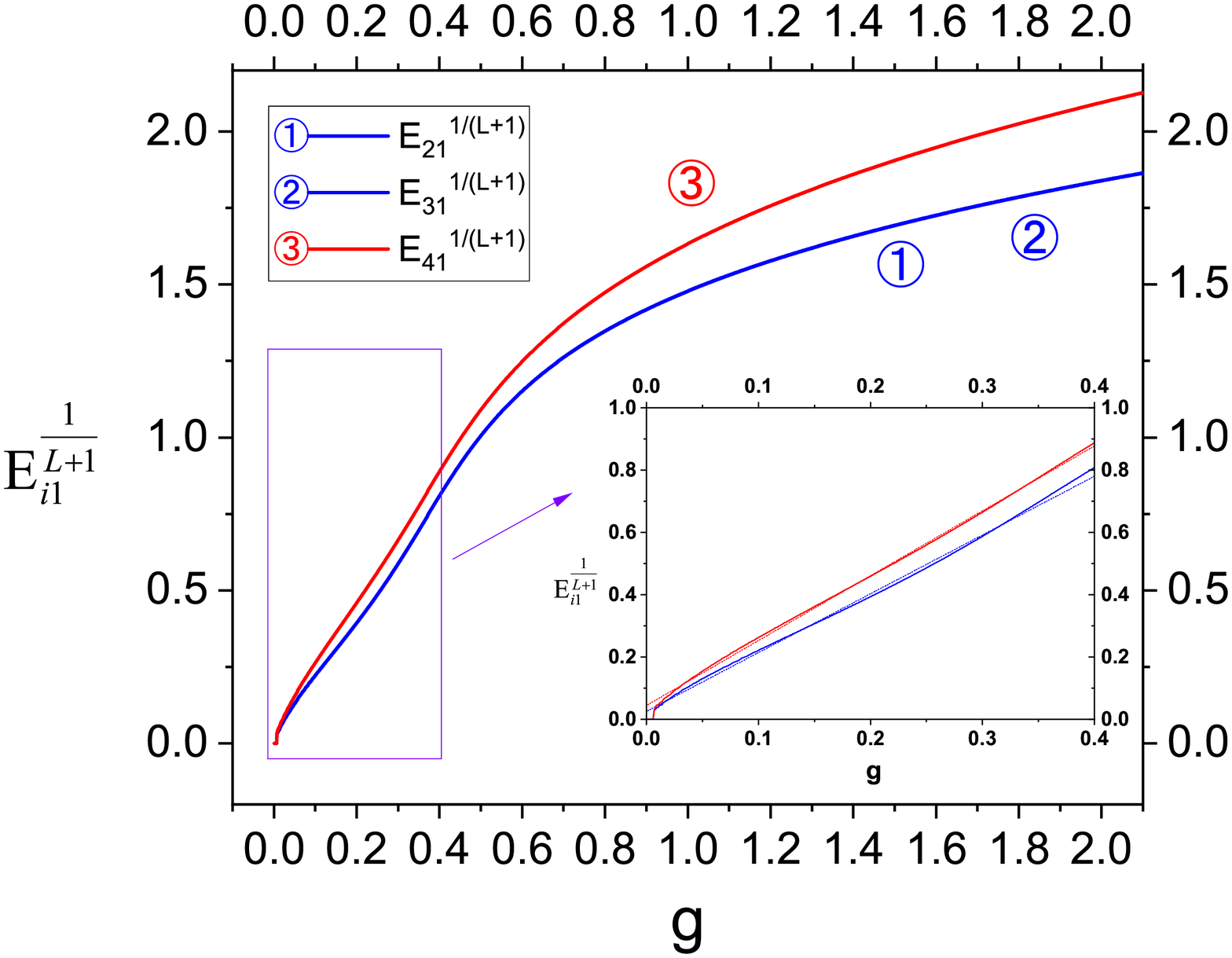}

\caption{(a) $E(V_x,V_y,V_z)$  on $d=3$ $2\times2\times2$ lattice.   $E_1=E(1,1,1)$, $E_2=E(-1,1,1)=E(1,-1,1)=E(1,1,-1)$, $E_3=E(-1,-1,1)=E(1,-1,-1)=E(-1,1,-1)$, $E_4=E(-1,-1,-1)$, as functions of $g$.   (b)  On $d=3$ $2\times2\times2$ lattice, $E_{i1}\equiv E_{i}-E_1$ as a function of $g$, $(i=2,3,4)$,
(c) $E(V_x,V_y)$ on $d=2$ $3\times 3$ lattice.  $E_1=E(1,,1)$, $E_2=E(-1,1)$, $E_3=E(1,-1)$, $E_4=E(-1,-1)$.       (d) On $d=2$ $3\times 3$ lattice, ${ E_{i1}  }^{ \frac{1}{L+1} } $,  $(i=2,3,4)$, as functions of $g$, where  $E_{i1}\equiv E_i-E_1$. }
\label{fig_torus}
\end{figure}

\section{Summary \label{summary} }

Implementing quantum adiabatic  algorithm in terms of  quantum circuit consisting of universal quantum gates of one or two qubits, we present  a digital  scheme of quantum adiabatic simulation of quantum  $\mathbb{Z}_2$ LGT, which is important in both high energy physics and condensed matter physics. Furthermore, we classically demonstrate this quantum simulation scheme by running the GPU  simulator QuEST on a Nvidia  GPU server, and obtain  results  useful in this field of physics.

In our algorithm, we have generalized Trotter decomposition and symmetrized Trotter decomposition such that the Hamiltonian adiabatically varies  during the decomposition.  Hence we have proposed a scheme of  digital adiabatic quantum simulation. Hopefully, this  method  can be used for various problems.

We have studied quantum  $\mathbb{Z}_2$ LGT  in lattices of spatial dimensions d=2 and d=3. Gauge invariance is realized in the initial state and is preserved during the adiabatic evolution. Since the lattices are very small,  singularities in QPT are rounded out. But key features   are  observed. It is indicated that QPT  is first-order in d=3  and is second-order in d=2, with critical point  consistent with previously known results. In d=3 and d=2 respectively, we also clearly observe  topological characteristics of QPT, including the vanishing of the ``magnetization'' for all values of the coupling constant, the excitation properties, the lowest energies in different topological sectors and the their splittings, which are proportional to $g^{L+1}$ in d=2. We have observed the change of the degeneracy caused by nonzero $g$, which is also an indication of topological order. To our knowledge, we have presented the unique numerical result on quantum  $\mathbb{Z}_2$ LGT  in 3 spatial dimensions.

This work  seems to be the first complete demonstration as a proof of principle, albeit in a classical simulator, carrying through  quantum simulation of a LGT  and observe the key features of  physics. Thereby  it  demonstrates that real quantum simulations of LGTs in future can be done in the proposed way.

On the other hand, this work also shows that high-performance classical demonstration  of quantum simulation,  which may be  dubbed  pseudoquantum simulation, represents a new way of computation, in addition to facilitating the development of quantum software for experimental quantum simulation.

As shown in the comparison with previous results from tensor network calculation,  it is a basic element of  our adiabatic approach  that the parameter in the Hamiltonian is varied, so it is very natural and convenient to obtain various quantities as functions of this parameter, which may not be  convenient in other methods,  hence adiabatic quantum simulation and pseudoquantum simulation are  convenient tools  of QPT  study.

The classical demonstration proceeds according to rules of quantum mechanics, therefore pseudoquantum simulation is legitimately a reliable approach, as far as computational resources allow. Compared with other computational approaches, it can be used without intricate algorithmic design depending on the details of the computed problem, as in many other numerical methods, which are often applicable only to one or two dimensions.

As the next step regarding quantum and pseudoquantum simulations of LGT, we shall study Fermions coupled with the $\mathbb{Z}_2$  gauge field using our method, on which the results can be  compared  with the existing results of QMC calculations, which are free of Fermion sign problem. Thereby the method can be further benchmarked, which can then be applied to other LGTs, for which Fermion sign problem exists in MC-based methods, as well as the LGTs that recently have been studied by using tensor network methods~\cite{Tagliacozzo2,Celi2}.

After the release of  the present work  as a preprint (arXiv:1910.08020),  there appeared more recent progress in the related fields, including DMRG study of the one-dimensional spinless Fermions coupled with $\mathbb{Z}_2$  gauge field~\cite{Borla},    schemes of measuring   nonlocal observables in the quantum simulation of a general LGT~\cite{Zohar4},  experimental observation of gauge invariance in a 71-site quantum simulator of an extended U(1) LGT~\cite{Yang}, and a sign-free QMC study of $\mathbb{Z}_2$  gauge field coupled with both Fermions and Bosons, demonstrating the transition from conventional metal to orthogonal metal~\cite{Qi}.

\begin{acknowledgments}
This work was supported by National Science Foundation of China (Grant No. 11574054).
\end{acknowledgments}

\end{document}